\documentclass[a4paper,fleqn,usenatbib,useAMS]{mnras}

\usepackage{graphicx}   
\usepackage{color}
\usepackage{lscape}
\usepackage{txfonts}
\usepackage{amsfonts}

\def\bfnabla{{\mbox{\boldmath $\bfnabla$}}}

\def\mbh{M_{\rm{BH}}}

\newcommand\bv{{\mbox{\boldmath $v$}}}

\newcommand\bn{{\mbox{\boldmath $n$}}}
\newcommand\bH{{\mbox{\boldmath $H$}}}

\newcommand\bF{{\mbox{\boldmath $F$}}}

\def\<{\,\langle\langle}
\def\>{\,\rangle\rangle}

\title[Early evolution of super-Eddington accretion flow in TDEs]
{Early evolution of super-Eddington accretion flow in tidal disruption events}
\author[Erlin Qiao et al.]{Erlin Qiao $^{1,2}$\thanks{E-mail: qiaoel@nao.cas.cn}, Yongxin Wu$^{1,2}$, Yiyang Lin$^{1,2}$, 
Meng Guo$^{3,4}$\thanks{E-mail: guomeng@sdas.org}, Jifeng Liu$^{1,2}$\thanks{E-mail: jfliu@nao.cas.cn}, 
\newauthor
Chenlei Guo$^{1,2}$,
Chichuan Jin$^{1,2,7}$, 
and Ning Jiang$^{5,6}$ \\
$^{1}$National Astronomical Observatories, Chinese Academy of
Sciences, Beijing 100101, China \\
$^{2}$School of Astronomy and
Space Sciences, University of Chinese Academy of Sciences, 19A Yuquan Road, Beijing 100049, China\\
$^{3}$Key Laboratory of Computing Power Network and Information Security, Ministry of Education, Shandong Computer 
Science Center (National \\
Supercomputing Center in Jinan), Qilu University of Technology (Shandong Academy of Sciences), Jinan, Shandong 250013, China\\
$^{4}$Jinan Institute of Supercomputing Technology, Jinan, Shandong 250103, China \\
$^{5}$CAS Key Laboratory for Research in Galaxies and Cosmology, Department of Astronomy, University of Science and Technology of China, Hefei, \\
Anhui 230026, China\\
$^{6}$School of Astronomy and Space Science, University of Science and Technology of China, Hefei 230026, China\\
$^{7}$Institute for Frontier in Astronomy and Astrophysics, Beijing Normal University, Beijing 102206, China
}

\date{Accepted XXX. Received YYY; in original form ZZZ}
\pubyear{2023}

\begin{document}
\label{firstpage}
\pagerange{\pageref{firstpage}--\pageref{lastpage}}
\maketitle

\begin{abstract}
Tidal disruption events (TDEs) are luminous black hole (BH) transient sources, which are detected mainly in X-ray and optical bands. 
It is generally believed that the X-ray emission in TDEs is produced by an accretion disc formed as the stellar debris accreted onto the 
central BH. The origin of the optical emission is not determined, but could be explained by the `reprocessing' model with the X-ray emission  
reprocessed into optical band by a surrounding optically thick envelope or outflow. 
In this paper, we performed radiation hydrodynamic simulations of super-Eddington accretion flow with Athena++ code in the environment of TDEs, i.e., injecting 
a continuous mass flow rate  at the circularization radius in the form of $\dot M_{\rm inject} \propto t^{-5/3}$ for the mass supply rate.  We show that a significant 
fraction of the matter in the accretion inflow are blowed off forming outflow, and the properties of the outflow are viewing-angle dependent. 
We further calculate the emergent spectra of such an inflow/outflow system for different viewing angles
with the method of Monto Carlo radiative transfer. Based on the emergent spectra, we show that the observed features of TDEs, 
such as the X-ray and optical luminosities, the blackbody temperature of X-ray and optical emission and the corresponding emission radii, 
the ratio of X-ray luminosity to optical luminosity, as well as the evolution of these quantities can be explained in the framework 
of viewing-angle effect of super-Eddington accretion around a BH. 
\end{abstract}



\section{Introduction}\label{Introduction}
A star can be tidally disrupted by a supermassive black hole (BH) in the center of a galaxy as the star approaches to  
a critical radius from the BH, which has been investigated in the astrophysical context since 1970s \citep[][]{Hills1975,Frank1976,Gurzadian1979,Carter1982,Luminet1985,Hills1988,Rees1988}.
The critical radius is the so-called tidal disruption radius, which can be expressed as $R_{\rm T}\approx R_{*}(M/M_{*})^{1/3}$
(with $R_{*}$ being the radius of the star, $M$ being the mass of the BH, and $M_{*}$ being the mass of the star) \citep[][]{Hills1975}. 
TDEs can produce an observable flare only if the tidal disruption radius $R_{\rm T}$ is greater than the radius of the event horizon of the BH.
After the disruption of the star, in general, half of the stellar debris will escape from the BH, and the other half of the stellar debris will be bound and fallback towards the BH.
The fallback rate is derived to be super-Eddington initially for a typical BH mass of $\sim 10^{6-7}M_{\odot}$, and decays with a form of `$t^{-5/3}$' assuming a flat orbital 
energy distribution among the bound debris \citep[][for review]{Rees1988,Evans1989,Phinney1989,Rossi2021}.

The flare is expected in soft X-rays as predicted from the accretion disk around a BH with a mass of $\sim 10^{6-7} M_{\odot}$ at a super-Eddington accretion 
rate \citep[e.g.][]{Ulmer1999,Miller2015}. The observational evidence for the presence of TDEs is from the $\emph ROSAT$ all-sky survey in soft X-rays (0.5-4 keV) 
of 1990–1991, including four main sources NGC 5905, RXJ1242-1119, RXJ1624+7554, and RXJ1420+5334 \citep[][]{Bade1996,Komossa1999b,Komossa1999a,Grupe1999}. 
The decline of the long-term X-ray light curves is well consistent with 
the decay law of `$t^{-5/3}$', which are first reported based on the ROSAT data of NGC5905 \citep[][]{Komossa1999b}, and later complemented
by combing the ROSAT data of NGC 5905, RXJ1242-1119, RXJ1624+7554, and RXJ1420+5334 \citep[][for review]{Komossa2004IAUS}.
Combing the TDEs detected by $ROSAT$, and the TDEs detected later by XMM-{\emph Newton} and $Chandra$,  currently, the number of X-ray detected TDEs
roughly twenty. The main properties of X-ray detected TDEs are summarized several excellent reviews \citep[e.g.][]{Komossa2015,Saxton2021SSRv}. 

The flares are also detected in optical band \citep[][for review]{Gezari2006,Gezari2008,Gezari2009,Komossa_etal2008,vanVelzen2011a,Gezari2012,Holoien2014,Arcavi2014,
Chornock2014,Vinko2015,Holoien2016,vanVelzen2020SSRv}. Actually, almost two thirds of the reported TDEs to date are detected in wide-field optical 
surveys \citep[][for review]{Gezari2021ARAA}.
The origin of the optical emission in TDEs is still in debate, which could be powered by the shock induced by the collision of the debris at  the self-intersection radius
from the BH \citep[e.g.][]{Piran2015,Jiang2016,Bonnerot2017,Steinberg2024}, or could be produced by the ‘reprocessing’ model with the 
accretion produced X-ray emission reprocessed into optical band by a surrounding optically thick envelope or disk
outflow \citep[e.g.][]{Loeb1997,Strubbe2009,Coughlin2014,Lodato2011,Roth2016,Metzger2016,Dai2018,Curd2019,Metzger2022}.

\begin{figure*}
\includegraphics[width=177mm,height=84.9mm,angle=0.0]{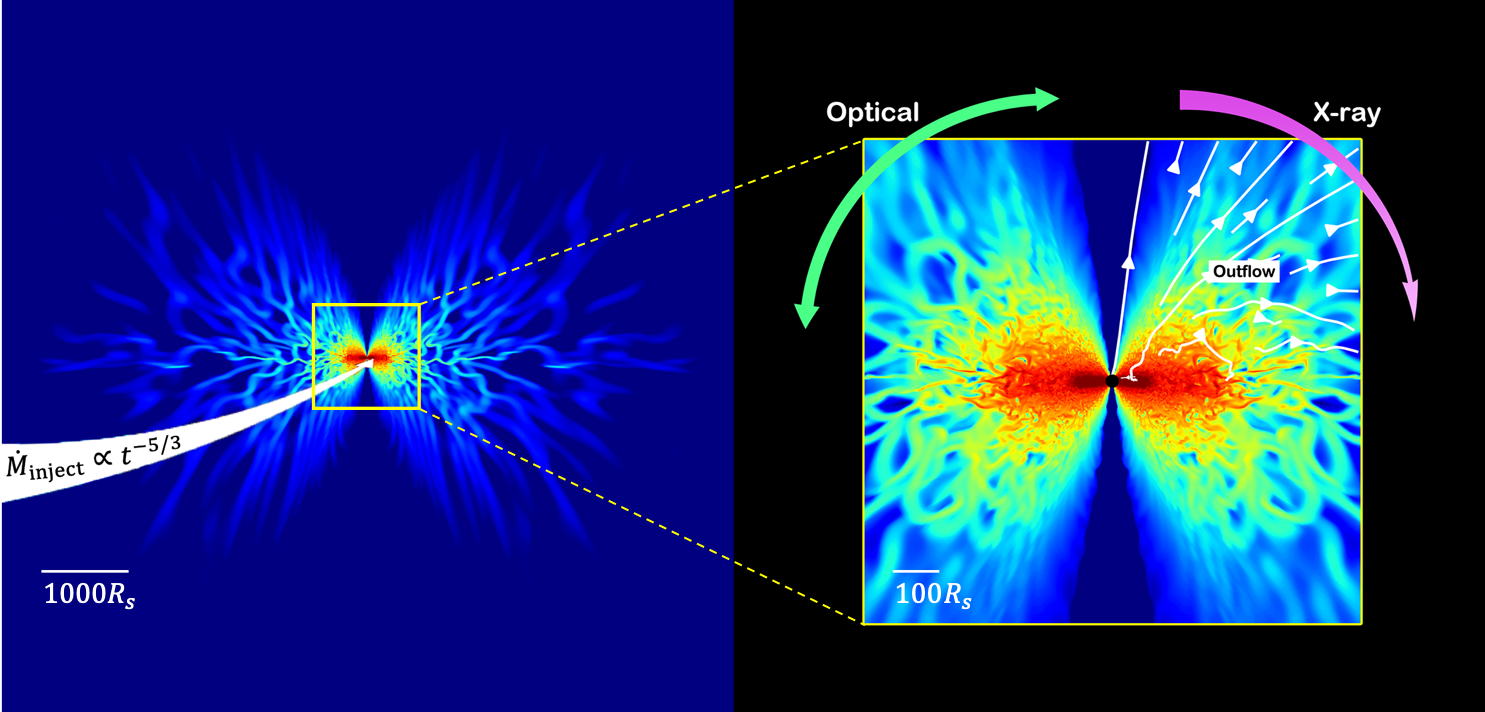}
\caption{\label{f:illustration} Left panel: An panoramic snapshot of the inflow/outflow for super-Eddington accretion around a BH in the environment of a TDE with a continuous injection 
rate of $\dot M_{\rm inject} \propto t^{-5/3}$ at the circularization radius. The snapshot is based on the real numerical simulations, in which BH mass is set to be ${10^6M_{\odot}}$
and the circularization radius is set to be $\sim 47$ Schwarzschild radius (see the last two paragraphs of Section \ref{method:simulation} for the details of the simulation setup).
In the snapshot, gas has filled the region between the circularization radius and the BH, and the stable accretion is ongoing. 
The different colors represent the different gas densities, which decrease gradually from dark red, red, orange, yellow, cyan to blue.
Right panel: An inset of the inflow/outflow region.
The white arrows are the streamlines indicating the inflow and the outflow of the accretion.
The pink arc-shaped arrow in the top-right corner indicates that the observed X-ray luminosity decreases with increasing the viewing angle. 
The green arc-shaped arrow in the top-left corner indicates that the observed optical luminosity is roughly isotropic with different viewing angles.
}
\end{figure*}

Since the pioneering paper of \citet[][]{Rees1988}, the feature of `$t^{-5/3}$'  of the light curve has been suggested to be one of the 
most prominent criterions for searching for TDEs for a long time \citep[][for review]{Komossa2015,Saxton2021SSRv,vanVelzen2020SSRv}. 
However, the theoretically derived `$t^{-5/3}$' refers to the evolution of the fallback rate, which may not be equivalent to the light curve in 
some specific band \citep[e.g.][]{Lodato2009,Lodato2011}. The physics of TDEs is very complicated. 
As far as the accretion process is concerned, TDE is an ideal laboratory to study the evolution of the accretion flow around a supermassive BH.
In order to more realistically simulate the evolution of the accretion flow around a supermassive BH in a TDE, in this paper, we conduct hydrodynamic simulations of 
super-Eddington accretion flow with Athena++ code, injecting a mass flow rate at the circularization radius in the form of $\dot M_{\rm inject} \propto t^{-5/3}$ as the 
mass supply rates. The simulation is performed until 32 day since injecting mass at the circularization radius.
We show that a significant fraction of the matter in the accretion flow are blowed off forming the outflow, and the properties of the 
outflow are viewing-angle dependent, as has been shown in many 
hydrodynamic simulations or magnetohydrodynamic simulations of super-Eddington accretion flow around either a stellar-mass BH or a super-massive 
BH \citep[e.g.][]{Ohsuga2005,Ohsuga2011,Jiang2014ApJ...796..106J,McKinney2014,Sadowski2014,Jiang2019,Dai2018,Curd2019,Thomsen2022}.
An illustration of our scenario for the accretion flow in a TDE is shown in Fig. \ref {f:illustration}.
We then post-process the simulation data for the emergent spectra for different viewing angles with the Monte Carlo radiation transfer code PYTHON.
Based on the emergent spectra, we show that the observed 
X-ray and optical luminosity, the blackbody temperature and radius of X-ray and optical emission, the ratio of X-ray luminosity to 
optical luminosity, as well as the evolution of these quantities of TDEs at the early phase can be explained in the framework of viewing-angle effect of 
super-Eddington accretion around a BH.

This paper is organized as follows. In Section \ref{Numerical method}, we present the numerical method, including, in section \ref{method:simulation}, the setup for the radiation hydrodynamic simulations of
super-Eddington accretion flow in TDE environment with Athena++ code, and in section \ref{subsec:Monte Carlo}, the setup for the Monte Carlo radiative transfer for the emergent spectra. 
The results are presented in Section \ref{Results}, summary and discussion are presented in Section \ref{Summary and Discussion}.

\section{Numerical method}\label{Numerical method}
By using the Athena++ code \citep[][]{Stone2020,athena21}, we perform two-dimensional axisymmetric radiation hydrodynamic simulations of super-Eddington
accretion around a  BH in the environment of a TDE. In the simulation, we adopt spherical coordinates ($r$, $\theta$, $\phi$).
One can refer to Section \ref{method:simulation} for the 
simulation setup. Based on the simulation data, we then calculate the corresponding emergent spectra of the inflow/outflow system of 
super-Eddington accretion with the method of Monte Carlo radiative transfer, which is summarized in Section \ref{subsec:Monte Carlo}.

\subsection{Simulation setup }\label{method:simulation}
The equations of hydrodynamics coupled with the time dependent radiative transfer equations we solve are \footnote{ One can also 
refer to \citet[][]{Jiang2014ApJ...796..106J,2021ApJS..253...49J} for the equations and the numerical algorithm.},
\begin{eqnarray}
\frac{\partial\rho}{\partial t}+\nabla\cdot(\rho \bv)&=&0, \nonumber \\
\frac{\partial(\rho \textbf{\textit{v}})}{\partial t}+\nabla\cdot(\rho \textbf{\textit{v}} \textbf{\textit{v}}+\mathsf{P}^{*})&=&-\textbf{\textit{S}}_{\textbf{\textit{r}}}(\textbf{\textit{P}})-\rho\nabla\phi+\nabla\cdot\mathsf{T},\  \nonumber \\
\frac{\partial{E}}{\partial t}+\nabla\cdot\left[(E+P^{\ast})\bv-\sf T \cdot \bv\right]&=&-cS_r(E)-\rho\bv\cdot\nabla\phi.
\label{MHDEquation}
\end{eqnarray}
\begin{eqnarray}
 \frac{\partial I}{\partial t}+c\bn\cdot\nabla I&=&c\sigma_a\left(\frac{a_rT^4}{4\pi}-I\right)+c\sigma_s(J-I) \nonumber \\
 &+&3\bn\cdot\bv\sigma_a\left(\frac{a_rT^4}{4\pi}-J\right)
 \nonumber \\
 &+&\bn\cdot\bv(\sigma_a+\sigma_s)\left(I+3J \right)
 -2\sigma_s\bv\cdot\bH \nonumber \\
 &-&(\sigma_a-\sigma_s)\frac{\bv\cdot\bv}{c}J- (\sigma_a-\sigma_s)\frac{\bv\cdot(\bv\cdot{\sf K})}{c}.\nonumber \\
 \label{RTequation}
 \end{eqnarray}
 \begin{eqnarray}\label{Radsource}
S_r(E)&=&\sigma_a\left(a_rT^4-E_r\right)\nonumber \\
&+&\left(\sigma_a-\sigma_s\right)\frac{\bv}{c^2}\cdot\left[
\bF_r-\left(\bv E_r+\bv\cdot{\sf P_r}\right)\right], \nonumber\\
\textbf{\textit{S}}_{\textbf{\textit {r}}}(\textbf{\textit{P}})&=&-\frac{(\sigma_{s}+\sigma_{a})}{c}\left[
\bF_r-\left(\bv E_r+\bv\cdot{\sf P_r}\right)\right] \nonumber\\
&+&\frac{\bv}{c}\sigma_a\left(a_rT^4-E_r\right).
\end{eqnarray}
Here, $\rho$ and $\bv$ are density and velocity of the accretion flow, ${\sf P}^{\ast}\equiv P_{g}{\sf I}$ (with ${\sf I}$
the unit tensor), $P_{g}$ is the gas pressure. $\sf T$ is the anomalous stress
tensor, which is responsible for angular momentum transfer. 
In this paper, we only consider $r\phi$ component of the stress tensor, i.e., $\sf T_{r\phi}$=$\eta r {\partial\over \partial r} {v_{\phi}\over r}$,
with $\eta$ being the dynamical viscosity parameter. 

The total gas energy density is
\begin{eqnarray}
E=E_g+\frac{1}{2}\rho v^2,
\end{eqnarray}
where $E_g$ is the internal gas energy density.   We adopt an equation of state
for an ideal gas with adiabatic index $\gamma=5/3$, thus
$E_g=P_{g}/(\gamma-1)$ for $\gamma\neq 1$ and gas temperature 
$T=P_{g}\mu/R_{ideal}\rho$, where 
$R_{ideal}$ is the ideal gas constant. Mean molecular weight $\mu$ is assume to be $0.6$. The radiation momentum and energy source terms are 
$\textbf{\textit{S}}_{\textbf{\textit{r}}}(\textbf{\textit{P}}),\ S_r(E)$ 
and $I$ is the specific intensity along the direction with unit vector $\bn$. 
$\sigma_a=\sigma_{ff}+\sigma_{bf}$ is the absorption opacities, with $\sigma_{ff}=1.7\times 10^{-25}T^{-7/2}({\rho\over m_{p}})^2\ {\rm cm^{-1}}$ being the free-free absorption opacity
and $\sigma_{bf}=4.8\times 10^{-24}T^{-7/2}({\rho\over m_{p}})^2({Z\over Z_{\odot}})\ {\rm cm^{-1}}$
being bound-free absorption opacity respectively, where $m_{p}$ is the proton mass, $\rm Z$ is the metallicity and $\rm Z_{\odot}$ is the solar metallicity respectively.
$\sigma_s=0.34\rho\ {\rm cm^{-1}}$ is the scattering opacities, with the solar chemical abundance assumed.  
$a_r$ is the radiation constants and $c$ is the speed of light. The radiation energy density $E_r$, radiation flux $\bF_r$ and radiation 
pressure tensor ${\sf P_r}$ are defined through the angular quadrature of the specific intensity as,
\begin{eqnarray} \label{integrateangle}
 J&\equiv& \int Id\Omega,\nonumber \\ 
 \bH&\equiv& \int \bn Id \Omega, \nonumber \\ 
 {\sf K}&\equiv& \int \bn\bn I d\Omega, \nonumber \\
 E_r&=&4\pi J, \ \bF_r=4\pi c\bH, \ {\sf P_r}=4\pi{\sf K}.
\end{eqnarray}
The source terms in equation (\ref{Radsource}) are written in mixed frames. Specifically, radiation is described in the Eulerian (lab) frame
and the opacities are calculated in the fluid (comoving) frame \citep[][]{Lowrie1999,Jiang2012,Jiang2014ApJ...796..106J}.
The fluid frame radiation flux $\bF_{r,0}$ is related to Eulerian frame flux $\bF_r$ as $\bF_{r,0}=\bF_r-\left(\bv E_r+\bv\cdot{\sf P_r}\right)$.
The radiation pressure tensor $\sf P_{r}$ is directly calculated with expression in equation (\ref{integrateangle}) without 
need any assumption or independent equation to calculate the Eddington tensor.
With $\sf P_{r}$, we can directly calculate dynamical viscosity parameter $\eta$, which can be expressed as 
$\eta=\alpha {{p_{g}+{\sf P_{r,rr}(\sigma_{a}+\sigma_{s}) {\vert{\nabla} r}\vert} \over {v_{\phi}/r}}}$.
In this paper, $\alpha=0.1$ is assumed in the simulations.

A pseudo-Newtonian potential is adopted to mimic the effects of general relativity around a Schwarzschild BH, i.e., 
\begin{eqnarray}
\phi=-\frac{G\mbh}{r-R_{\rm S}},
\end{eqnarray}
where $G$ is the gravitational constant, $\mbh$ is the BH mass, $r$ is the distance to the BH and 
$R_{\rm S}\equiv 2G\mbh/c^2$ is the Schwarzschild radius. 

In this paper, we consider the TDE environment for the simulations. Specifically, we start the simulation by continuously injecting a mass rate with the form 
of $\dot M_{\rm inject}(t)=\dot M_{\rm fb}(t)={1\over 3}(M_{*}/t_{\rm fb})(1+t/t_{\rm fb})^{-5/3}$ at the circularization $R_{\rm c}$ along the $\phi$ direction (azimuth direction) 
with the local Keplerian velocity, where $\dot M_{\rm fb}$ is
the fallback rate, and $t_{\rm fb}\approx 41.3({M_{\rm BH}\over 10^{6}M_{\odot}})^{1/2}({M_{*}\over M_{\odot}})^{-1}({R_{*}\over R_{\odot}})^{3/2} \rm \ days $ is the 
fallback timescale of the most bounded debris. $M_{*}$  and $R_{*}$ are the mass and the radius of the disrupted star, which are scaled with solar mass  
$M_{\odot}$ and radius $R_{\odot}$ respectively. Here  $t_{\rm fb}$ is derived by assuming the parabolic orbit of the star, $R_{\rm T}/R_{\rm p}=1$ (with $R_{\rm p}$ being the 
pericenter of the star orbit), and flat distribution of the orbit energy among the debris. The circularization radius is derived as $R_{\rm c}=2R_{\rm T}$ according to 
the angular momentum conservation. 

We fix $M_{\rm BH}=10^{6}M_{\odot}$, $M_{*}=M_{\odot}$ and $R_{*}=R_{\odot}$, with which the initial injecting mass
accretion rate ${1\over 3}(M_{*}/t_{\rm fb})$ is calculated to be $\sim 133.8 \dot M_{\rm Edd}$ (with $\dot M_{\rm Edd}$ = $1.39 \times 10^{18} M/M_{\rm \odot} \rm \ g s^{-1}$),
and $R_{\rm c}$ is calculated to be $\sim 47R_{\rm S}$. 
In the simulation, we inject gas in the region $45R_{\rm S} \leqslant r \leqslant 49R_{\rm S}$ and $\frac{9}{20}\pi \leqslant \theta \leqslant \frac{11}{20}\pi$.
Our simulation has a computational domain of $2R_{\rm S} \leqslant r \leqslant 10^5 R_{\rm S}$ in the radial direction, and 
$0 \leqslant \theta \leqslant \pi$ in the $\theta$ direction. The computational resolution is $N_r \times N_{\theta} = 768 \times 256$.
Specifically, in the region $2R_{\rm S} \leqslant r \leqslant 3R_{\rm S}$, we set 16 uniformly spaced cells. We set 240 logarithmically spaced cells in the 
region $3R_{\rm S} \leqslant r \leqslant 50R_{\rm S}$. In the region $50R_{\rm S} \leqslant r \leqslant 200R_{\rm S}$, we set 128 logarithmically spaced cells, 
and in the region $200R_{\rm S} \leqslant r \leqslant 93405R_{\rm S}$, we set 380 logarithmically spaced cells. Finally, in the 
region $93405R_{\rm S} \leqslant r \leqslant 10^5 R_{\rm S}$, we set 4 uniformly spaced cells. 
In the $\theta$ direction, we set 256 non-uniformly spaced cells. The size of the cells is set to be smaller close to the mid-plane and the distribution of the cells 
is symmetric to the mid-plane. Outflow boundary conditions are considered at inner and outer radial boundary, while at $\theta =0$  and at $\theta = \pi$ we use the reflecting 
boundary conditions. In this paper, the simulation is performed until $t=32$ day since injecting mass at the circularization radius.

\subsection{Monte Carlo radiative transfer setup}\label{subsec:Monte Carlo}
We calculate the emergent spectrum of the super-Eddington accretion flow (inflow/outflow) for different viewing angles $\theta$
with the Monte Carlo radiation transfer code PYTHON. 
PYTHON code is first designed to solve radiation transfer in winds emitted from accretion disks of cataclysmic variables \citep[][]{Long2002},
and is developed to enable its application to active galactic nuclei \citep[e.g.][]{Sim2010MNRAS.404.1369S,Sim2010MNRAS.408.1396S,Higginbottom2013,Matthews2016,Matthews2017}.
PYTHON code can trace the photon propagation in 3D, in the present paper, however, the gas density, temperature and the velocity are based on
the simulation data, which are symmetric in the azimuth direction.

In order to get a self-consistent spectrum of the inflow/outflow system, the gas temperature and radiation temperature are recalculated with the method of Monte Carlo calculation 
under the assumption of radiative equilibrium. In PYTHON code, an initial gas temperature (based on the simulation data) is input for determining the ionization 
state of the gas,
and then the gas temperature is solved 
iteratively until a stable state is reached. In the Monte Carlo simulation, a set of photon bundles over a wide frequency are generated and propagated until 
leave the computational domain or absorbed by the gas. The Compton scattering, free-free absorption, bound-free absorption and bound-bound absorption 
processed are included in the radiation transfer calculation. 

We select the radius where the $\theta$-direction averaged velocity field transits from outflow 
to inflow as the inner boundary of the computational domain, which is found to be just the gas injection point in the simulation (i.e., $47R_{\rm S}$).
The outer boundary is set at 5000$R_{\rm S}$, much larger than the electron scattering photosphere. 
For each cell in the domain, the location and size are same as the cell setup in the hydrodynamic simulation.
The gas comprises of H, He and O with solar abundance. The photons injected from the inner boundary in every iteration is assumed to a 
single blackbody with a temperature of  a few $\times 10^5 \rm K$, which is calculated by averaging the radiation temperature 
of the hydrodynamic simulation.

\section{Results}\label{Results}
\subsection{Results of radiation hydrodynamic simulations}\label{subsec:simulations}
\begin{figure*}
\includegraphics[width=65mm,height=86.3mm,angle=0.0]{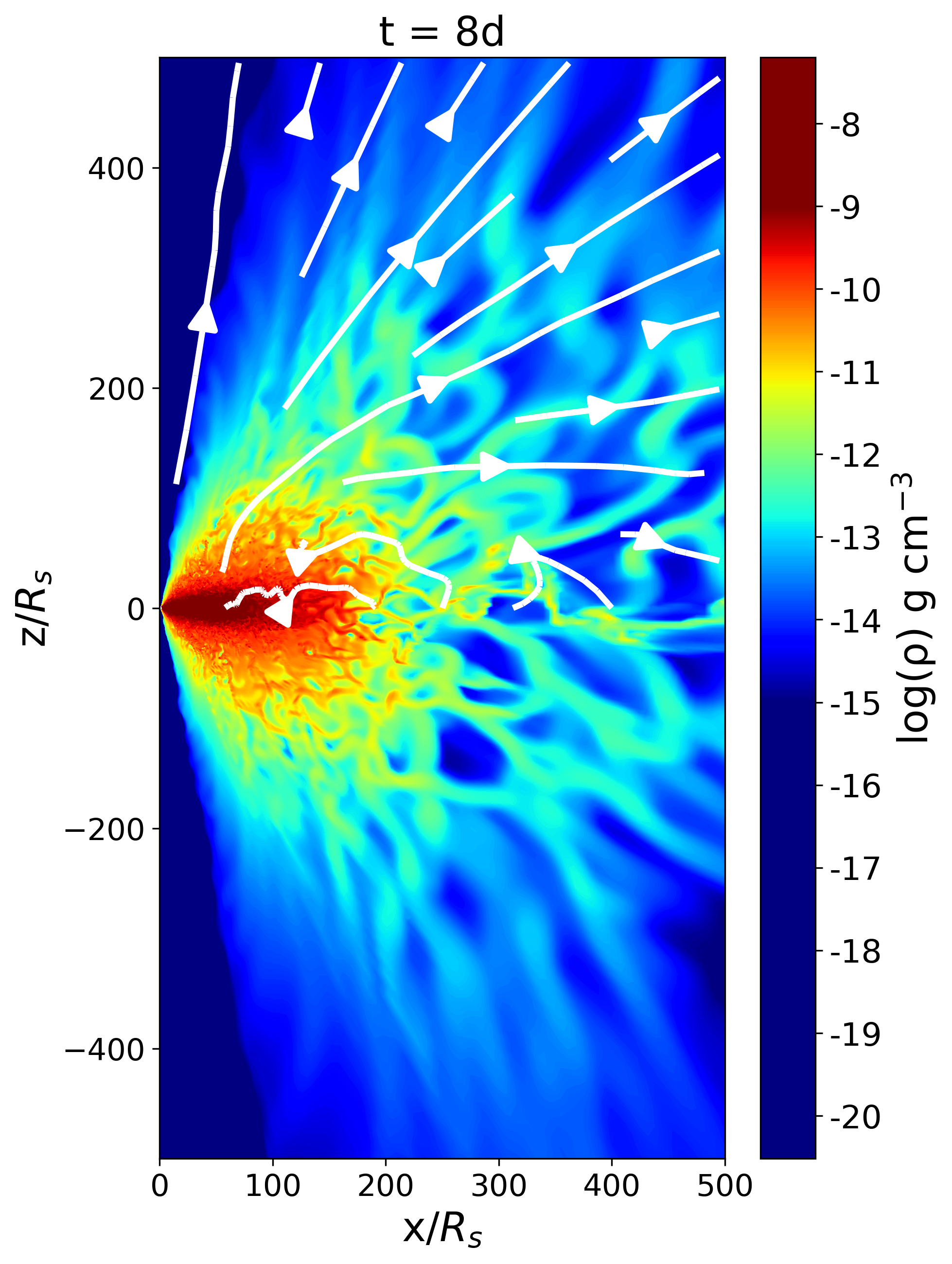}
\includegraphics[width=65mm,height=86.3mm,angle=0.0]{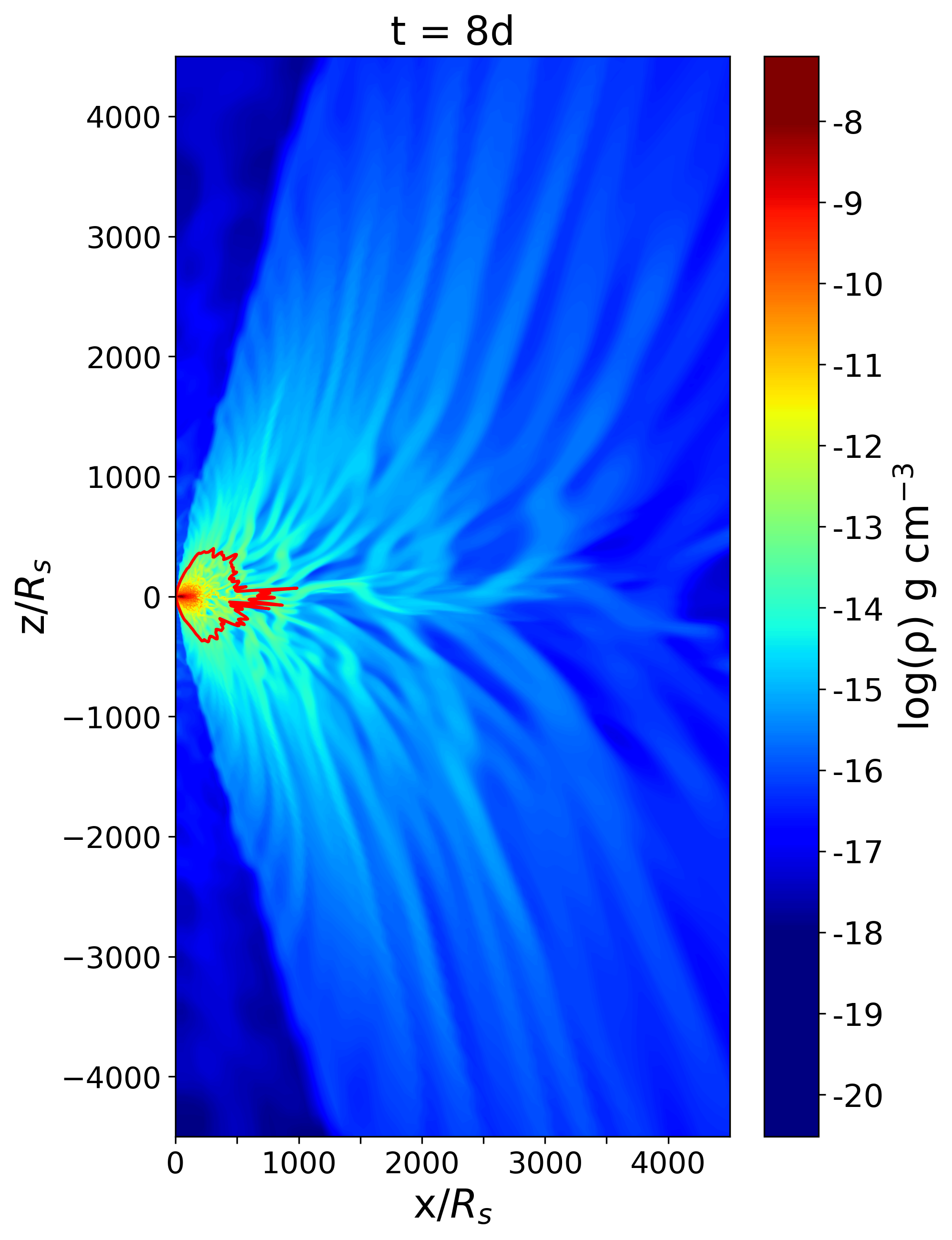}
\caption{\label{f:8-1-density}  Snapshots of gas density (in $r-\theta$ plane) at $t=8$ day since the injection of matter at the circularization radius. 
Left panel:  A zoom-in snapshot of the  gas density. The white arrows indicates the streamlines of the gas velocity. Right panel:  A zoom-out snapshot of the  gas density.
The red thick line indicates the photosphere of electron scattering.}
\end{figure*}

In Fig. \ref{f:8-1-density}, we plot a snapshot of gas density at $t=8$ day since the injection of matter at the circularization radius.
Here, we take $t=8$ day as an example since at $t=8$ day the injected gas has securely filled the region between the circularization radius
and the BH, and the stable accretion is ongoing. In the left panel of Fig. \ref{f:8-1-density}, we present a zoom-in snapshot of the gas density at $t=8$ day, in which 
the gas density decreases gradually from the color of dark red, red, orange, yellow, cyan to blue. It can be seen that the gas densities are viewing angle $\theta$ dependent.
The white arrows indicate the streamlines of the gas velocity in $r-\theta$ plane. 

In the right panel of Fig. \ref{f:8-1-density}, we plot a zoom-out snapshot of the gas density for clearly showing the overall picture of the accretion flow,
in which the photosphere of the electron scattering is added. The radius of the photosphere of the electron scattering $R_{\rm ph}$ for each $\theta$ is calculated by setting the 
scattering optical depth $\tau_{\rm es}=-\int_{\infty}^{R_{\rm ph}} 0.34\rho \ dr=1$. One can see the red thick line in the right panel of Fig. \ref{f:8-1-density} for clarity.
We define a critical angle $\theta_{\rm crit}$, where the optical depth of the electron scattering $\tau_{\rm es}$ is just 1 with increasing $\theta$.
$\theta_{\rm crit}$ is calculated to be $\sim 15^{\circ}$ in our simulation. For $\theta<\theta_{\rm crit}$, there is no photosphere found.
For $\theta>\theta_{\rm crit}$, the outflow gradually becomes optically thick, and the photosphere is formed.
As we can see in the left panel of Fig. \ref{f:8-1-density}, when $\theta<\theta_{\rm crit}$, the density of the outflow is very low, roughly in the range of $10^{-15}$ to $10^{-18}\rm g\ cm^{-3}$. 
The density of the outflow gradually increases as  $\theta>\theta_{\rm crit}$, and reaches a maximum value of $\sim 10^{-8} \rm g\ cm^{-3}$ at $\theta=90^{\circ}$. 

\begin{figure*}
\includegraphics[width=65mm,height=86.3mm,angle=0.0]{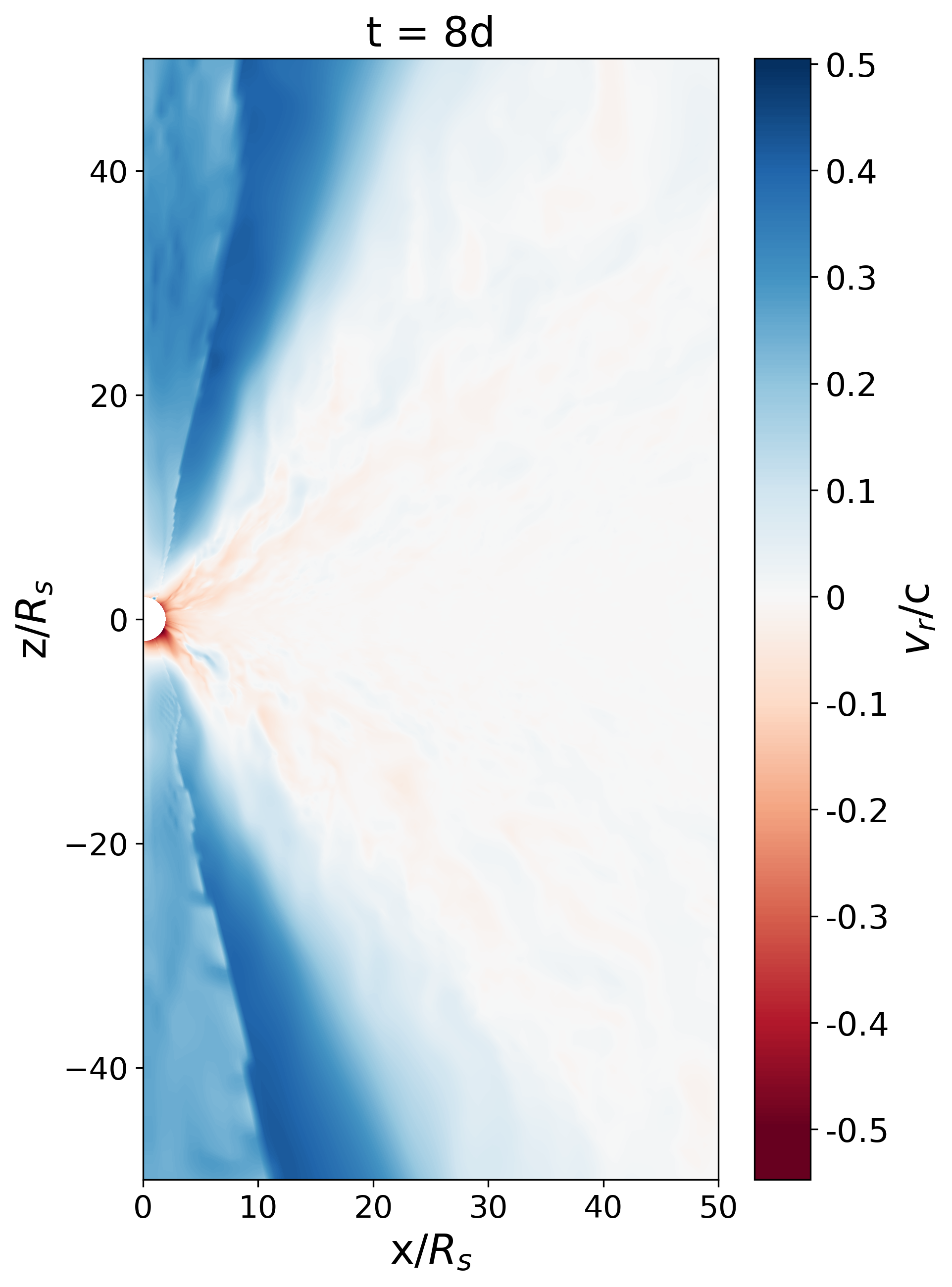}
\includegraphics[width=65mm,height=86.3mm,angle=0.0]{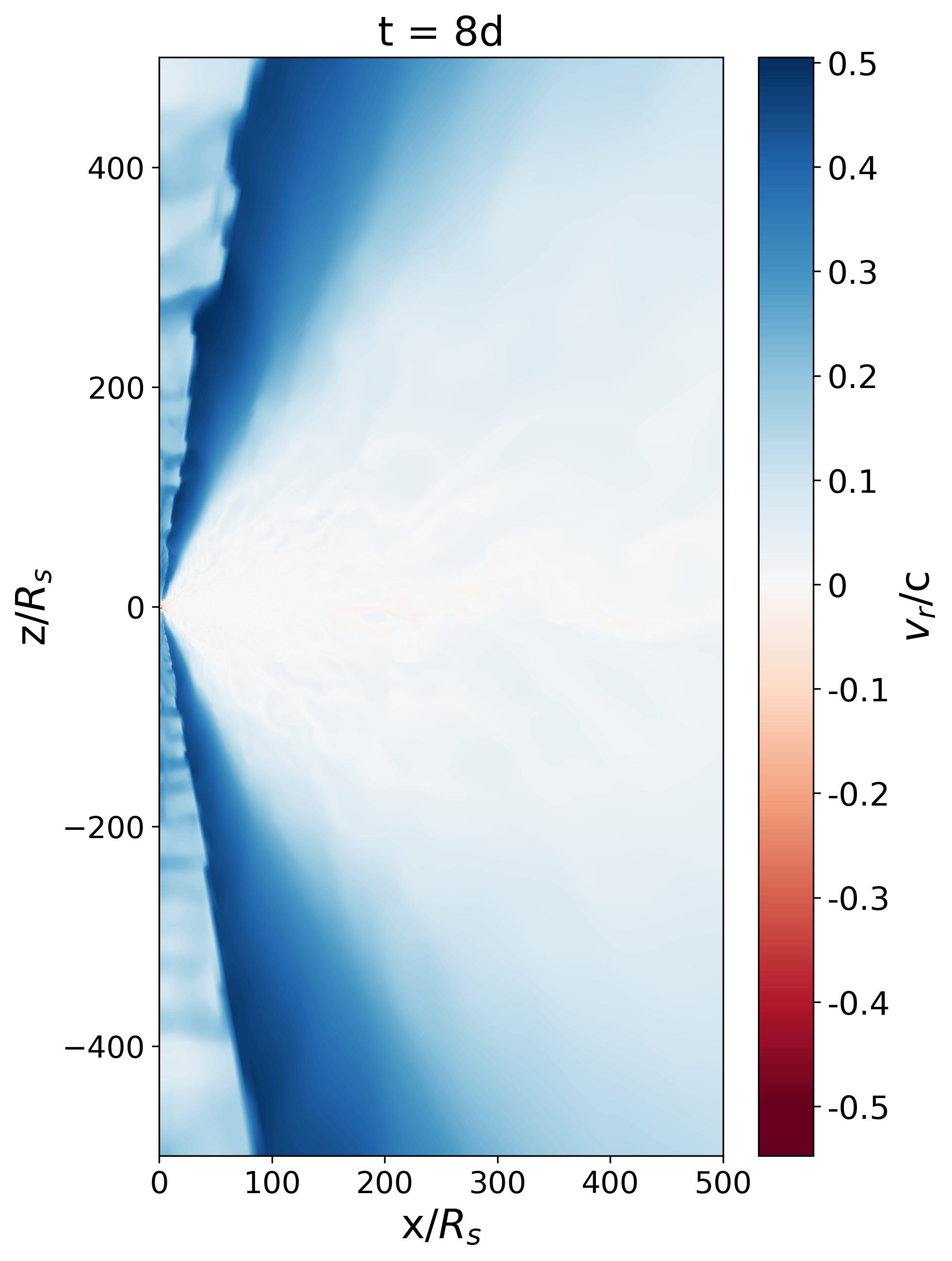}
\caption{\label{f:8-1-velocity}  Snapshots of gas velocity in the radial direction  (in $r-\theta$ plane) at $t=8$ day since the injection of matter at the circularization radius. $v_{r}/c>0$ 
indicates outflow and $v_{r}/c<0$ indicates inflow. Left panel:  A zoom-in snapshot of gas velocity in the radial direction. Right panel:  A zoom-out snapshot of gas velocity in the radial direction.}
\end{figure*}

In order to more clearly show the distribution of the gas velocity, in particular the gas velocity in the radial direction, we plot a snapshot of 
gas velocity in the radial direction at $t=8$ day in the left panel of Fig. \ref{f:8-1-velocity}. As we can see, the gas velocity in the radial direction $v_{r}/c$ (scaled with
speed of light) is also $\theta$ dependent. Specifically, $v_{r}/c$ reaches maximum values of $\sim 0.4-0.5$ in the region between $\theta \sim 15^{\circ}-30^{\circ}$,
and gradually decreases to $\sim 0.1$ or even lower with increasing $\theta$.
We also plot a zoom-out snapshot of the gas velocity in the radial direction in the right panel of Fig. \ref{f:8-1-density}
as a comparison with that of the zoom-in snapshot. It is found that the overall properties of $v_{r}/c$  does not change much for a fixed $\theta$ with 
increasing the distance from the innermost region of the BH to thousands of Schwarzschild radius.

\begin{figure*}
\includegraphics[width=65mm,height=86.3mm,angle=0.0]{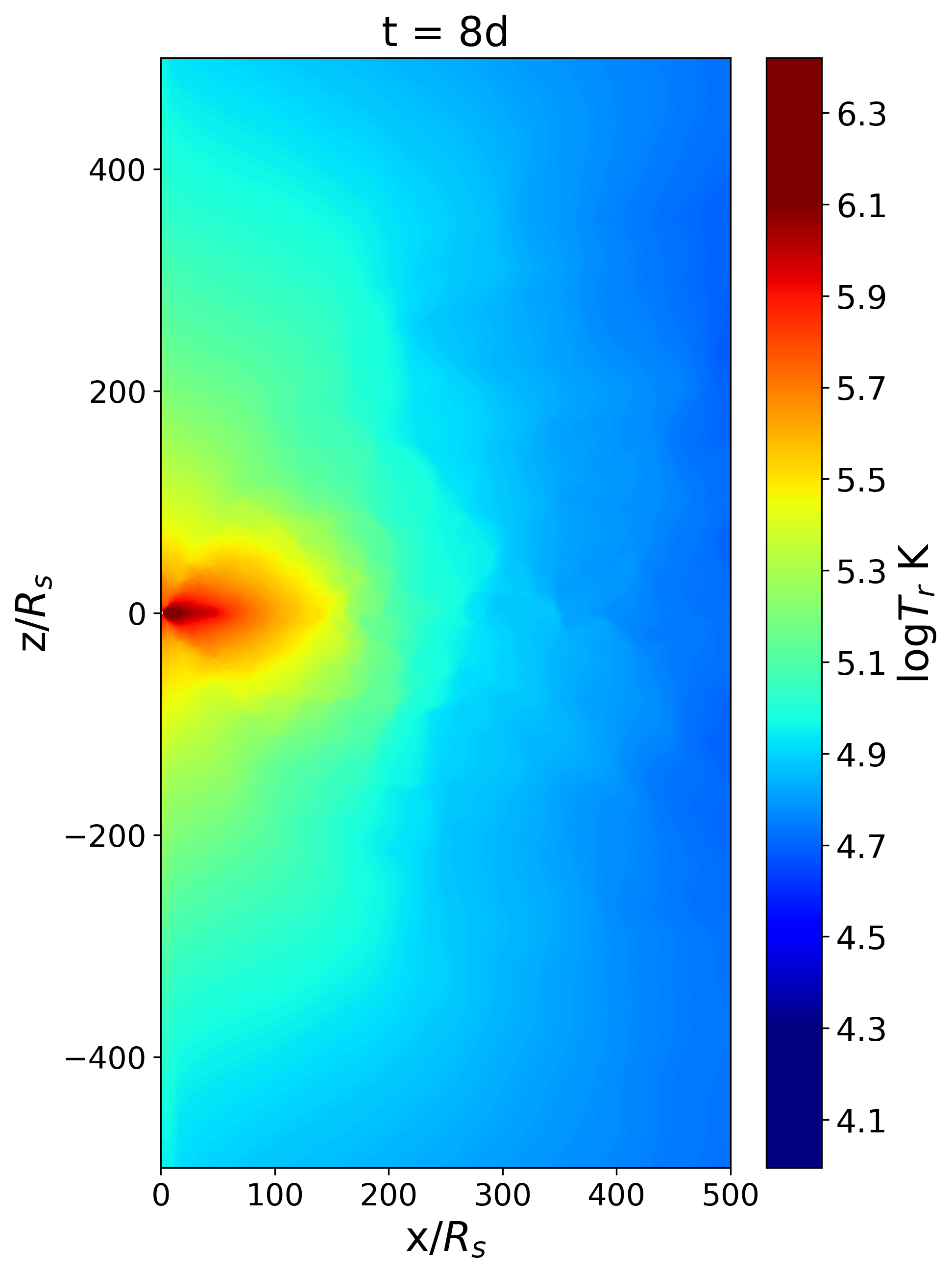}
\includegraphics[width=65mm,height=86.3mm,angle=0.0]{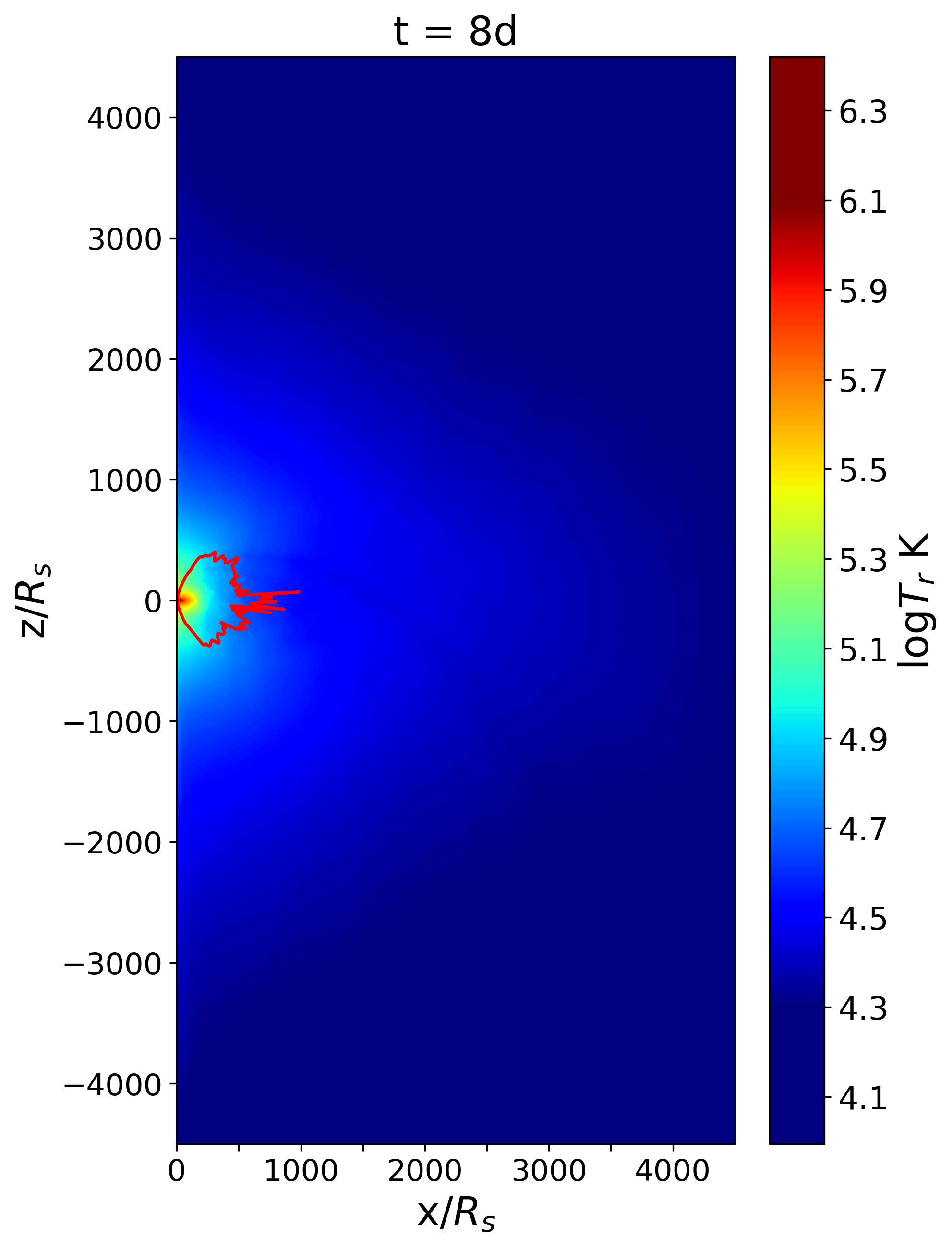}
\caption{\label{f:8-1-Tr}  Snapshots of radiation temperature (in $r-\theta$ plane) at $t=8$ day since the injection of matter at the circularization radius.
Left panel:  A zoom-in snapshot of radiation temperature. Right panel:  A zoom-out snapshot of radiation temperature. The red thick line indicates the photosphere of electron scattering.}
\end{figure*}

In the left panel of Fig. \ref{f:8-1-Tr}, we plot a  zoom-in snapshot of the radiation temperature $T_{\rm r}$ at $t=8$ day. We define the radiation temperature as 
$T_{\rm r}=(E_{r}/a_{r})^{1/4}$ with $E_{r}$ being the radiation energy density and $a_{r}$ being the radiation constant. In general,  $T_{\rm r}$
increases with increasing $\theta$ and decreases with increasing the distance from the BH. In the right panel of Fig. \ref{f:8-1-Tr}, 
we further plot a  zoom-out snapshot of $T_{\rm r}$ for the purpose that the photosphere of the electron scattering can be added. 
It can be seen that the value of $T_{\rm r}$ is roughly a few times $10^{4}$ K at the photosphere, which means that radiation of the optically thick outflow will be 
dominated in UV band as will be shown Section \ref{subsec:spectra} for the emergent spectra.

In order to show the evolution of the properties of the accretion flow, we plot the gas density, gas velocity in the radial direction and the radiation temperature at  $t=16, 24, 32$ day 
($t=8$ day added again for comparison) since the injection of matter at the circularization radius in Fig. \ref{f:App_density}, Fig. \ref{f:App_velocity} and Fig. \ref{f:App_Tr} respectively. 
In Fig. \ref{f:App_mdotin}, we also plot the scaled mass inflow rate $\dot M_{\rm in}/\dot M_{\rm Edd}$ as a function of $t$ at the inner boundary, i.e., 
 $r=2R_{\rm S}$, of the accretion flow, which intrinsically reflects the accretion into the BH with time.

\subsection{Emergent spectra}\label{subsec:spectra}
\begin{figure*}
\includegraphics[width=160mm,height=83mm,angle=0.0]{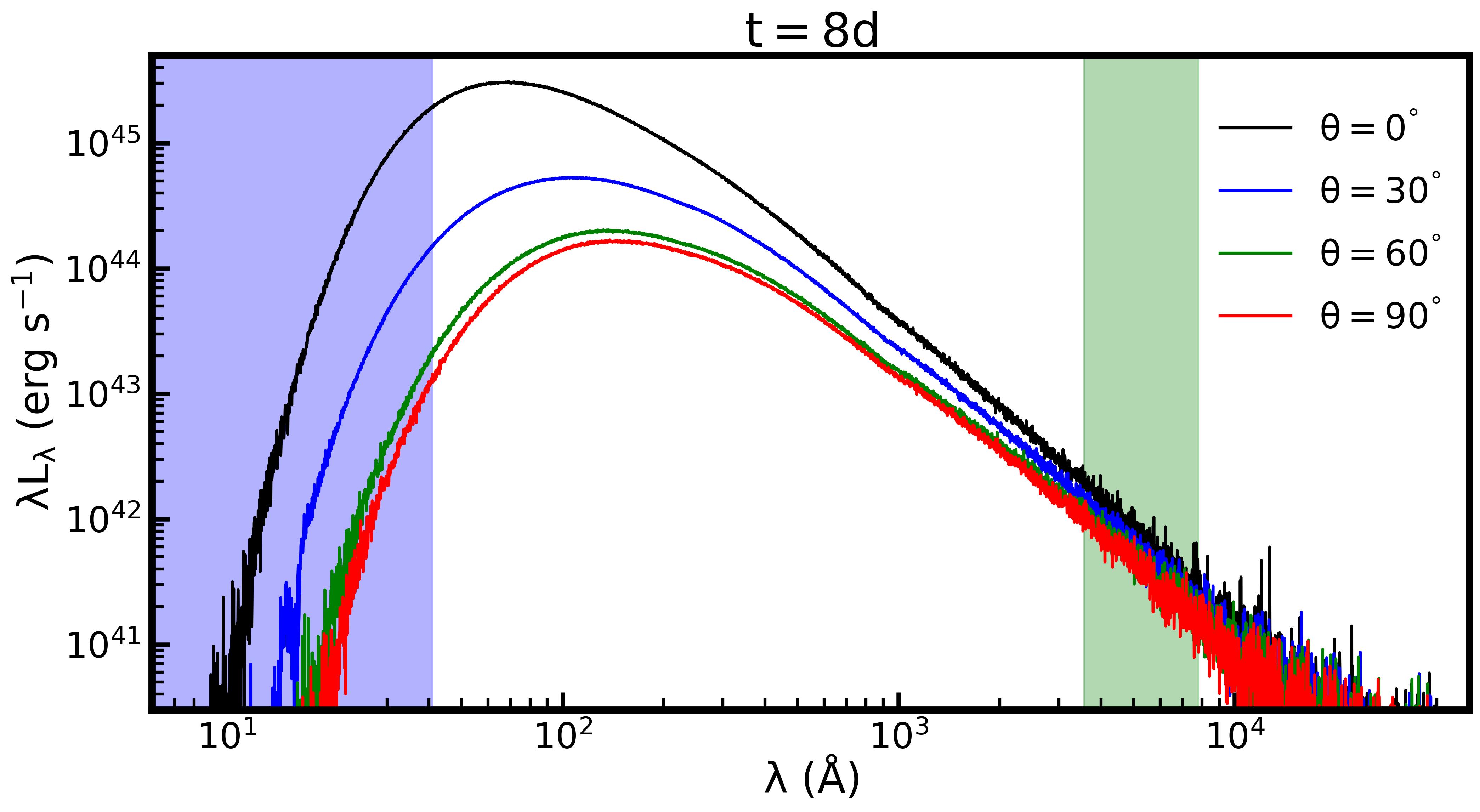}
\caption{\label{f:sp8day}Emergent spectra for different viewing angles,  i.e., $\rm \theta=0^{\rm o}, 30^{\rm o}, 60^{\rm o}$, $90^{\rm o}$ respectively, at $t=8$ day since the injection of 
matter at the circularization radius. The purple shaded region indicates X-ray band of $0.2-3\rm keV$ and the green  shaded region indicates optical band of $3500-7800\rm \AA$.
}
\end{figure*}

With the method of Monte Carlo radiative transfer described in Section \ref{subsec:Monte Carlo}, we post-process the simulation data for the emergent spectrum. 
In Fig. \ref{f:sp8day}, we plot the emergent spectra for different viewing angles, i.e., $\rm \theta=0^{\rm o}, 30^{\rm o}, 60^{\rm o}$, $90^{\rm o}$ respectively, at $t=8$ day since the injection 
of matter at the circularization radius. In general, it can be seen that, the theoretical luminosity of the accretion flow decreases with increasing $\theta$
as shown in Fig.12 of \citet[][]{Curd2019}.
The decreases of the luminosity with increasing $\theta$ are resulted by several reasons, such as the decrease of the subtended solid angle of the accretion flow as seen by the 
distant observers and the decrease of the beaming effect of the outflow \citep[][]{Sadowski2015}.

The purple shaded region indicates X-ray band of $0.2-3\rm keV$ and the green shaded region indicates optical band of $3500-7800\rm \AA$.
It can be seen that the X-ray luminosity $L_{\rm X,\ 0.3-2keV}$ drops roughly two orders of magnitude with increasing $\theta$ from $0^{\circ}-90^{\circ}$.
This can be understood as follows. As has been shown in Section \ref{subsec:simulations}, there is a critical angle $\theta_{\rm crit}$, 
the gas density of the outflow in the region from $0^{\circ}-\theta_{\rm crit}$ is very low. Thus most of the emission (in soft X-rays) from the innermost region of the accretion flow 
along the direction of $0^{\circ}-\theta_{\rm crit}$ can directly escape from the BH, and only a very small fraction of the soft X-ray emission
along these directions will be scattered by the electrons in the outflow to other directions.
The magnitude of this fraction to be scattered depends on the electron scattering optical depth along the radial direction.
On the other hand, if the soft X-ray emission from the innermost region of the accretion flow initially emits along the direction of $\theta>\theta_{\rm crit}$, 
due to the optical depth of the outflow gradually becoming optically thick with increasing $\theta$, nearly all the soft X-ray emission along these directions will be reprocessed via 
scattering, free-free, bound-free and bound-bound absorption, and re-emits in the optical and ultraviolet (UV) bands.
In summary, if we observe the accretion flow around the BH along the direction of $\theta<\theta_{\rm crit}$, we can see a stronger X-ray emission. If we 
observe the accretion flow along the direction of $\theta>\theta_{\rm crit}$, the X-ray emission will be significantly reduced, and the X-ray luminosity decreases with
increasing the $\theta$. As for the optical emission, since the photosphere is nearly isotropic, the observed optical emission is expected not to change much
for different $\theta$, as can be seen in Fig. \ref{f:sp8day}.
We also note that the emergent spectra is dominated the UV emission for different $\theta$, and the luminosity of UV emission gradually decreases with 
increasing $\theta$, which physically depends on the distribution of the gas temperature and density in the outflow.  

\begin{figure*}
\includegraphics[width=160mm,height=81.3mm,angle=0.0]{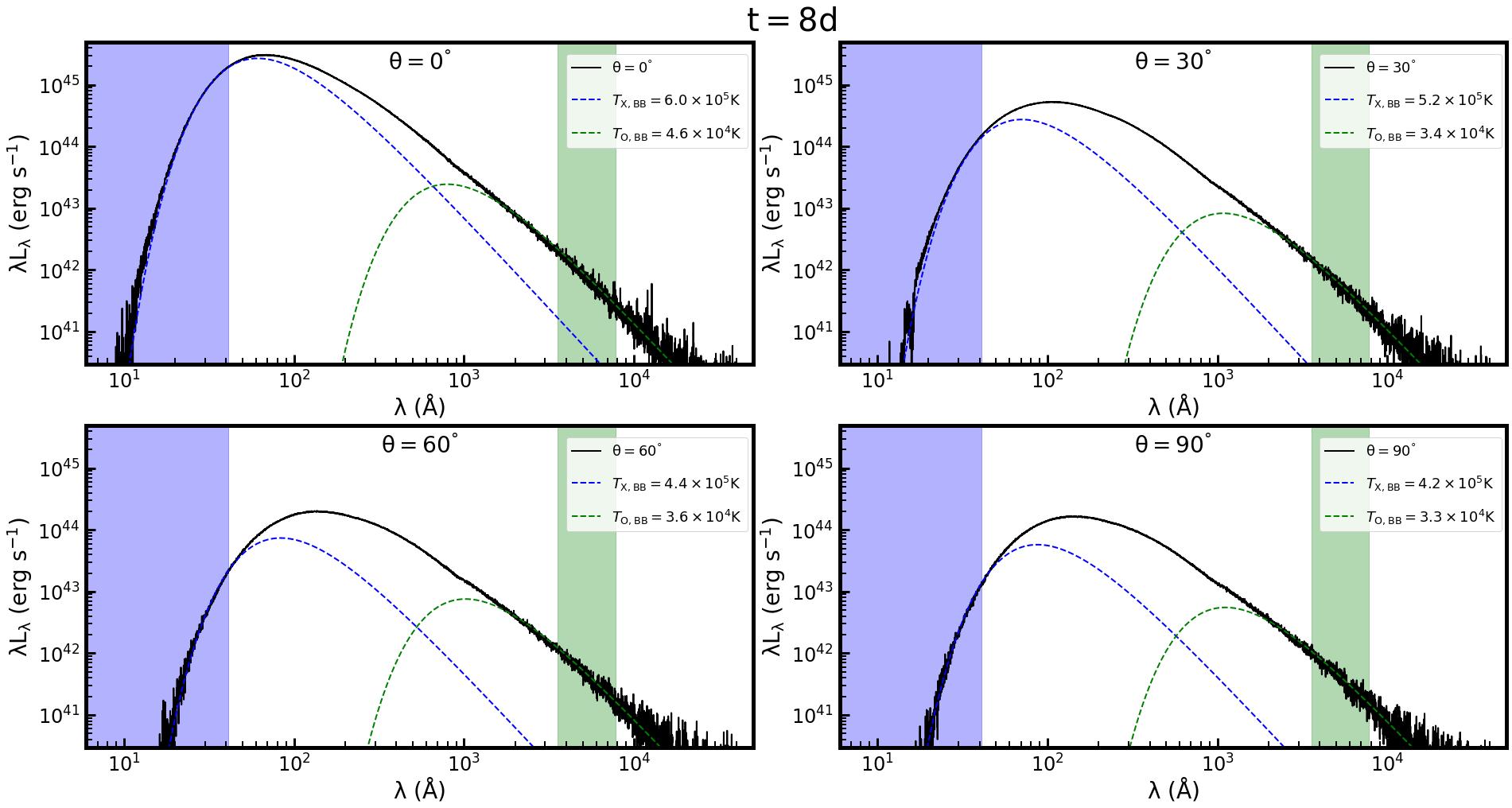}
\caption{\label{f:sp_fit}Emergent spectra for different viewing angles, i.e., $\rm \theta=0^{\rm o}, 30^{\rm o}, 60^{\rm o}$, $90^{\rm o}$ respectively, at $t=8$ day since the injection of matter 
at the circularization radius. The spectra are fitted with a single blackbody for X-ray (purple dashed line) and a single blackbody for optical band (green dashed line) respectively.
The purple shaded region indicates X-ray band of $0.2-3\rm keV$ and the green  shaded region indicates optical band of $3500-7800\rm \AA$.
}
\end{figure*}

\begin{figure*}
\includegraphics[width=160mm,height=105.86mm,angle=0.0]{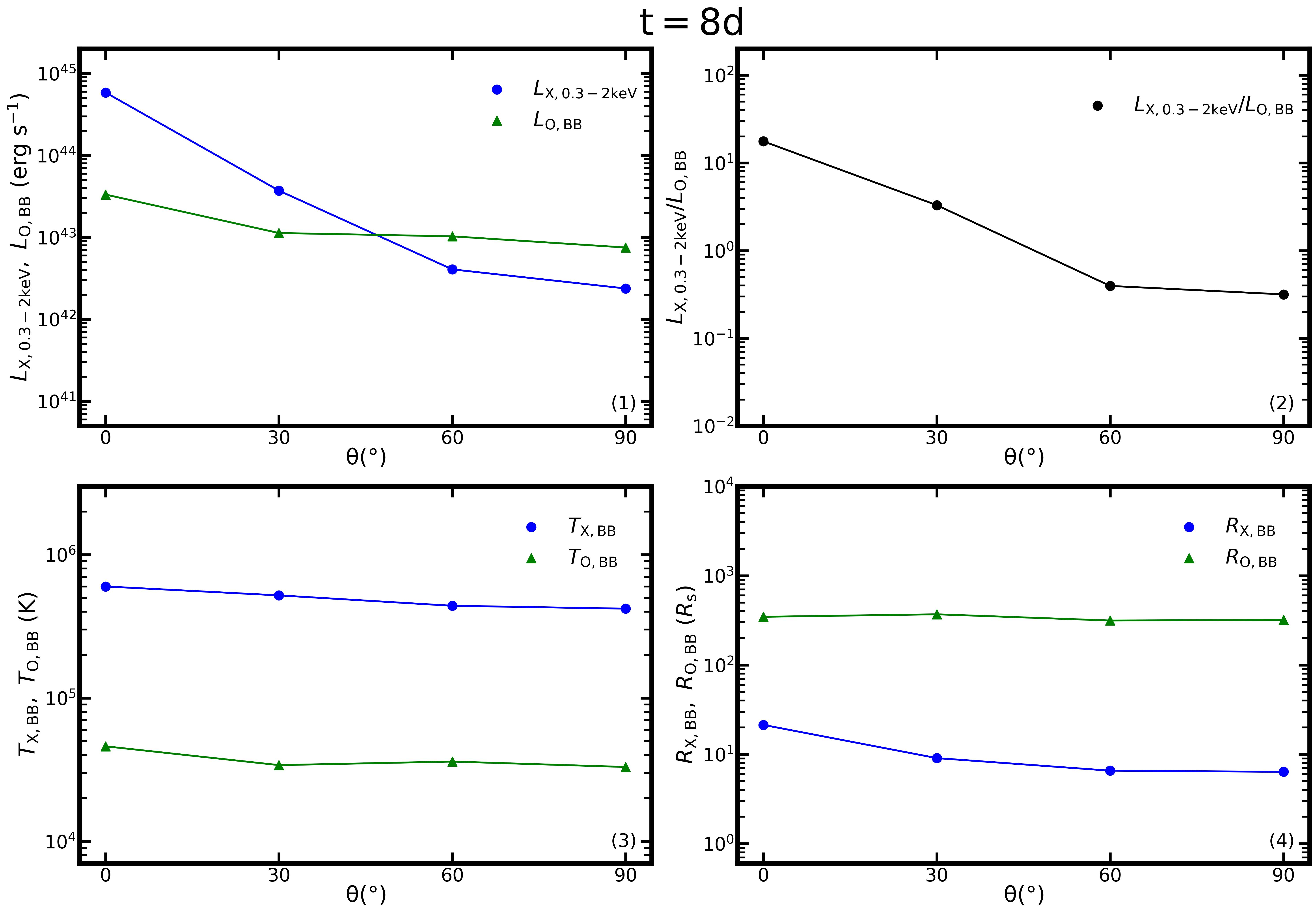}
\caption{\label{f:sp_parameter} Fitting parameters based on the emergent spectra as a function of viewing angle $\theta$
at $t=8$ day since the injection of matter at the circularization radius.
Panel (1): X-ray luminosity $L_{\rm X, \ 0.3-2 keV}$ and the optical blackbody luminosity $L_{\rm O, \ BB}$ as a function of $\theta$.
Panel (2): $L_{\rm X, \ 0.3-2 keV}/L_{\rm O, \ BB}$ as a function of  $\theta$.
Panel (3): X-ray blackbody temperature $T_{\rm X, \ BB}$ and optical black body temperature $T_{\rm O, \ BB}$ as a function of  $\theta$. 
Panel (4): X-ray blackbody radius $R_{\rm X, \ BB}$ and the optical blackbody radius $R_{\rm O, \ BB}$ (in units of $R_{\rm S}$) 
as a function of $\theta$.
}
\end{figure*}

Further, we fit the emergent spectra with a single blackbody for X-rays ($0.2-3\rm keV$) and a single blackbody for optical emission ($3500-7800\rm \AA$) 
respectively for different viewing angel $\theta$.
One can see the fitting results in Fig. \ref{f:sp_fit} for details. It is found that the X-ray blackbody temperatures $T_{\rm X,\ BB}$ are $6.0\times 10^{5}$ ,
$5.2\times 10^{5}$, $4.4\times 10^{5}$ and $4.2\times 10^{5}$ K for viewing angle $\rm \theta=0^{\rm o}, 30^{\rm o}, 60^{\rm o}$ and $90^{\rm o}$ respectively,
and the optical blackbody temperatures $T_{\rm O,\ BB}$ are $4.6\times 10^{4}$ ,
$3.4\times 10^{4}$, $3.6\times 10^{4}$ and $3.3\times 10^{4}$ K for viewing angle $\rm \theta=0^{\rm o}, 30^{\rm o}, 60^{\rm o}$ and $90^{\rm o}$ respectively.
One can also see panel (3) of Fig. \ref{f:sp_parameter} for $T_{\rm X,\ BB}$ and  $T_{\rm O,\ BB}$ as a function of $\theta$ respectively for clarity. 
It can be seen that both $T_{\rm X,\ BB}$ $\sim$ a few $\times 10^{5}$ K and  $T_{\rm O,\ BB}$ $\sim$ a few $\times 10^{4}$ K are consistent with 
either simultaneous observations for some specific sources, such as ASASSN-14li \citep[e.g.][]{vanVelzen2016} or statistically for different sources at the early stage of the outburst.
One can also see the review for $T_{\rm X,\ BB}$ in \citet[][]{Saxton2021SSRv} and for $T_{\rm O,\ BB}$ in \citet[][]{vanVelzen2020SSRv}.

Based on the emergent spectra, in Fig. \ref{f:sp_parameter} we plot  $L_{\rm X, \ 0.3-2 keV}$ and the corresponding blackbody luminosity of the optical emission
$L_{\rm O, \ BB}$, $L_{\rm X, \ 0.3-2 keV}/L_{\rm O, \ BB}$, $T_{\rm X,\ BB}$ and $T_{\rm O,\ BB}$, as well as $R_{\rm X,\ BB}$ and $R_{\rm O,\ BB}$
as a function of $\theta$ respectively. Here  $R_{\rm X, \ BB}$ and $R_{\rm O, \ BB}$ are the derived radius of the X-ray emission region
and the optical emission region respectively with  $R_{\rm X,\ BB}= \big({L_{\rm X, \ BB}\over 4\pi \sigma T_{\rm X, \ BB}^4}\big)^{1/2}$
and  $R_{\rm O,\ BB}= \big({L_{\rm O, \ BB}\over 4\pi \sigma T_{\rm O, \ BB}^4}\big)^{1/2}$, where $L_{\rm X, \ BB}$ is the corresponding blackbody luminosity of the X-ray emission. 
As has been analyzed in the last paragraph, $L_{\rm X, \ 0.3-2 keV}$ decreases with increasing $\theta$, and $L_{\rm O, \ BB}$ is roughly constant with $\theta$ with very
small variations.  Specifically, $L_{\rm X, \ 0.3-2 keV}$ decreases from $\sim 6\times 10^{44} \rm erg/s$ to  $\sim 2\times 10^{42} \rm erg/s$ with $\theta$ increasing from 
 $0^{\rm o}$ to $90^{\rm o}$, indicating a $\sim 300$-times decrease of the X-ray luminosity with increasing $\theta$ from  $0^{\rm o}$ to $90^{\rm o}$.
 The X-ray luminosity  range can well cover the observed X-ray luminosity of $\sim 10^{42-44} \rm erg/s$ \citep[][see Table 1 for summary]{Saxton2021SSRv}. 
We propose that the effect of the viewing angle is one the possibilities to explain the diverse X-ray luminosity at the early phase of TDEs
 as also suggested by several other authors \citep[][]{Dai2018,Curd2019,Thomsen2022,Guolo2024}.
Other factors, such as BH mass may also play a role in the X-ray luminosity, which is needed to be tested in the future work. 
 $L_{\rm O, \ BB}$ is $\sim 10^{43} \rm erg/s$ with a maximum variation of $\sim 4$ times with changing $\theta$ from $0^{\rm o}$ to $90^{\rm o}$,
 which is consistent the expectations of many theoretical studies that the optical emission is nearly isotropic \citep[e.g.,][]{Curd2019}.
 One can see panel (1) of Fig. \ref{f:sp_parameter}  for $L_{\rm X, \ 0.3-2 keV}$ and $L_{\rm O, \ BB}$ for different $\theta$ respectively for details.  

In panel (2) of Fig. \ref{f:sp_parameter}, we plot $L_{\rm X, \ 0.3-2 keV}/L_{\rm O, \ BB}$  as a function of $\theta$.
It can be seen that $L_{\rm X, \ 0.3-2 keV}/L_{\rm O, \ BB}$ decreases from $\sim 20$ to $\sim 0.3$ with $\theta$ increasing from 
$0^{\rm o}$ to $90^{\rm o}$, which means that if we observe the accretion flow for different viewing angle $\theta$, the ratio between $L_{\rm X, \ 0.3-2 keV}$ and $L_{\rm O, \ BB}$
can reach a difference of nearly two orders of magnitude.  
The derived $L_{\rm X, \ 0.3-2 keV}/L_{\rm O, \ BB}$ from our calculations for different viewing angle can match most of the observed  
$L_{\rm X, \ 0.3-2 keV}/L_{\rm O, \ BB}$ in TDEs \citep[e.g.,][]{Auchettl2017,vanVelzen2021}. 
In panel (4) of Fig. \ref{f:sp_parameter}, we plot $R_{\rm X,\ BB}$ and $R_{\rm O,\ BB}$ as a function of $\theta$ respectively.
It can be seen that $R_{\rm X,\ BB}$ decreases from $\sim 20R_{\rm S}$ to $\sim 5R_{\rm S}$ with $\theta$ increasing from $0^{\rm o}$ to $90^{\rm o}$,
and $R_{\rm O,\ BB}$ is roughly a constant of $\sim 3\times 10^{2} R_{\rm S}$ with different $\theta$.
The derived $R_{\rm O,\ BB}$ is roughly consistent with observations, and the derived $R_{\rm X,\ BB}$ is roughly one order of
magnitude larger than observations. The calculated $R_{\rm X,\ BB}$ from observations is even
smaller than the event horizon for a non-rotating BH, which is indeed problematic, and need to be understood in the future \citep[][for review]{Gezari2021ARAA}.
 

In order to show the evolution of the spectral features, 
we present the emergent spectra for different $\theta$ at  $t=16, 24, 32$ day ($t=8$ day added again for comparison) respectively since the injection of matter
at the circularization radius in Fig. \ref{f:App_sp} and Fig. \ref{f:App_sp_fit}. In Fig. \ref{f:App_sp_fit}, the spectra at different times are fitted with a single blackbody for X-ray (purple dashed line) 
and a single blackbody for optical band (green dashed line) respectively as that of in Fig. \ref{f:sp_fit}.
We present $L_{\rm X, \ 0.3-2 keV}$ and $L_{\rm O, \ BB}$, $L_{\rm X, \ 0.3-2 keV}/L_{\rm O, \ BB}$, $T_{\rm X,\ BB}$ and $T_{\rm O,\ BB}$, as well as $R_{\rm X,\ BB}$ and $O_{\rm X,\ BB}$
as a function of $\theta$ at $t=16, 24, 32$ day ($t=8$ day added again for comparison) respectively since the injection of matter at the circularization radius in Fig. \ref{f:App_sp_parameter}.


\subsection{Spectral evolution}\label{subsec:evolution}
\begin{figure*}
\includegraphics[width=80mm,height=52.11mm,angle=0.0]{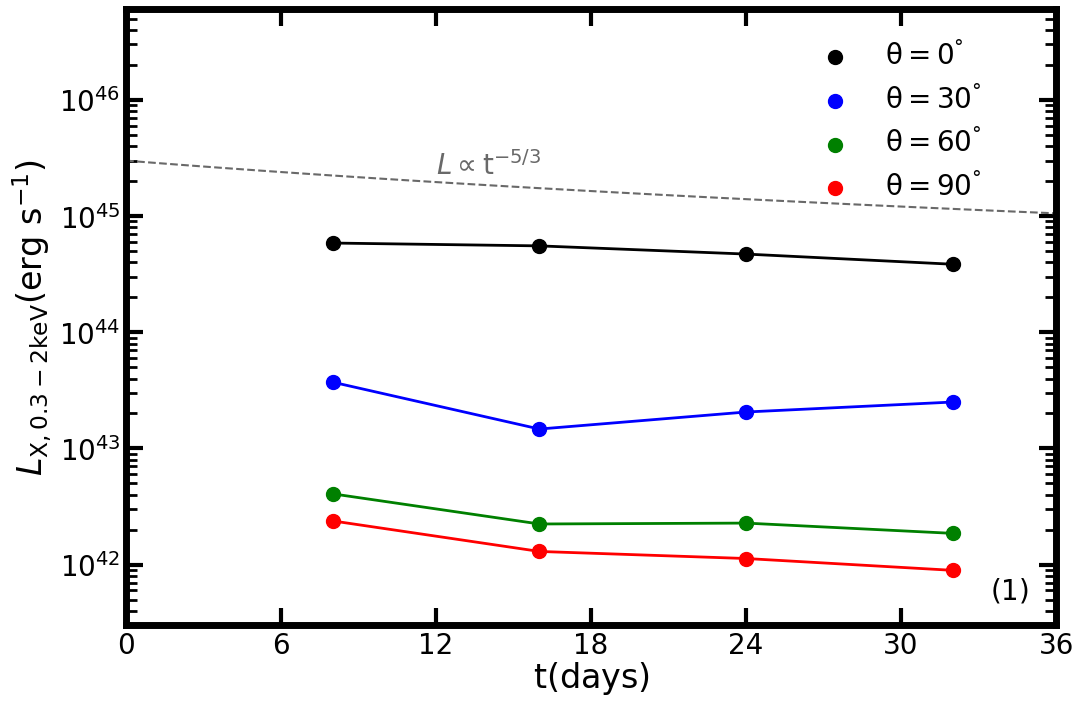}
\includegraphics[width=80mm,height=52.11mm,angle=0.0]{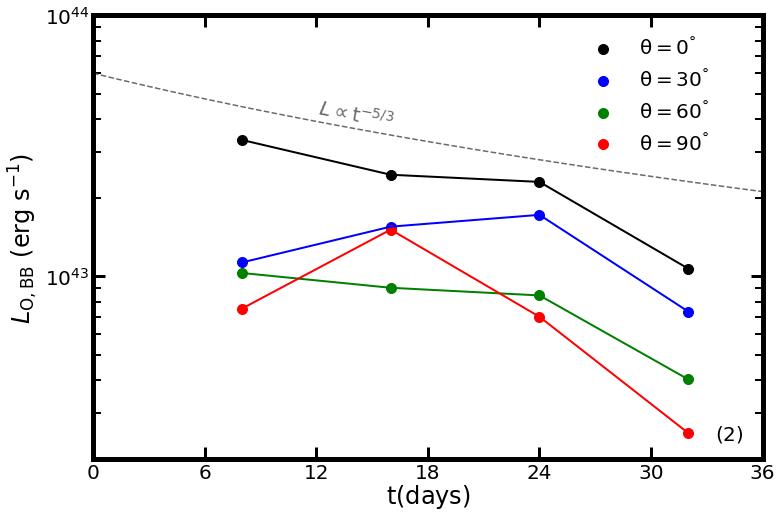}
\includegraphics[width=80mm,height=52.11mm,angle=0.0]{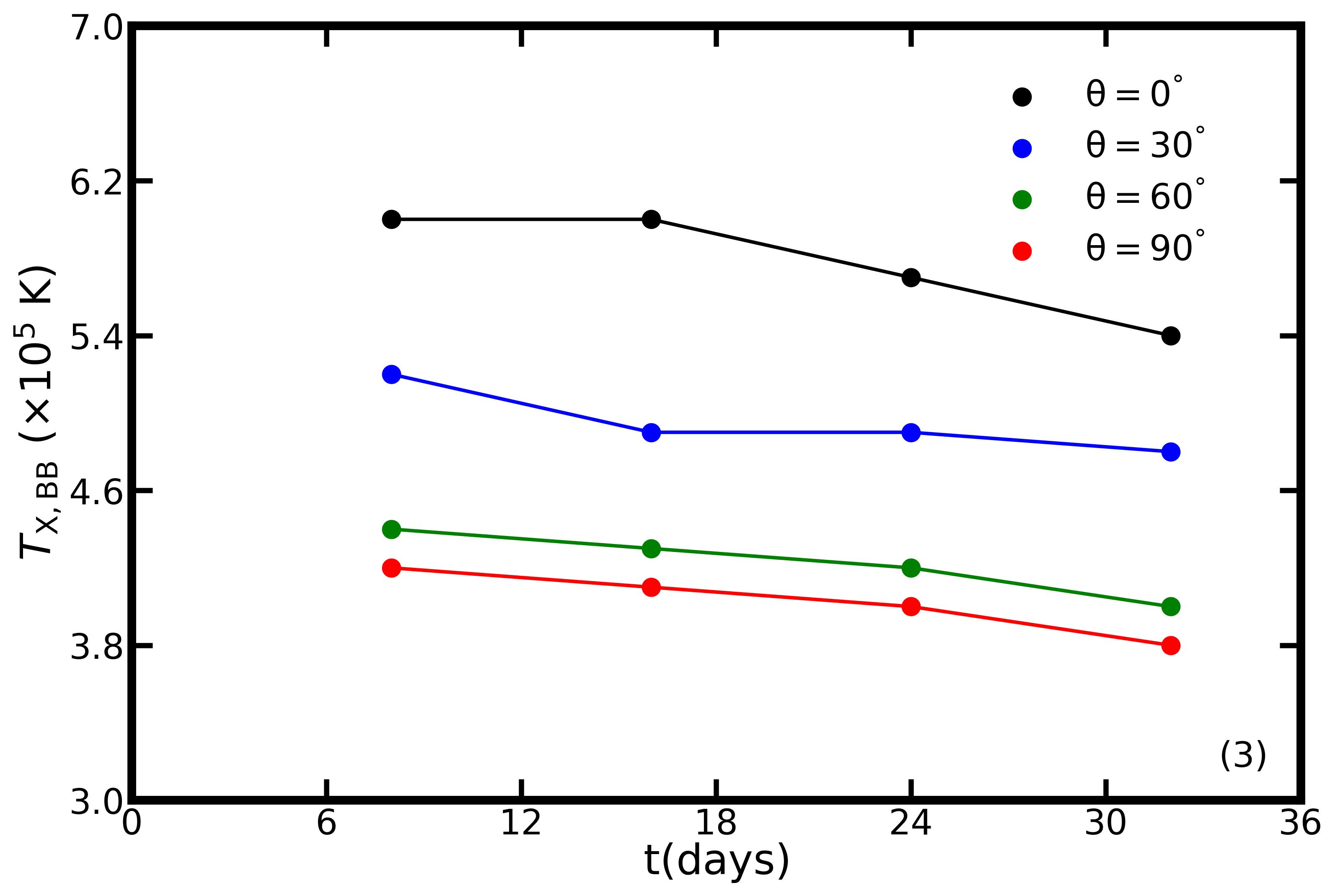}
\includegraphics[width=80mm,height=52.11mm,angle=0.0]{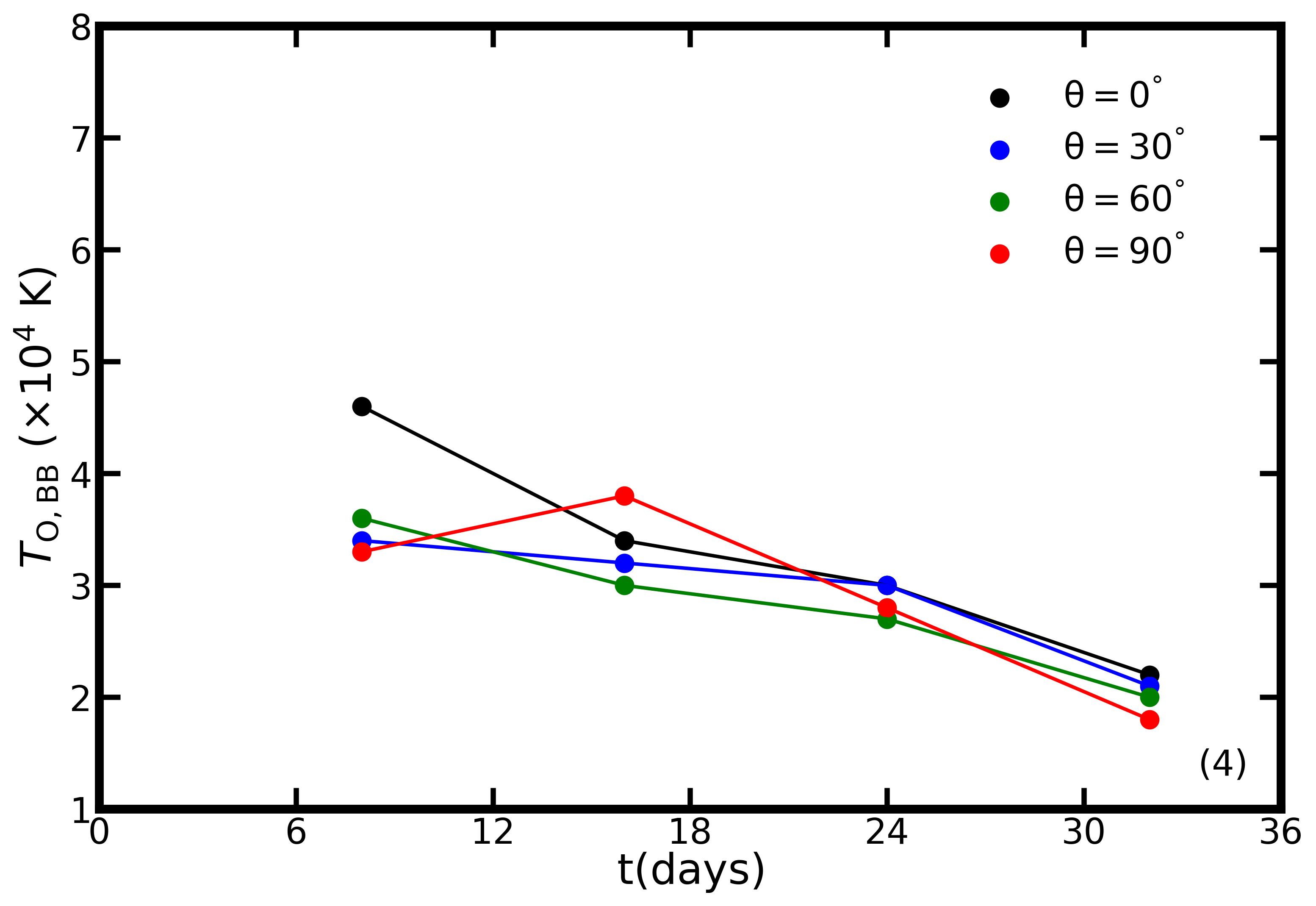}
\includegraphics[width=80mm,height=52.11mm,angle=0.0]{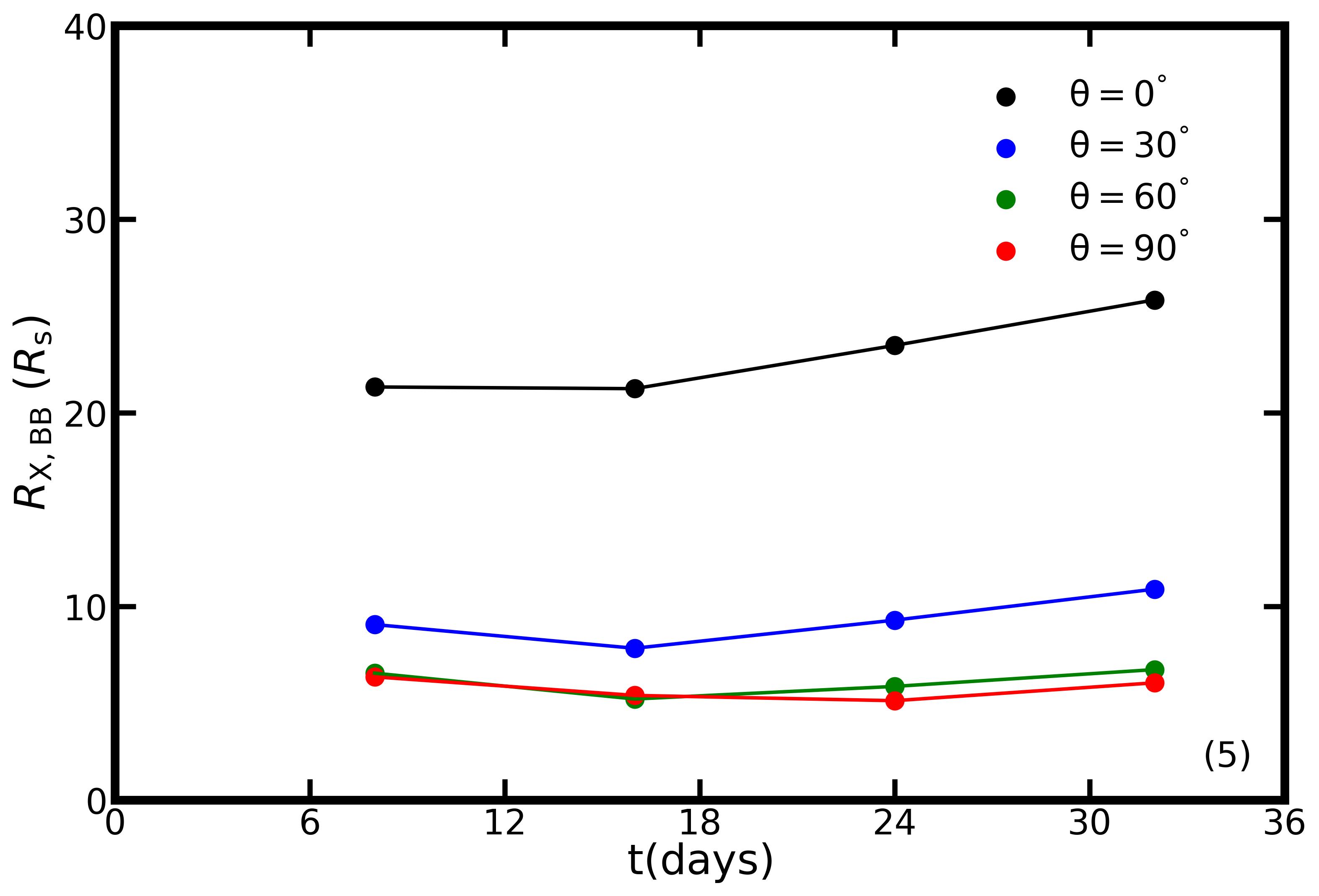}
\includegraphics[width=80mm,height=52.11mm,angle=0.0]{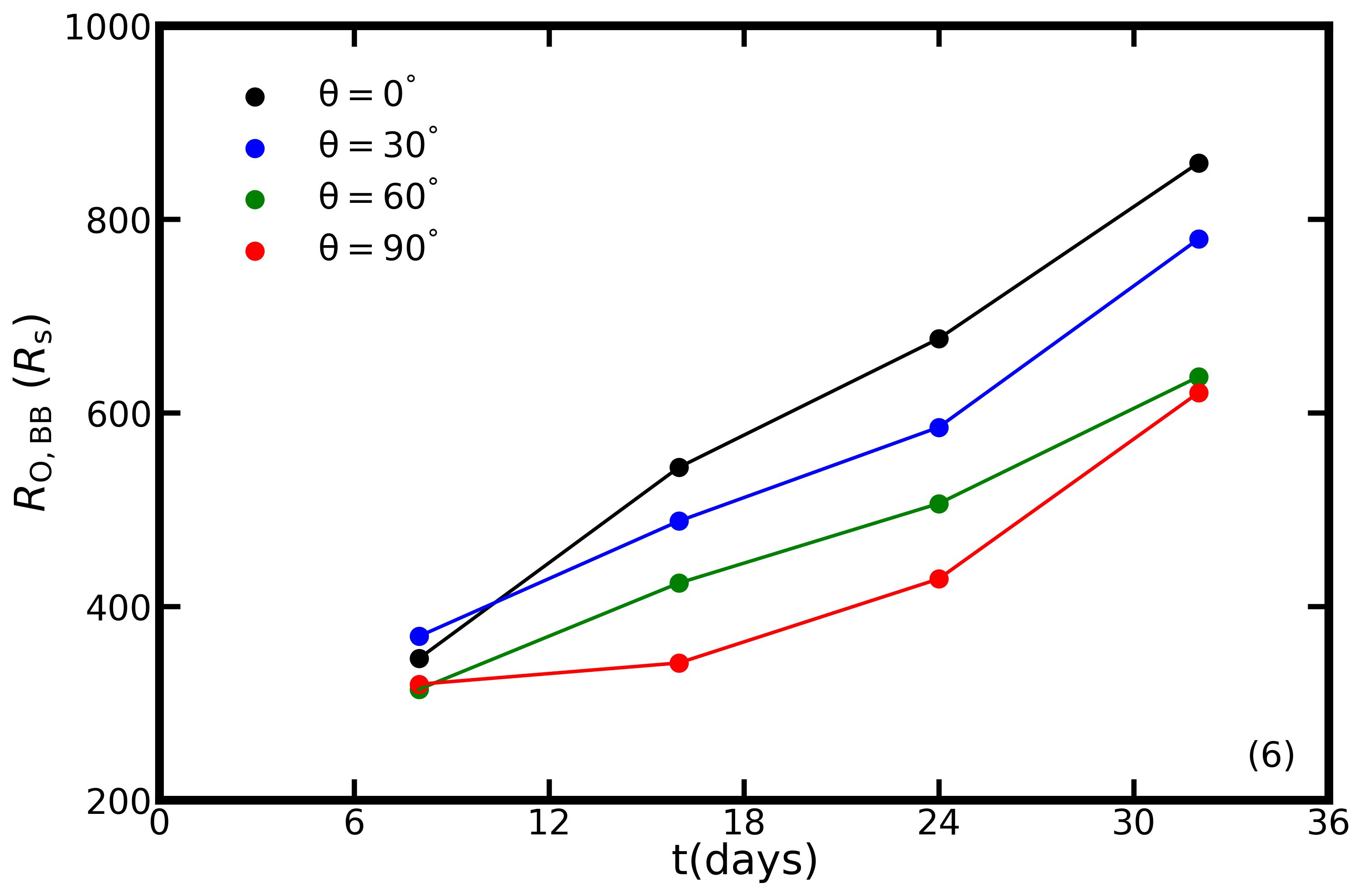}
\includegraphics[width=80mm,height=52.11mm,angle=0.0]{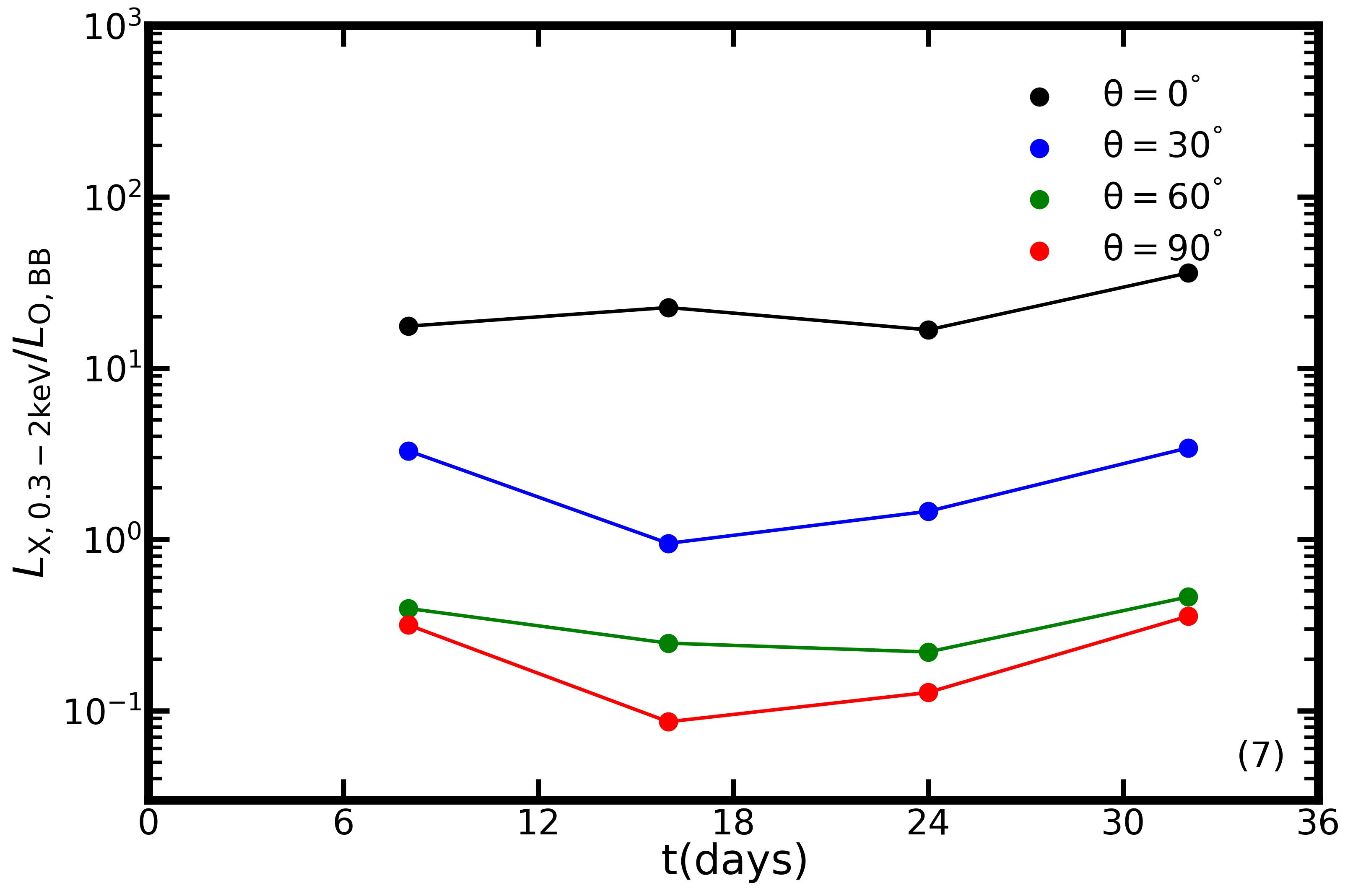}
\caption{\label{f:evolution1} $L_{\rm X, \ 0.3-2 keV}$,  $L_{\rm O, \ BB}$, $T_{\rm X,\ BB}$, $T_{\rm O,\ BB}$, $R_{\rm X,\ BB}$, $R_{\rm O,\ BB}$ 
and $L_{\rm X, \ 0.3-2 keV}/L_{\rm O, \ BB}$ as a function of $t$ for different viewing angles $\theta$. }
\end{figure*}

Combing the results presented in Section \ref{subsec:spectra}, we plot  $L_{\rm X, \ 0.3-2 keV}$, $L_{\rm O, \ BB}$, $T_{\rm X,\ BB}$, $T_{\rm O,\ BB}$,  $R_{\rm X,\ BB}$, $R_{\rm O,\ BB}$
and $L_{\rm X, \ 0.3-2 keV}/L_{\rm O, \ BB}$ as a function of $t$ in Fig. \ref{f:evolution1}.
In panel (1) of Fig. \ref{f:evolution1}, we plot  $L_{\rm X, \ 0.3-2 keV}$ as a function of $t$ for $\theta=0^{\rm o}, 30^{\rm o}, 60^{\rm o}$ and $90^{\rm o}$ respectively.
It can be seen that the evolution of $L_{\rm X, \ 0.3-2 keV}$ roughly follows $t^{-5/3}$ as the fallback rate presumed in the present paper. 
This is because the evolution of the X-ray emission intrinsically reflects the evolution of the accretion inflow rates, and further reflects the evolution of the fallback rates assuming a rapid circularization
 \citep[][]{Rees1988,Komossa2004IAUS,Komossa2015,Saxton2021SSRv}.
The corresponding $T_{\rm X,\ BB}$ and $R_{\rm X,\ BB}$ are plotted in panel (3) and panel (5) in Fig. \ref{f:evolution1} respectively.      
In general, it can be seen that  $L_{\rm X, \ 0.3-2 keV}$, $T_{\rm X,\ BB}$ and $R_{\rm X,\ BB}$ show very show evolution in timescale $\sim 32$ days.
Meanwhile, we should note that the effects of viewing angle $\theta$ on $L_{\rm X, \ 0.3-2 keV}$ and $R_{\rm X,\ BB}$ are significant as discussed in 
Section \ref{subsec:spectra}. 
A further detailed study, such as constraints on the viewing angle $\theta$ of TDEs by fitting the X-ray spectra is necessary in the future,
which however exceeds the scope of the present paper.
In panel (2) of Fig. \ref{f:evolution1}, we plot $L_{\rm O, \ BB}$ as a function of $t$.
As discussed in Section \ref{subsec:spectra}, the optical emission is produced via the reprocess of the soft X-ray emission in the optically thick outflows.
So, the evolution of the optical emission is closely related to the formation and evolution of the photosphere of the outflows. 
We can imagine that  $L_{\rm O, \ BB}$ increases with time initially corresponding to the gradual formation of the photosphere, then decreases with time corresponding to
the decrease of content of matter in the outflow since the decrease of the mass supply rate, i.e. the fallback rate, to the BH.
The evolution of $L_{\rm O, \ BB}$ in panel (2) of Fig. \ref{f:evolution1} roughly reflects this trend. 
Generally speaking, the photosphere expands with time as shown in Fig. \ref{f:App_density} and Fig. \ref{f:App_Tr}.
The expansion of the photosphere will result in the decrease of the temperature. One can see panel (4) and panel (6) in Fig. \ref{f:evolution1}
for the evolution of  $T_{\rm O,\ BB}$ and $R_{\rm O,\ BB}$ with $t$ for clarity.
In summary, the theoretical evolution of $L_{\rm O, \ BB}$ is relatively complicated, which depends on the detailed radiative transfer, and the formation of the photosphere etc.
Further, if we try to exactly match the observations, we have to consider other possible influences, such as the problem of radiation zero time, as discussed in the 
third paragraph of Section \ref{Summary and Discussion}. 

In panel (7) of Fig. \ref{f:evolution1}, we plot $L_{\rm X, \ 0.3-2 keV}/L_{\rm O, \ BB}$  as a function of $t$ for $\theta=0^{\rm o}, 30^{\rm o}, 60^{\rm o}$, $90^{\rm o}$ respectively.
In general, the evolution of  $L_{\rm X, \ 0.3-2 keV}/L_{\rm O, \ BB}$ with $t$ does not change much for a fixed $\theta$ in the timescale of $\sim 32$ days in the present simulations.
As we can see, the magnitude of $L_{\rm X, \ 0.3-2 keV}/L_{\rm O, \ BB}$ strongly depends on viewing angle $\theta$. 
So if possible, we expect that the observed diverse value of $L_{\rm X, \ 0.3-2 keV}/L_{\rm O, \ BB}$ can be explained for different viewing angles $\theta$  in the framework of the 
inflow/outflow reprocessing model for super-Eddington accretion.

\section{Summary and Discussion}\label{Summary and Discussion}
TDEs are highly nonlinear systems, including hydrodynamics/magnetohydrodynamics, radiation transfer, self-gravity, general relativity, magnetic field etc.
A typical TDE often has three phases, i.e., disruption of the star, evolution of the stream and the accretion processes \citep[][]{Rees1988}.
Numerical simulation is the only way, based on which we can fully understand the physics in TDEs.
So far, a great number of efforts with the method of numerical simulations have been made for understanding the physical processes in TDEs. 
However, since the complexity of the physics and large computational resources required, currently, there is no simulations, which
can simulate the three phases of a TDE as a whole self-consistently. Instead, the three phases of a TDE are simulated separately, e.g., for 
disruption of the star \citep[e.g.][for review]{Evans1989,Lodato2009,Guillochon2013,Law-Smith2019,Steinberg2019,Golightly2019ApJ...872..163G,Golightly2019ApJ...882L..26G,
Sacchi2019,Ryu2020ApJ...904...98R,Ryu2020ApJ...904...99R,Ryu2020ApJ...904..100R,Rossi2021}, 
for evolution of the stream \citep[e.g.][for review]{Hayasaki2013,Shiokawa2015,Jiang2016,Luwenbin2020,Bonnerot2020,Bonnerot2021SSRv}, and for accretion
processes \citep[e.g.][for review]{Dai2018,Curd2019,Bu2022,Bu2023,Thomsen2022,Dai2021SSRv}.
Although the separate treatments the three phases of a TDE may result in some uncertainties and miss some potential informations, it is still a very important
way for understanding some key observational aspects of TDEs. 

In this paper, we focused on the radiation hydrodynamic simulations for the accretion process of TDEs with Athena++ code.
There are two key points for the setup of the simulation, i.e., (1) assuming the mass accretion rate injected with the form of $\dot M_{\rm inject} \propto t^{-5/3}$, 
(2) assuming the injecting point for the accreted mass at $2R_{\rm T}$, which is the theoretical circularization radius of the debris gas of disrupting a star with parabolic orbit.
We then post-processed the simulation data with the method of Monto Carlo radiative transfer for the emergent spectra for different viewing angles. 
Based on the emergent spectra, we show that the X-ray luminosity $L_{\rm X, \ 0.3-2 keV}$, optical blackbody luminosity $L_{\rm O, \ BB}$,
X-ray blackbody temperature $T_{\rm X,\ BB}$, optical blackbody temperature $T_{\rm O,\ BB}$, 
X-ray emission radius $R_{\rm X,\ BB}$, optical emission radius $R_{\rm O,\ BB}$, and $L_{\rm X, \ 0.3-2 keV}/L_{\rm O, \ BB}$, as well as the evolution 
of all these quantities are crudely consistent with observations in the framework of viewing-angle effect of super-Eddington accretion around a BH.

Finally, we would like to address that in our simulation the zero time point is the time starting to inject matter at the circularization radius.
As for observations, actually, it is difficult to track the zero time point of the radiation after the disruption of the star since the following processes, such as 
the evolution and the circularization of the debris stream, are still unclear. So when the simulation results are applied to compare with observations, we assume that the 
debris stream is circularized very quickly, and neglect the corresponding radiation. The study for the circularization process has been studied by several authors, 
however, it is still one of the most important questions that is not fully understood in TDEs.  
In addition, we neglect the fallback matter during the rising phase, i.e., the fallback matter before the peak fallback rate, to be injected to feed the BH in the simulation. 
The amount of this mass can be $\sim 10\%$ of the total stellar mass \citep[][]{Metzger2022}, the effect of which to the simulation is possible to be significant,
and will be considered in the future.
The exact match of the zero time point in our simulation to observations is a challenging work, the study of which may alleviate some inconsistencies between
simulations and observations, such as the evolution of the size of the photosphere and the corresponding temperature etc. \citep[][]{vanVelzen2021}. 

\subsection{Long-term evolution of the simulations}
In this paper, we performed a 32-day long simulation since injecting mass at the circularization radius. 
For the fixing parameters in the present paper, i.e.,  $M_{\rm BH}=10^{6}M_{\odot}$, $M_{*}=M_{\odot}$ and $R_{*}=R_{\odot}$, at $t=0$ the 
injecting mass accretion rate is $133.8 \dot M_{\rm Edd}$, and at $t=32$ day the injecting mass accretion 
rate is $51.4 \dot M_{\rm Edd}$. So it can be seen that at $t=32$ day, the injecting mass accretion rate is still significantly higher than $\dot M_{\rm Edd}$.
Generally speaking, if mass accretion rate is much higher than $\dot M_{\rm Edd}$, the photo trapping effect will make the variation of the
luminosity as a function of mass accretion rate being very flat, which actually means that the evolution of the accretion flow with time is very slow.
Due to the computational resources limitation, in this paper, we stop the simulation at 32 day since injecting mass at the circularization radius.

It is expected that some more key physics for the evolution of the accretion flow around a supermassive BH could be revealed in a longer-time 
simulation, e.g., in the timescale of years.
Compared with the traditional AGN, TDE has a relatively smaller accretion disk, so the change of the mass supply rate 
can be more easily reflected on the change of the spectrum and the luminosity. 
So TDEs are believed to be ideal laboratories to study the evolution of the accretion flow around a supermassive BH.
For example, if we still follow $\dot M_{\rm inject} \propto t^{-5/3}$, at $t\sim 362-472$ day, $\dot M_{\rm inject}$ will evolve to $\sim 2-3 \dot M_{\rm Edd}$.
At this time, it is expected that the accretion flow will begin to suffer a radiation pressure instability, which theoretically will predict a 
signifiant oscillation of the luminosity \citep[][]{Abramowicz1988}.
At $t\sim 1540$ day, $\dot M_{\rm inject}$ will evolve to $\sim 0.3 \dot M_{\rm Edd}$, it is expected that the accretion will gradually leave 
the region of the radiation pressure instability, and the accretion flow will become stable gradually.  Further, at $t\sim 6000$ day, 
$\dot M_{\rm inject}$ will evolve to $\sim 0.03 \dot M_{\rm Edd}$. At this time, it is expected that the accretion flow will transit to the advection dominated
accretion flow (ADAF). ADAF is radiatively inefficient, and a significant drop of the luminosity will be 
expected \citep[][for review]{Narayan1994,Narayan1995b,Yuan2014}.
Combing observational data, all of these mentioned above and some other aspects for evolution of the accretion flow around a supermassive BH 
predicted by semi-analytic accretion theory could be tested by longer-time scale simulations in the future.  

Besides, a comparative study of the evolution of the accretion flow in TDEs and BH low-mass X-ray binaries can be used to test the universality of accretion theory 
for BHs with different mass scales. Previous study suggested that there are some correspondence  of the accretion flow (such as the accretion geometry) between BH low-mass 
X-ray binaries and AGNs \citep[][]{Falcke2004,Kording2006,Wu2008,Moravec2022}. 
For example,  the low/hard spectra state in BH low-mass X-ray binaries may correspond to the low-luminosity 
AGNs \citep[e.g.][for review]{Merloni2003,Taam2012,Liu2022iSci...25j3544L},
the high/soft spectra state in BH low-mass X-ray binaries may correspond to the luminous AGNs\citep[e.g.][]{Maccarone2003}, and the very high state in 
BH low-mass X-ray binaries may correspond to the Narrow line Seyfert galaxies etc.
One of the key parameters that can connect the accretion process between BH low-mass X-ray binaries and AGNs is the dimensionless 
mass accretion rate $\dot m$ (with $\dot m \equiv \dot M/\dot M_{\rm Edd}$) .
In general, it is believed that the evolution of the accretion flow in BH low-mass X-ray binaries is determined by $\dot m$  \citep[][]{Esin1997}. In most of the lifetime, 
BH low-mass X-ray binary is in the quiescent state, in which the accretion rate is very low, and the emission is very weak, sometimes even can not be detected. 
With the increase  of $\dot m$ to $\sim 0.01$, BH low-mass X-ray binary will enter the outburst state, during which BH low-mass X-ray binary exhibits significantly rise and 
decay of the luminosity and the spectral evolution with increasing or decreasing $\dot m$ \citep[][for review]{,Meyer2000a,Meyer2000b,Qiao2013,Remillard2006}. 
In many cases, we can observe a complete evolution of the accretion flow with $\dot m$ during the outburst, which can last a few months to a year.
As for AGN, we can not see the significant evolution of the accretion flow with $\dot m$ in the timescale of human. 
The suggested correspondence between the different spectral state in BH low-mass X-ray binaries and the different types of AGNs is actually uncertain. 
TDE intrinsically is an ideal laboratory to study the complete evolution of accretion flow, formation of the jet around a supermassive BH. The comparative study of 
the evolution of the accretion flow in TDEs and the evolution of the accretion flow in BH low-mass X-ray binaries is believed to be an effective approach to test the 
universality of BH accretion theory for different mass scales. Since the large change of $\dot m$ in BH low-mass X-ray binaries, a longer-time scale simulation in TDEs is 
needed for such a study. 

\subsection{On comparison with observations}
Based on the radiation hydrodynamic simulations, one of the key progresses in this paper is that we calculate the emergent spectrum of the super-Eddington accretion 
flow around a supermassive BH in a TDE, which definitely will have some potential applications for TDE research. 
Here, we give an example such as for understanding the so-called missing energy problem in TDEs \citep[][]{Piran2015}.
In \citet[][]{Piran2015}, the author proposed a mechanism, i.e., the orbital energy of the debris dissipated by shocks at the apocenter from the BH 
can power the observed optical emission of some TDEs. This process can well explain the observed temperatures, emission radii, and line widths of serval TDEs. 
Meanwhile, the decline rate of the optical light curve is also roughly consistent with the heating rate due to the shocks. 
However, this process can produce only a small fraction  ($\sim 0.01$) of the total energy that is expected from accretion onto the central supermassive 
BH, leading to the missing energy problem. 

On the other hand, the missing energy problem are also reflected from the X-ray light curves. Specifically, the total accreted mass $\Delta M$ in a complete 
TDE in general can be calculated by integrating the X-ray light curve, i.e.,
\begin{eqnarray}\label{e:deltaM}
\Delta M=k_{\rm bol}/\epsilon c^2\int_{t}^{\infty}L_{\rm x}(t)dt.
\end{eqnarray}
As we can see two factors, i.e., the correction factor $k_{\rm bol}$ and the radiative efficiency $\epsilon$ are needed for estimating $\Delta M$.
In \citet[][]{Lilixin2002ApJ}, the authors constructed the X-ray light curve of the prototype TDE NGC 5905  from the estimated disruption time 1990.54 year to 
1996.89 year, and calculated  $\Delta M$ with equation (\ref {e:deltaM}) by assuming $k_{\rm bol}=1.0$ and $\epsilon=0.1$.
 It is found that $\Delta M \sim (2.5\pm 0.5) \time 10^{-4} M_{\odot}$. The derived $\Delta M$ is significantly lower than the predicted value 
 to be $\sim 0.5\dot M_{\odot}$ if a sun-like main sequence star is disrupted, which however was interpreted as the disruption of a brown dwarf or a planet. 
The same method is used for estimating $\Delta M$ in several TDE sources, such as in RX J1242.61119 and SDSS J131122.15-012345.6  etc.
$\Delta M$ is calculated to be $\sim 0.01M_{\odot}$ for RX J1242.61119 by assuming $k_{\rm bol}=1$ and $\epsilon=0.1$ \citep[][]{Komossaetal.2004}.
$\Delta M$ is calculated also to be $\sim 0.01M_{\odot}$ for SDSS J131122.15-012345.6 by assuming $k_{\rm bol}=1.4$ and $\epsilon=0.1$ \citep[][]{Maksym2010}.

However, we should note that both the value of $k_{\rm bol}$ and  $\epsilon$ are uncertain, which can significantly affect the calculation of $\Delta M$.
According to the emergent spectrum presented in this paper, $k_{\rm bol}$ and  $\epsilon$ actually can be calculated precisely.
From Fig. \ref{f:sp8day} we can see, the emergent spectra are dominant by extreme ultraviolet (EUV) no matter the viewing angle $\theta$.
The value of $k_{\rm bol}$ can be from a few to a few hundred with increasing $\theta$. So we think that in several previous studies the adoption of $k_{\rm bol} \sim 1$
significantly underestimates the bolometric luminosity and further for $\Delta M$. Meanwhile, since the photon trapping effect, the radiative efficiency $\epsilon$
is expected to be less than 0.1, which theoretically can be calculated with the formula of $\epsilon={L_{\rm bol}/\dot M c^2}$. $L_{\rm bol}$ is bolometric luminosity,
which can be calculated by integrating the theoretical emergent spectra. Combing the correct $k_{\rm bol}$ and $\epsilon$, it is expected that $\Delta M$
could reach a reasonable value if assuming that a sun-like main sequence star is disrupted, consequently alleviating the missing energy problem.

The calculation of the emergent spectra depends on the data from the hydrodynamic simulations. In this paper, we perform the simulation by 
assuming $M_{\rm BH}=10^{6}M_{\odot}$, $M_{*}=M_{\odot}$, $R_{*}=R_{\odot}$, the injecting point at $R_{\rm c}=2R_{\rm T}$, and 
the injecting rate as $\dot M_{\rm inject}(t) \propto t^{-5/3}$, all of which can affect the results of the hydrodynamic simulation and further can affect the emergent spectra.
It is very necessary to test the effect of BH mass $M_{\rm BH}$, the type of the disrupted stars (i.e., $M_{*}$ and $R_{*}$), 
the mass injecting point  $R_{\rm inject}$, and the form of the  injecting rate $\dot M_{\rm inject}(t)$ to the emergent spectra. 
Further, we expect to construct a simulation database of the emergent spectra, which can be used to quantitatively study the origin and 
evolution of optical/UV-X-ray emission in TDEs.  
This study has  exceeded the research in the present paper, and definitely will be done in detail in the future.

\section*{Acknowledgments}
We thank the very useful discussions with Yan-Fei Jiang from Flatiron Institute for performing Athena++ code with radiative transfer. 
Erlin Qiao thank the referee for the very valuable comments, which make the discussions to the main results of the paper more complete.
This work is supported by the Strategic Priority Research Program of the Chinese Academy of Science (Grant No. XDB0550200),
Shandong Provincial Key Research and Development Program (No. 2022CXGC020106), National Supercomputing Center in Jinan for computing resource,
National Key R\&D Program of China (No. 2023YFA1607903) and National Natural Science Foundation of China (Grant No. 12333004, 12173048, 12473016).


\section*{Data availability}
The data underlying this article will be shared on reasonable request to the
corresponding author.

\bibliographystyle{mnras}
\bibliography{qiaoel_TDE}

\begin{thebibliography}{}
\makeatletter
\relax
\def\mn@urlcharsother{\let\do\@makeother \do\$\do\&\do\#\do\^\do\_\do\%\do\~}
\def\mn@doi{\begingroup\mn@urlcharsother \@ifnextchar [ {\mn@doi@}
  {\mn@doi@[]}}
\def\mn@doi@[#1]#2{\def\@tempa{#1}\ifx\@tempa\@empty \href
  {http://dx.doi.org/#2} {doi:#2}\else \href {http://dx.doi.org/#2} {#1}\fi
  \endgroup}
\def\mn@eprint#1#2{\mn@eprint@#1:#2::\@nil}
\def\mn@eprint@arXiv#1{\href {http://arxiv.org/abs/#1} {{\tt arXiv:#1}}}
\def\mn@eprint@dblp#1{\href {http://dblp.uni-trier.de/rec/bibtex/#1.xml}
  {dblp:#1}}
\def\mn@eprint@#1:#2:#3:#4\@nil{\def\@tempa {#1}\def\@tempb {#2}\def\@tempc
  {#3}\ifx \@tempc \@empty \let \@tempc \@tempb \let \@tempb \@tempa \fi \ifx
  \@tempb \@empty \def\@tempb {arXiv}\fi \@ifundefined
  {mn@eprint@\@tempb}{\@tempb:\@tempc}{\expandafter \expandafter \csname
  mn@eprint@\@tempb\endcsname \expandafter{\@tempc}}}

\bibitem[\protect\citeauthoryear{{Abramowicz}, {Czerny}, {Lasota}  \&
  {Szuszkiewicz}}{{Abramowicz} et~al.}{1988}]{Abramowicz1988}
{Abramowicz} M.~A.,  {Czerny} B.,  {Lasota} J.~P.,   {Szuszkiewicz} E.,  1988,
  \mn@doi [\apj] {10.1086/166683}, \href
  {http://adsabs.harvard.edu/abs/1988ApJ...332..646A} {332, 646}

\bibitem[\protect\citeauthoryear{{Arcavi} et~al.,}{{Arcavi}
  et~al.}{2014}]{Arcavi2014}
{Arcavi} I.,  et~al., 2014, \mn@doi [\apj] {10.1088/0004-637X/793/1/38}, \href
  {https://ui.adsabs.harvard.edu/abs/2014ApJ...793...38A} {793, 38}

\bibitem[\protect\citeauthoryear{{Auchettl}, {Guillochon}  \&
  {Ramirez-Ruiz}}{{Auchettl} et~al.}{2017}]{Auchettl2017}
{Auchettl} K.,  {Guillochon} J.,   {Ramirez-Ruiz} E.,  2017, \mn@doi [\apj]
  {10.3847/1538-4357/aa633b}, \href
  {https://ui.adsabs.harvard.edu/abs/2017ApJ...838..149A} {838, 149}

\bibitem[\protect\citeauthoryear{{Bade}, {Komossa}  \& {Dahlem}}{{Bade}
  et~al.}{1996}]{Bade1996}
{Bade} N.,  {Komossa} S.,   {Dahlem} M.,  1996, \aap, \href
  {https://ui.adsabs.harvard.edu/abs/1996A&A...309L..35B} {309, L35}

\bibitem[\protect\citeauthoryear{{Bonnerot} \& {Lu}}{{Bonnerot} \&
  {Lu}}{2020}]{Bonnerot2020}
{Bonnerot} C.,  {Lu} W.,  2020, \mn@doi [\mnras] {10.1093/mnras/staa1246},
  \href {https://ui.adsabs.harvard.edu/abs/2020MNRAS.495.1374B} {495, 1374}

\bibitem[\protect\citeauthoryear{{Bonnerot} \& {Stone}}{{Bonnerot} \&
  {Stone}}{2021}]{Bonnerot2021SSRv}
{Bonnerot} C.,  {Stone} N.~C.,  2021, \mn@doi [\ssr]
  {10.1007/s11214-020-00789-1}, \href
  {https://ui.adsabs.harvard.edu/abs/2021SSRv..217...16B} {217, 16}

\bibitem[\protect\citeauthoryear{{Bonnerot}, {Rossi}  \& {Lodato}}{{Bonnerot}
  et~al.}{2017}]{Bonnerot2017}
{Bonnerot} C.,  {Rossi} E.~M.,   {Lodato} G.,  2017, \mn@doi [\mnras]
  {10.1093/mnras/stw2547}, \href
  {https://ui.adsabs.harvard.edu/abs/2017MNRAS.464.2816B} {464, 2816}

\bibitem[\protect\citeauthoryear{{Bu}, {Qiao}, {Yang}, {Liu}, {Chen}  \&
  {Wu}}{{Bu} et~al.}{2022}]{Bu2022}
{Bu} D.-F.,  {Qiao} E.,  {Yang} X.-H.,  {Liu} J.,  {Chen} Z.,   {Wu} Y.,  2022,
  \mn@doi [\mnras] {10.1093/mnras/stac2399}, \href
  {https://ui.adsabs.harvard.edu/abs/2022MNRAS.516.2833B} {516, 2833}

\bibitem[\protect\citeauthoryear{{Bu}, {Qiao}  \& {Yang}}{{Bu}
  et~al.}{2023}]{Bu2023}
{Bu} D.-F.,  {Qiao} E.,   {Yang} X.-H.,  2023, \mn@doi [\mnras]
  {10.1093/mnras/stad1696}, \href
  {https://ui.adsabs.harvard.edu/abs/2023MNRAS.523.4136B} {523, 4136}

\bibitem[\protect\citeauthoryear{{Carter} \& {Luminet}}{{Carter} \&
  {Luminet}}{1982}]{Carter1982}
{Carter} B.,  {Luminet} J.~P.,  1982, \mn@doi [\nat] {10.1038/296211a0}, \href
  {https://ui.adsabs.harvard.edu/abs/1982Natur.296..211C} {296, 211}

\bibitem[\protect\citeauthoryear{{Chornock} et~al.,}{{Chornock}
  et~al.}{2014}]{Chornock2014}
{Chornock} R.,  et~al., 2014, \mn@doi [\apj] {10.1088/0004-637X/780/1/44},
  \href {https://ui.adsabs.harvard.edu/abs/2014ApJ...780...44C} {780, 44}

\bibitem[\protect\citeauthoryear{{Coughlin} \& {Begelman}}{{Coughlin} \&
  {Begelman}}{2014}]{Coughlin2014}
{Coughlin} E.~R.,  {Begelman} M.~C.,  2014, \mn@doi [\apj]
  {10.1088/0004-637X/781/2/82}, \href
  {https://ui.adsabs.harvard.edu/abs/2014ApJ...781...82C} {781, 82}

\bibitem[\protect\citeauthoryear{{Curd} \& {Narayan}}{{Curd} \&
  {Narayan}}{2019}]{Curd2019}
{Curd} B.,  {Narayan} R.,  2019, \mn@doi [\mnras] {10.1093/mnras/sty3134},
  \href {https://ui.adsabs.harvard.edu/abs/2019MNRAS.483..565C} {483, 565}

\bibitem[\protect\citeauthoryear{{Dai}, {McKinney}, {Roth}, {Ramirez-Ruiz}  \&
  {Miller}}{{Dai} et~al.}{2018}]{Dai2018}
{Dai} L.,  {McKinney} J.~C.,  {Roth} N.,  {Ramirez-Ruiz} E.,   {Miller} M.~C.,
  2018, \mn@doi [\apjl] {10.3847/2041-8213/aab429}, \href
  {https://ui.adsabs.harvard.edu/abs/2018ApJ...859L..20D} {859, L20}

\bibitem[\protect\citeauthoryear{{Dai}, {Lodato}  \& {Cheng}}{{Dai}
  et~al.}{2021}]{Dai2021SSRv}
{Dai} J.~L.,  {Lodato} G.,   {Cheng} R.,  2021, \mn@doi [\ssr]
  {10.1007/s11214-020-00747-x}, \href
  {https://ui.adsabs.harvard.edu/abs/2021SSRv..217...12D} {217, 12}

\bibitem[\protect\citeauthoryear{{Esin}, {McClintock}  \& {Narayan}}{{Esin}
  et~al.}{1997}]{Esin1997}
{Esin} A.~A.,  {McClintock} J.~E.,   {Narayan} R.,  1997, \mn@doi [\apj]
  {10.1086/304829}, \href {http://adsabs.harvard.edu/abs/1997ApJ...489..865E}
  {489, 865}

\bibitem[\protect\citeauthoryear{{Evans} \& {Kochanek}}{{Evans} \&
  {Kochanek}}{1989}]{Evans1989}
{Evans} C.~R.,  {Kochanek} C.~S.,  1989, \mn@doi [\apjl] {10.1086/185567},
  \href {https://ui.adsabs.harvard.edu/abs/1989ApJ...346L..13E} {346, L13}

\bibitem[\protect\citeauthoryear{{Falcke}, {K{\"o}rding}  \&
  {Markoff}}{{Falcke} et~al.}{2004}]{Falcke2004}
{Falcke} H.,  {K{\"o}rding} E.,   {Markoff} S.,  2004, \mn@doi [\aap]
  {10.1051/0004-6361:20031683}, \href
  {https://ui.adsabs.harvard.edu/abs/2004A&A...414..895F} {414, 895}

\bibitem[\protect\citeauthoryear{{Frank} \& {Rees}}{{Frank} \&
  {Rees}}{1976}]{Frank1976}
{Frank} J.,  {Rees} M.~J.,  1976, \mn@doi [\mnras] {10.1093/mnras/176.3.633},
  \href {https://ui.adsabs.harvard.edu/abs/1976MNRAS.176..633F} {176, 633}

\bibitem[\protect\citeauthoryear{{Gezari}}{{Gezari}}{2021}]{Gezari2021ARAA}
{Gezari} S.,  2021, \mn@doi [\araa] {10.1146/annurev-astro-111720-030029},
  \href {https://ui.adsabs.harvard.edu/abs/2021ARA&A..59...21G} {59, 21}

\bibitem[\protect\citeauthoryear{{Gezari} et~al.,}{{Gezari}
  et~al.}{2006}]{Gezari2006}
{Gezari} S.,  et~al., 2006, \mn@doi [\apjl] {10.1086/509918}, \href
  {https://ui.adsabs.harvard.edu/abs/2006ApJ...653L..25G} {653, L25}

\bibitem[\protect\citeauthoryear{{Gezari} et~al.,}{{Gezari}
  et~al.}{2008}]{Gezari2008}
{Gezari} S.,  et~al., 2008, \mn@doi [\apj] {10.1086/529008}, \href
  {https://ui.adsabs.harvard.edu/abs/2008ApJ...676..944G} {676, 944}

\bibitem[\protect\citeauthoryear{{Gezari} et~al.,}{{Gezari}
  et~al.}{2009}]{Gezari2009}
{Gezari} S.,  et~al., 2009, \mn@doi [\apj] {10.1088/0004-637X/698/2/1367},
  \href {https://ui.adsabs.harvard.edu/abs/2009ApJ...698.1367G} {698, 1367}

\bibitem[\protect\citeauthoryear{{Gezari} et~al.,}{{Gezari}
  et~al.}{2012}]{Gezari2012}
{Gezari} S.,  et~al., 2012, \mn@doi [\nat] {10.1038/nature10990}, \href
  {https://ui.adsabs.harvard.edu/abs/2012Natur.485..217G} {485, 217}

\bibitem[\protect\citeauthoryear{{Golightly}, {Coughlin}  \&
  {Nixon}}{{Golightly} et~al.}{2019a}]{Golightly2019ApJ...872..163G}
{Golightly} E. C.~A.,  {Coughlin} E.~R.,   {Nixon} C.~J.,  2019a, \mn@doi
  [\apj] {10.3847/1538-4357/aafd2f}, \href
  {https://ui.adsabs.harvard.edu/abs/2019ApJ...872..163G} {872, 163}

\bibitem[\protect\citeauthoryear{{Golightly}, {Nixon}  \&
  {Coughlin}}{{Golightly} et~al.}{2019b}]{Golightly2019ApJ...882L..26G}
{Golightly} E.~C.~A.,  {Nixon} C.~J.,   {Coughlin} E.~R.,  2019b, \mn@doi
  [\apjl] {10.3847/2041-8213/ab380d}, \href
  {https://ui.adsabs.harvard.edu/abs/2019ApJ...882L..26G} {882, L26}

\bibitem[\protect\citeauthoryear{{Grupe}, {Beuermann}, {Mannheim}  \&
  {Thomas}}{{Grupe} et~al.}{1999}]{Grupe1999}
{Grupe} D.,  {Beuermann} K.,  {Mannheim} K.,   {Thomas} H.~C.,  1999, \mn@doi
  [\aap] {10.48550/arXiv.astro-ph/9908347}, \href
  {https://ui.adsabs.harvard.edu/abs/1999A&A...350..805G} {350, 805}

\bibitem[\protect\citeauthoryear{{Guillochon} \& {Ramirez-Ruiz}}{{Guillochon}
  \& {Ramirez-Ruiz}}{2013}]{Guillochon2013}
{Guillochon} J.,  {Ramirez-Ruiz} E.,  2013, \mn@doi [\apj]
  {10.1088/0004-637X/767/1/25}, \href
  {https://ui.adsabs.harvard.edu/abs/2013ApJ...767...25G} {767, 25}

\bibitem[\protect\citeauthoryear{{Guolo}, {Gezari}, {Yao}, {van Velzen},
  {Hammerstein}, {Cenko}  \& {Tokayer}}{{Guolo} et~al.}{2024}]{Guolo2024}
{Guolo} M.,  {Gezari} S.,  {Yao} Y.,  {van Velzen} S.,  {Hammerstein} E.,
  {Cenko} S.~B.,   {Tokayer} Y.~M.,  2024, \mn@doi [\apj]
  {10.3847/1538-4357/ad2f9f}, \href
  {https://ui.adsabs.harvard.edu/abs/2024ApJ...966..160G} {966, 160}

\bibitem[\protect\citeauthoryear{{Gurzadian} \& {Ozernoi}}{{Gurzadian} \&
  {Ozernoi}}{1979}]{Gurzadian1979}
{Gurzadian} V.~G.,  {Ozernoi} L.~M.,  1979, \mn@doi [\nat] {10.1038/280214a0},
  \href {https://ui.adsabs.harvard.edu/abs/1979Natur.280..214G} {280, 214}

\bibitem[\protect\citeauthoryear{{Hayasaki}, {Stone}  \& {Loeb}}{{Hayasaki}
  et~al.}{2013}]{Hayasaki2013}
{Hayasaki} K.,  {Stone} N.,   {Loeb} A.,  2013, \mn@doi [\mnras]
  {10.1093/mnras/stt871}, \href
  {https://ui.adsabs.harvard.edu/abs/2013MNRAS.434..909H} {434, 909}

\bibitem[\protect\citeauthoryear{{Higginbottom}, {Knigge}, {Long}, {Sim}  \&
  {Matthews}}{{Higginbottom} et~al.}{2013}]{Higginbottom2013}
{Higginbottom} N.,  {Knigge} C.,  {Long} K.~S.,  {Sim} S.~A.,   {Matthews}
  J.~H.,  2013, \mn@doi [\mnras] {10.1093/mnras/stt1658}, \href
  {https://ui.adsabs.harvard.edu/abs/2013MNRAS.436.1390H} {436, 1390}

\bibitem[\protect\citeauthoryear{{Hills}}{{Hills}}{1975}]{Hills1975}
{Hills} J.~G.,  1975, \mn@doi [\nat] {10.1038/254295a0}, \href
  {https://ui.adsabs.harvard.edu/abs/1975Natur.254..295H} {254, 295}

\bibitem[\protect\citeauthoryear{{Hills}}{{Hills}}{1988}]{Hills1988}
{Hills} J.~G.,  1988, \mn@doi [\nat] {10.1038/331687a0}, \href
  {https://ui.adsabs.harvard.edu/abs/1988Natur.331..687H} {331, 687}

\bibitem[\protect\citeauthoryear{{Holoien} et~al.,}{{Holoien}
  et~al.}{2014}]{Holoien2014}
{Holoien} T.~W.~S.,  et~al., 2014, \mn@doi [\mnras] {10.1093/mnras/stu1922},
  \href {https://ui.adsabs.harvard.edu/abs/2014MNRAS.445.3263H} {445, 3263}

\bibitem[\protect\citeauthoryear{{Holoien} et~al.,}{{Holoien}
  et~al.}{2016}]{Holoien2016}
{Holoien} T.~W.~S.,  et~al., 2016, \mn@doi [\mnras] {10.1093/mnras/stv2486},
  \href {https://ui.adsabs.harvard.edu/abs/2016MNRAS.455.2918H} {455, 2918}

\bibitem[\protect\citeauthoryear{{Jiang}}{{Jiang}}{2021}]{2021ApJS..253...49J}
{Jiang} Y.-F.,  2021, \mn@doi [\apjs] {10.3847/1538-4365/abe303}, \href
  {https://ui.adsabs.harvard.edu/abs/2021ApJS..253...49J} {253, 49}

\bibitem[\protect\citeauthoryear{{Jiang}, {Stone}  \& {Davis}}{{Jiang}
  et~al.}{2012}]{Jiang2012}
{Jiang} Y.-F.,  {Stone} J.~M.,   {Davis} S.~W.,  2012, \mn@doi [\apjs]
  {10.1088/0067-0049/199/1/14}, \href
  {https://ui.adsabs.harvard.edu/abs/2012ApJS..199...14J} {199, 14}

\bibitem[\protect\citeauthoryear{{Jiang}, {Stone}  \& {Davis}}{{Jiang}
  et~al.}{2014}]{Jiang2014ApJ...796..106J}
{Jiang} Y.-F.,  {Stone} J.~M.,   {Davis} S.~W.,  2014, \mn@doi [\apj]
  {10.1088/0004-637X/796/2/106}, \href
  {https://ui.adsabs.harvard.edu/abs/2014ApJ...796..106J} {796, 106}

\bibitem[\protect\citeauthoryear{{Jiang}, {Guillochon}  \& {Loeb}}{{Jiang}
  et~al.}{2016}]{Jiang2016}
{Jiang} Y.-F.,  {Guillochon} J.,   {Loeb} A.,  2016, \mn@doi [\apj]
  {10.3847/0004-637X/830/2/125}, \href
  {https://ui.adsabs.harvard.edu/abs/2016ApJ...830..125J} {830, 125}

\bibitem[\protect\citeauthoryear{{Jiang}, {Stone}  \& {Davis}}{{Jiang}
  et~al.}{2019}]{Jiang2019}
{Jiang} Y.-F.,  {Stone} J.~M.,   {Davis} S.~W.,  2019, \mn@doi [\apj]
  {10.3847/1538-4357/ab29ff}, \href
  {https://ui.adsabs.harvard.edu/abs/2019ApJ...880...67J} {880, 67}

\bibitem[\protect\citeauthoryear{{Komossa}}{{Komossa}}{2004}]{Komossa2004IAUS}
{Komossa} S.,  2004, in {Storchi-Bergmann} T.,  {Ho} L.~C.,   {Schmitt} H.~R.,
  eds,  Vol. 222, The Interplay Among Black Holes, Stars and ISM in Galactic
  Nuclei. pp 45--48, \mn@doi{10.1017/S1743921304001425}

\bibitem[\protect\citeauthoryear{{Komossa}}{{Komossa}}{2015}]{Komossa2015}
{Komossa} S.,  2015, \mn@doi [Journal of High Energy Astrophysics]
  {10.1016/j.jheap.2015.04.006}, \href
  {https://ui.adsabs.harvard.edu/abs/2015JHEAp...7..148K} {7, 148}

\bibitem[\protect\citeauthoryear{{Komossa} \& {Bade}}{{Komossa} \&
  {Bade}}{1999}]{Komossa1999b}
{Komossa} S.,  {Bade} N.,  1999, \mn@doi [\aap]
  {10.48550/arXiv.astro-ph/9901141}, \href
  {https://ui.adsabs.harvard.edu/abs/1999A&A...343..775K} {343, 775}

\bibitem[\protect\citeauthoryear{{Komossa} \& {Greiner}}{{Komossa} \&
  {Greiner}}{1999}]{Komossa1999a}
{Komossa} S.,  {Greiner} J.,  1999, \mn@doi [\aap]
  {10.48550/arXiv.astro-ph/9908216}, \href
  {https://ui.adsabs.harvard.edu/abs/1999A&A...349L..45K} {349, L45}

\bibitem[\protect\citeauthoryear{{Komossa}, {Halpern}, {Schartel}, {Hasinger},
  {Santos-Lleo}  \& {Predehl}}{{Komossa} et~al.}{2004}]{Komossaetal.2004}
{Komossa} S.,  {Halpern} J.,  {Schartel} N.,  {Hasinger} G.,  {Santos-Lleo} M.,
    {Predehl} P.,  2004, \mn@doi [\apjl] {10.1086/382046}, \href
  {https://ui.adsabs.harvard.edu/abs/2004ApJ...603L..17K} {603, L17}

\bibitem[\protect\citeauthoryear{{Komossa} et~al.,}{{Komossa}
  et~al.}{2008}]{Komossa_etal2008}
{Komossa} S.,  et~al., 2008, \mn@doi [\apjl] {10.1086/588281}, \href
  {https://ui.adsabs.harvard.edu/abs/2008ApJ...678L..13K} {678, L13}

\bibitem[\protect\citeauthoryear{{K{\"o}rding}, {Jester}  \&
  {Fender}}{{K{\"o}rding} et~al.}{2006}]{Kording2006}
{K{\"o}rding} E.~G.,  {Jester} S.,   {Fender} R.,  2006, \mn@doi [\mnras]
  {10.1111/j.1365-2966.2006.10954.x}, \href
  {https://ui.adsabs.harvard.edu/abs/2006MNRAS.372.1366K} {372, 1366}

\bibitem[\protect\citeauthoryear{{Law-Smith}, {Guillochon}  \&
  {Ramirez-Ruiz}}{{Law-Smith} et~al.}{2019}]{Law-Smith2019}
{Law-Smith} J.,  {Guillochon} J.,   {Ramirez-Ruiz} E.,  2019, \mn@doi [\apjl]
  {10.3847/2041-8213/ab379a}, \href
  {https://ui.adsabs.harvard.edu/abs/2019ApJ...882L..25L} {882, L25}

\bibitem[\protect\citeauthoryear{{Li}, {Narayan}  \& {Menou}}{{Li}
  et~al.}{2002}]{Lilixin2002ApJ}
{Li} L.-X.,  {Narayan} R.,   {Menou} K.,  2002, \mn@doi [\apj]
  {10.1086/341890}, \href
  {https://ui.adsabs.harvard.edu/abs/2002ApJ...576..753L} {576, 753}

\bibitem[\protect\citeauthoryear{{Liu} \& {Qiao}}{{Liu} \&
  {Qiao}}{2022}]{Liu2022iSci...25j3544L}
{Liu} B.~F.,  {Qiao} E.,  2022, \mn@doi [iScience]
  {10.1016/j.isci.2021.103544}, \href
  {https://ui.adsabs.harvard.edu/abs/2022iSci...25j3544L} {25, 103544}

\bibitem[\protect\citeauthoryear{{Lodato} \& {Rossi}}{{Lodato} \&
  {Rossi}}{2011}]{Lodato2011}
{Lodato} G.,  {Rossi} E.~M.,  2011, \mn@doi [\mnras]
  {10.1111/j.1365-2966.2010.17448.x}, \href
  {https://ui.adsabs.harvard.edu/abs/2011MNRAS.410..359L} {410, 359}

\bibitem[\protect\citeauthoryear{{Lodato}, {King}  \& {Pringle}}{{Lodato}
  et~al.}{2009}]{Lodato2009}
{Lodato} G.,  {King} A.~R.,   {Pringle} J.~E.,  2009, \mn@doi [\mnras]
  {10.1111/j.1365-2966.2008.14049.x}, \href
  {https://ui.adsabs.harvard.edu/abs/2009MNRAS.392..332L} {392, 332}

\bibitem[\protect\citeauthoryear{{Loeb} \& {Ulmer}}{{Loeb} \&
  {Ulmer}}{1997}]{Loeb1997}
{Loeb} A.,  {Ulmer} A.,  1997, \mn@doi [\apj] {10.1086/304814}, \href
  {https://ui.adsabs.harvard.edu/abs/1997ApJ...489..573L} {489, 573}

\bibitem[\protect\citeauthoryear{{Long} \& {Knigge}}{{Long} \&
  {Knigge}}{2002}]{Long2002}
{Long} K.~S.,  {Knigge} C.,  2002, \mn@doi [\apj] {10.1086/342879}, \href
  {https://ui.adsabs.harvard.edu/abs/2002ApJ...579..725L} {579, 725}

\bibitem[\protect\citeauthoryear{{Lowrie}, {Morel}  \& {Hittinger}}{{Lowrie}
  et~al.}{1999}]{Lowrie1999}
{Lowrie} R.~B.,  {Morel} J.~E.,   {Hittinger} J.~A.,  1999, \mn@doi [\apj]
  {10.1086/307515}, \href
  {https://ui.adsabs.harvard.edu/abs/1999ApJ...521..432L} {521, 432}

\bibitem[\protect\citeauthoryear{{Lu} \& {Bonnerot}}{{Lu} \&
  {Bonnerot}}{2020}]{Luwenbin2020}
{Lu} W.,  {Bonnerot} C.,  2020, \mn@doi [\mnras] {10.1093/mnras/stz3405}, \href
  {https://ui.adsabs.harvard.edu/abs/2020MNRAS.492..686L} {492, 686}

\bibitem[\protect\citeauthoryear{{Luminet} \& {Marck}}{{Luminet} \&
  {Marck}}{1985}]{Luminet1985}
{Luminet} J.~P.,  {Marck} J.~A.,  1985, \mn@doi [\mnras]
  {10.1093/mnras/212.1.57}, \href
  {https://ui.adsabs.harvard.edu/abs/1985MNRAS.212...57L} {212, 57}

\bibitem[\protect\citeauthoryear{{Maccarone}}{{Maccarone}}{2003}]{Maccarone2003}
{Maccarone} T.~J.,  2003, \mn@doi [\aap] {10.1051/0004-6361:20031146}, \href
  {http://adsabs.harvard.edu/abs/2003A%26A...409..697M} {409, 697}

\bibitem[\protect\citeauthoryear{{Maksym}, {Ulmer}  \& {Eracleous}}{{Maksym}
  et~al.}{2010}]{Maksym2010}
{Maksym} W.~P.,  {Ulmer} M.~P.,   {Eracleous} M.,  2010, \mn@doi [\apj]
  {10.1088/0004-637X/722/2/1035}, \href
  {https://ui.adsabs.harvard.edu/abs/2010ApJ...722.1035M} {722, 1035}

\bibitem[\protect\citeauthoryear{{Matthews}, {Knigge}, {Long}, {Sim},
  {Higginbottom}  \& {Mangham}}{{Matthews} et~al.}{2016}]{Matthews2016}
{Matthews} J.~H.,  {Knigge} C.,  {Long} K.~S.,  {Sim} S.~A.,  {Higginbottom}
  N.,   {Mangham} S.~W.,  2016, \mn@doi [\mnras] {10.1093/mnras/stw323}, \href
  {https://ui.adsabs.harvard.edu/abs/2016MNRAS.458..293M} {458, 293}

\bibitem[\protect\citeauthoryear{{Matthews}, {Knigge}  \& {Long}}{{Matthews}
  et~al.}{2017}]{Matthews2017}
{Matthews} J.~H.,  {Knigge} C.,   {Long} K.~S.,  2017, \mn@doi [\mnras]
  {10.1093/mnras/stx231}, \href
  {https://ui.adsabs.harvard.edu/abs/2017MNRAS.467.2571M} {467, 2571}

\bibitem[\protect\citeauthoryear{{McKinney}, {Tchekhovskoy}, {Sadowski}  \&
  {Narayan}}{{McKinney} et~al.}{2014}]{McKinney2014}
{McKinney} J.~C.,  {Tchekhovskoy} A.,  {Sadowski} A.,   {Narayan} R.,  2014,
  \mn@doi [\mnras] {10.1093/mnras/stu762}, \href
  {https://ui.adsabs.harvard.edu/abs/2014MNRAS.441.3177M} {441, 3177}

\bibitem[\protect\citeauthoryear{{Merloni}, {Heinz}  \& {di Matteo}}{{Merloni}
  et~al.}{2003}]{Merloni2003}
{Merloni} A.,  {Heinz} S.,   {di Matteo} T.,  2003, \mn@doi [\mnras]
  {10.1046/j.1365-2966.2003.07017.x}, \href
  {http://adsabs.harvard.edu/abs/2003MNRAS.345.1057M} {345, 1057}

\bibitem[\protect\citeauthoryear{{Metzger}}{{Metzger}}{2022}]{Metzger2022}
{Metzger} B.~D.,  2022, \mn@doi [\apjl] {10.3847/2041-8213/ac90ba}, \href
  {https://ui.adsabs.harvard.edu/abs/2022ApJ...937L..12M} {937, L12}

\bibitem[\protect\citeauthoryear{{Metzger} \& {Stone}}{{Metzger} \&
  {Stone}}{2016}]{Metzger2016}
{Metzger} B.~D.,  {Stone} N.~C.,  2016, \mn@doi [\mnras]
  {10.1093/mnras/stw1394}, \href
  {https://ui.adsabs.harvard.edu/abs/2016MNRAS.461..948M} {461, 948}

\bibitem[\protect\citeauthoryear{{Meyer}, {Liu}  \& {Meyer-Hofmeister}}{{Meyer}
  et~al.}{2000a}]{Meyer2000a}
{Meyer} F.,  {Liu} B.~F.,   {Meyer-Hofmeister} E.,  2000a, \aap, \href
  {http://ads.bao.ac.cn/abs/2000A%26A...354L..67M} {354, L67}

\bibitem[\protect\citeauthoryear{{Meyer}, {Liu}  \& {Meyer-Hofmeister}}{{Meyer}
  et~al.}{2000b}]{Meyer2000b}
{Meyer} F.,  {Liu} B.~F.,   {Meyer-Hofmeister} E.,  2000b, \aap, \href
  {http://ads.bao.ac.cn/abs/2000A%26A...361..175M} {361, 175}

\bibitem[\protect\citeauthoryear{{Miller}}{{Miller}}{2015}]{Miller2015}
{Miller} M.~C.,  2015, \mn@doi [\apj] {10.1088/0004-637X/805/1/83}, \href
  {https://ui.adsabs.harvard.edu/abs/2015ApJ...805...83M} {805, 83}

\bibitem[\protect\citeauthoryear{{Moravec}, {Svoboda}, {Borkar}, {Boorman},
  {Kynoch}, {Panessa}, {Mingo}  \& {Guainazzi}}{{Moravec}
  et~al.}{2022}]{Moravec2022}
{Moravec} E.,  {Svoboda} J.,  {Borkar} A.,  {Boorman} P.,  {Kynoch} D.,
  {Panessa} F.,  {Mingo} B.,   {Guainazzi} M.,  2022, \mn@doi [\aap]
  {10.1051/0004-6361/202142870}, \href
  {https://ui.adsabs.harvard.edu/abs/2022A&A...662A..28M} {662, A28}

\bibitem[\protect\citeauthoryear{{Narayan} \& {Yi}}{{Narayan} \&
  {Yi}}{1994}]{Narayan1994}
{Narayan} R.,  {Yi} I.,  1994, \mn@doi [\apjl] {10.1086/187381}, \href
  {http://ads.bao.ac.cn/abs/1994ApJ...428L..13N} {428, L13}

\bibitem[\protect\citeauthoryear{{Narayan} \& {Yi}}{{Narayan} \&
  {Yi}}{1995}]{Narayan1995b}
{Narayan} R.,  {Yi} I.,  1995, \mn@doi [\apj] {10.1086/176343}, \href
  {http://ads.bao.ac.cn/abs/1995ApJ...452..710N} {452, 710}

\bibitem[\protect\citeauthoryear{{Ohsuga} \& {Mineshige}}{{Ohsuga} \&
  {Mineshige}}{2011}]{Ohsuga2011}
{Ohsuga} K.,  {Mineshige} S.,  2011, \mn@doi [\apj]
  {10.1088/0004-637X/736/1/2}, \href
  {https://ui.adsabs.harvard.edu/abs/2011ApJ...736....2O} {736, 2}

\bibitem[\protect\citeauthoryear{{Ohsuga}, {Mori}, {Nakamoto}  \&
  {Mineshige}}{{Ohsuga} et~al.}{2005}]{Ohsuga2005}
{Ohsuga} K.,  {Mori} M.,  {Nakamoto} T.,   {Mineshige} S.,  2005, \mn@doi
  [\apj] {10.1086/430728}, \href
  {https://ui.adsabs.harvard.edu/abs/2005ApJ...628..368O} {628, 368}

\bibitem[\protect\citeauthoryear{{Phinney}}{{Phinney}}{1989}]{Phinney1989}
{Phinney} E.~S.,  1989, in {Morris} M.,  ed.,  Vol. 136, The Center of the
  Galaxy. p.~543

\bibitem[\protect\citeauthoryear{{Piran}, {Svirski}, {Krolik}, {Cheng}  \&
  {Shiokawa}}{{Piran} et~al.}{2015}]{Piran2015}
{Piran} T.,  {Svirski} G.,  {Krolik} J.,  {Cheng} R.~M.,   {Shiokawa} H.,
  2015, \mn@doi [\apj] {10.1088/0004-637X/806/2/164}, \href
  {https://ui.adsabs.harvard.edu/abs/2015ApJ...806..164P} {806, 164}

\bibitem[\protect\citeauthoryear{{Qiao} \& {Liu}}{{Qiao} \&
  {Liu}}{2013}]{Qiao2013}
{Qiao} E.,  {Liu} B.~F.,  2013, \mn@doi [\apj] {10.1088/0004-637X/764/1/2},
  \href {http://ads.bao.ac.cn/abs/2013ApJ...764....2Q} {764, 2}

\bibitem[\protect\citeauthoryear{{Rees}}{{Rees}}{1988}]{Rees1988}
{Rees} M.~J.,  1988, \mn@doi [\nat] {10.1038/333523a0}, \href
  {https://ui.adsabs.harvard.edu/abs/1988Natur.333..523R} {333, 523}

\bibitem[\protect\citeauthoryear{{Remillard} \& {McClintock}}{{Remillard} \&
  {McClintock}}{2006}]{Remillard2006}
{Remillard} R.~A.,  {McClintock} J.~E.,  2006, \mn@doi [\araa]
  {10.1146/annurev.astro.44.051905.092532}, \href
  {http://adsabs.harvard.edu/abs/2006ARA%26A..44...49R} {44, 49}

\bibitem[\protect\citeauthoryear{{Rossi}, {Stone}, {Law-Smith}, {Macleod},
  {Lodato}, {Dai}  \& {Mandel}}{{Rossi} et~al.}{2021}]{Rossi2021}
{Rossi} E.~M.,  {Stone} N.~C.,  {Law-Smith} J.~A.~P.,  {Macleod} M.,  {Lodato}
  G.,  {Dai} J.~L.,   {Mandel} I.,  2021, \mn@doi [\ssr]
  {10.1007/s11214-021-00818-7}, \href
  {https://ui.adsabs.harvard.edu/abs/2021SSRv..217...40R} {217, 40}

\bibitem[\protect\citeauthoryear{{Roth}, {Kasen}, {Guillochon}  \&
  {Ramirez-Ruiz}}{{Roth} et~al.}{2016}]{Roth2016}
{Roth} N.,  {Kasen} D.,  {Guillochon} J.,   {Ramirez-Ruiz} E.,  2016, \mn@doi
  [\apj] {10.3847/0004-637X/827/1/3}, \href
  {https://ui.adsabs.harvard.edu/abs/2016ApJ...827....3R} {827, 3}

\bibitem[\protect\citeauthoryear{{Ryu}, {Krolik}, {Piran}  \& {Noble}}{{Ryu}
  et~al.}{2020a}]{Ryu2020ApJ...904...98R}
{Ryu} T.,  {Krolik} J.,  {Piran} T.,   {Noble} S.~C.,  2020a, \mn@doi [\apj]
  {10.3847/1538-4357/abb3cf}, \href
  {https://ui.adsabs.harvard.edu/abs/2020ApJ...904...98R} {904, 98}

\bibitem[\protect\citeauthoryear{{Ryu}, {Krolik}, {Piran}  \& {Noble}}{{Ryu}
  et~al.}{2020b}]{Ryu2020ApJ...904...99R}
{Ryu} T.,  {Krolik} J.,  {Piran} T.,   {Noble} S.~C.,  2020b, \mn@doi [\apj]
  {10.3847/1538-4357/abb3cd}, \href
  {https://ui.adsabs.harvard.edu/abs/2020ApJ...904...99R} {904, 99}

\bibitem[\protect\citeauthoryear{{Ryu}, {Krolik}, {Piran}  \& {Noble}}{{Ryu}
  et~al.}{2020c}]{Ryu2020ApJ...904..100R}
{Ryu} T.,  {Krolik} J.,  {Piran} T.,   {Noble} S.~C.,  2020c, \mn@doi [\apj]
  {10.3847/1538-4357/abb3ce}, \href
  {https://ui.adsabs.harvard.edu/abs/2020ApJ...904..100R} {904, 100}

\bibitem[\protect\citeauthoryear{{Sacchi} \& {Lodato}}{{Sacchi} \&
  {Lodato}}{2019}]{Sacchi2019}
{Sacchi} A.,  {Lodato} G.,  2019, \mn@doi [\mnras] {10.1093/mnras/stz981},
  \href {https://ui.adsabs.harvard.edu/abs/2019MNRAS.486.1833S} {486, 1833}

\bibitem[\protect\citeauthoryear{{Sadowski} \& {Narayan}}{{Sadowski} \&
  {Narayan}}{2015}]{Sadowski2015}
{Sadowski} A.,  {Narayan} R.,  2015, \mn@doi [\mnras] {10.1093/mnras/stv1802},
  \href {https://ui.adsabs.harvard.edu/abs/2015MNRAS.453.3213S} {453, 3213}

\bibitem[\protect\citeauthoryear{{S{{a}}dowski}, {Narayan}, {McKinney}  \&
  {Tchekhovskoy}}{{S{{a}}dowski} et~al.}{2014}]{Sadowski2014}
{S{{a}}dowski} A.,  {Narayan} R.,  {McKinney} J.~C.,   {Tchekhovskoy} A.,
  2014, \mn@doi [\mnras] {10.1093/mnras/stt2479}, \href
  {https://ui.adsabs.harvard.edu/abs/2014MNRAS.439..503S} {439, 503}

\bibitem[\protect\citeauthoryear{{Saxton}, {Komossa}, {Auchettl}  \&
  {Jonker}}{{Saxton} et~al.}{2021}]{Saxton2021SSRv}
{Saxton} R.,  {Komossa} S.,  {Auchettl} K.,   {Jonker} P.~G.,  2021,
  {Correction to: X-Ray Properties of TDEs}, Space Science Reviews, Volume 217,
  Issue 1, article id.18 (\mn@eprint {arXiv} {2103.15442}),
  \mn@doi{10.1007/s11214-020-00759-7}

\bibitem[\protect\citeauthoryear{{Shiokawa}, {Krolik}, {Cheng}, {Piran}  \&
  {Noble}}{{Shiokawa} et~al.}{2015}]{Shiokawa2015}
{Shiokawa} H.,  {Krolik} J.~H.,  {Cheng} R.~M.,  {Piran} T.,   {Noble} S.~C.,
  2015, \mn@doi [\apj] {10.1088/0004-637X/804/2/85}, \href
  {https://ui.adsabs.harvard.edu/abs/2015ApJ...804...85S} {804, 85}

\bibitem[\protect\citeauthoryear{{Sim}, {Miller}, {Long}, {Turner}  \&
  {Reeves}}{{Sim} et~al.}{2010a}]{Sim2010MNRAS.404.1369S}
{Sim} S.~A.,  {Miller} L.,  {Long} K.~S.,  {Turner} T.~J.,   {Reeves} J.~N.,
  2010a, \mn@doi [\mnras] {10.1111/j.1365-2966.2010.16396.x}, \href
  {https://ui.adsabs.harvard.edu/abs/2010MNRAS.404.1369S} {404, 1369}

\bibitem[\protect\citeauthoryear{{Sim}, {Proga}, {Miller}, {Long}  \&
  {Turner}}{{Sim} et~al.}{2010b}]{Sim2010MNRAS.408.1396S}
{Sim} S.~A.,  {Proga} D.,  {Miller} L.,  {Long} K.~S.,   {Turner} T.~J.,
  2010b, \mn@doi [\mnras] {10.1111/j.1365-2966.2010.17215.x}, \href
  {https://ui.adsabs.harvard.edu/abs/2010MNRAS.408.1396S} {408, 1396}

\bibitem[\protect\citeauthoryear{{Steinberg} \& {Stone}}{{Steinberg} \&
  {Stone}}{2024}]{Steinberg2024}
{Steinberg} E.,  {Stone} N.~C.,  2024, \mn@doi [\nat]
  {10.1038/s41586-023-06875-y}, \href
  {https://ui.adsabs.harvard.edu/abs/2024Natur.625..463S} {625, 463}

\bibitem[\protect\citeauthoryear{{Steinberg}, {Coughlin}, {Stone}  \&
  {Metzger}}{{Steinberg} et~al.}{2019}]{Steinberg2019}
{Steinberg} E.,  {Coughlin} E.~R.,  {Stone} N.~C.,   {Metzger} B.~D.,  2019,
  \mn@doi [\mnras] {10.1093/mnrasl/slz048}, \href
  {https://ui.adsabs.harvard.edu/abs/2019MNRAS.485L.146S} {485, L146}

\bibitem[\protect\citeauthoryear{{Stone}, {Tomida}, {White}  \&
  {Felker}}{{Stone} et~al.}{2020}]{Stone2020}
{Stone} J.~M.,  {Tomida} K.,  {White} C.~J.,   {Felker} K.~G.,  2020, \mn@doi
  [\apjs] {10.3847/1538-4365/ab929b}, \href
  {https://ui.adsabs.harvard.edu/abs/2020ApJS..249....4S} {249, 4}

\bibitem[\protect\citeauthoryear{{Strubbe} \& {Quataert}}{{Strubbe} \&
  {Quataert}}{2009}]{Strubbe2009}
{Strubbe} L.~E.,  {Quataert} E.,  2009, \mn@doi [\mnras]
  {10.1111/j.1365-2966.2009.15599.x}, \href
  {https://ui.adsabs.harvard.edu/abs/2009MNRAS.400.2070S} {400, 2070}

\bibitem[\protect\citeauthoryear{{Taam}, {Liu}, {Yuan}  \& {Qiao}}{{Taam}
  et~al.}{2012}]{Taam2012}
{Taam} R.~E.,  {Liu} B.~F.,  {Yuan} W.,   {Qiao} E.,  2012, \mn@doi [\apj]
  {10.1088/0004-637X/759/1/65}, \href
  {http://ads.bao.ac.cn/abs/2012ApJ...759...65T} {759, 65}

\bibitem[\protect\citeauthoryear{{Thomsen}, {Kwan}, {Dai}, {Wu}, {Roth}  \&
  {Ramirez-Ruiz}}{{Thomsen} et~al.}{2022}]{Thomsen2022}
{Thomsen} L.~L.,  {Kwan} T.~M.,  {Dai} L.,  {Wu} S.~C.,  {Roth} N.,
  {Ramirez-Ruiz} E.,  2022, \mn@doi [\apjl] {10.3847/2041-8213/ac911f}, \href
  {https://ui.adsabs.harvard.edu/abs/2022ApJ...937L..28T} {937, L28}

\bibitem[\protect\citeauthoryear{{Ulmer}}{{Ulmer}}{1999}]{Ulmer1999}
{Ulmer} A.,  1999, \mn@doi [\apj] {10.1086/306909}, \href
  {https://ui.adsabs.harvard.edu/abs/1999ApJ...514..180U} {514, 180}

\bibitem[\protect\citeauthoryear{{Vink{\'o}} et~al.,}{{Vink{\'o}}
  et~al.}{2015}]{Vinko2015}
{Vink{\'o}} J.,  et~al., 2015, \mn@doi [\apj] {10.1088/0004-637X/798/1/12},
  \href {https://ui.adsabs.harvard.edu/abs/2015ApJ...798...12V} {798, 12}

\bibitem[\protect\citeauthoryear{{Wu} \& {Gu}}{{Wu} \& {Gu}}{2008}]{Wu2008}
{Wu} Q.,  {Gu} M.,  2008, \mn@doi [\apj] {10.1086/588187}, \href
  {http://adsabs.harvard.edu/abs/2008ApJ...682..212W} {682, 212}

\bibitem[\protect\citeauthoryear{{Yuan} \& {Narayan}}{{Yuan} \&
  {Narayan}}{2014}]{Yuan2014}
{Yuan} F.,  {Narayan} R.,  2014, \mn@doi [\araa]
  {10.1146/annurev-astro-082812-141003}, \href
  {http://adsabs.harvard.edu/abs/2014ARA%26A..52..529Y} {52, 529}

\bibitem[\protect\citeauthoryear{development team}{development
  team}{2021}]{athena21}
development team A.,  2021, {PrincetonUniversity/athena-public-version:
  Athena++ v21.0}, \mn@doi{10.5281/zenodo.4455880}, \url
  {https://doi.org/10.5281/zenodo.4455880}

\bibitem[\protect\citeauthoryear{{van Velzen} et~al.,}{{van Velzen}
  et~al.}{2011}]{vanVelzen2011a}
{van Velzen} S.,  et~al., 2011, \mn@doi [\apj] {10.1088/0004-637X/741/2/73},
  \href {https://ui.adsabs.harvard.edu/abs/2011ApJ...741...73V} {741, 73}

\bibitem[\protect\citeauthoryear{{van Velzen} et~al.,}{{van Velzen}
  et~al.}{2016}]{vanVelzen2016}
{van Velzen} S.,  et~al., 2016, \mn@doi [Science] {10.1126/science.aad1182},
  \href {https://ui.adsabs.harvard.edu/abs/2016Sci...351...62V} {351, 62}

\bibitem[\protect\citeauthoryear{{van Velzen}, {Holoien}, {Onori}, {Hung}  \&
  {Arcavi}}{{van Velzen} et~al.}{2020}]{vanVelzen2020SSRv}
{van Velzen} S.,  {Holoien} T. W.~S.,  {Onori} F.,  {Hung} T.,   {Arcavi} I.,
  2020, \mn@doi [\ssr] {10.1007/s11214-020-00753-z}, \href
  {https://ui.adsabs.harvard.edu/abs/2020SSRv..216..124V} {216, 124}

\bibitem[\protect\citeauthoryear{{van Velzen} et~al.,}{{van Velzen}
  et~al.}{2021}]{vanVelzen2021}
{van Velzen} S.,  et~al., 2021, \mn@doi [\apj] {10.3847/1538-4357/abc258},
  \href {https://ui.adsabs.harvard.edu/abs/2021ApJ...908....4V} {908, 4}

\makeatother
\end{thebibliography}



\appendix
\section{Full results of the simulations and the emergent spectra}

\begin{figure*}
\includegraphics[width=40mm,height=53.1mm,angle=0.0]{8-1.png}
\includegraphics[width=40mm,height=53.1mm,angle=0.0]{8-2.png}
\includegraphics[width=40mm,height=53.1mm,angle=0.0]{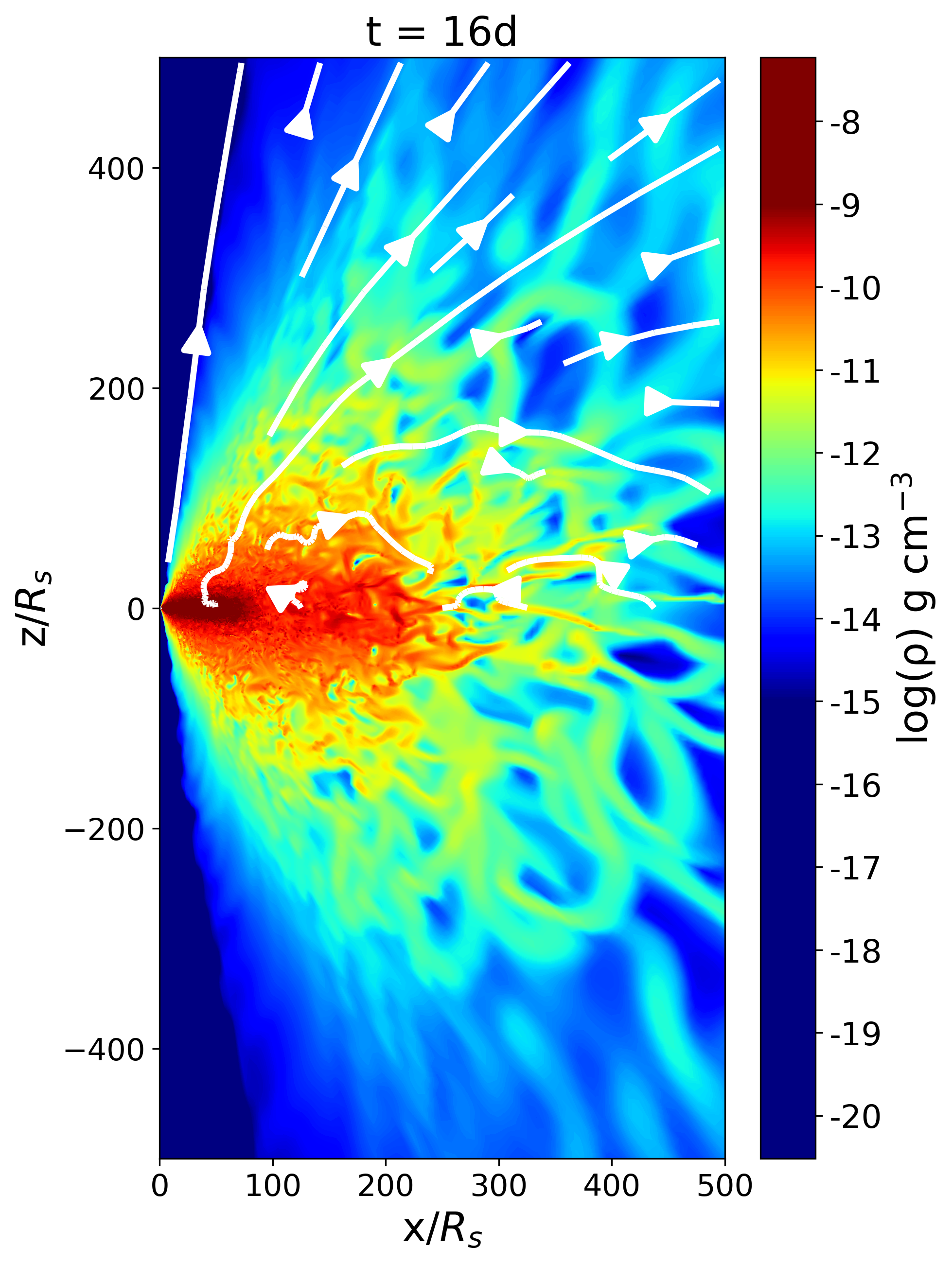}
\includegraphics[width=40mm,height=53.1mm,angle=0.0]{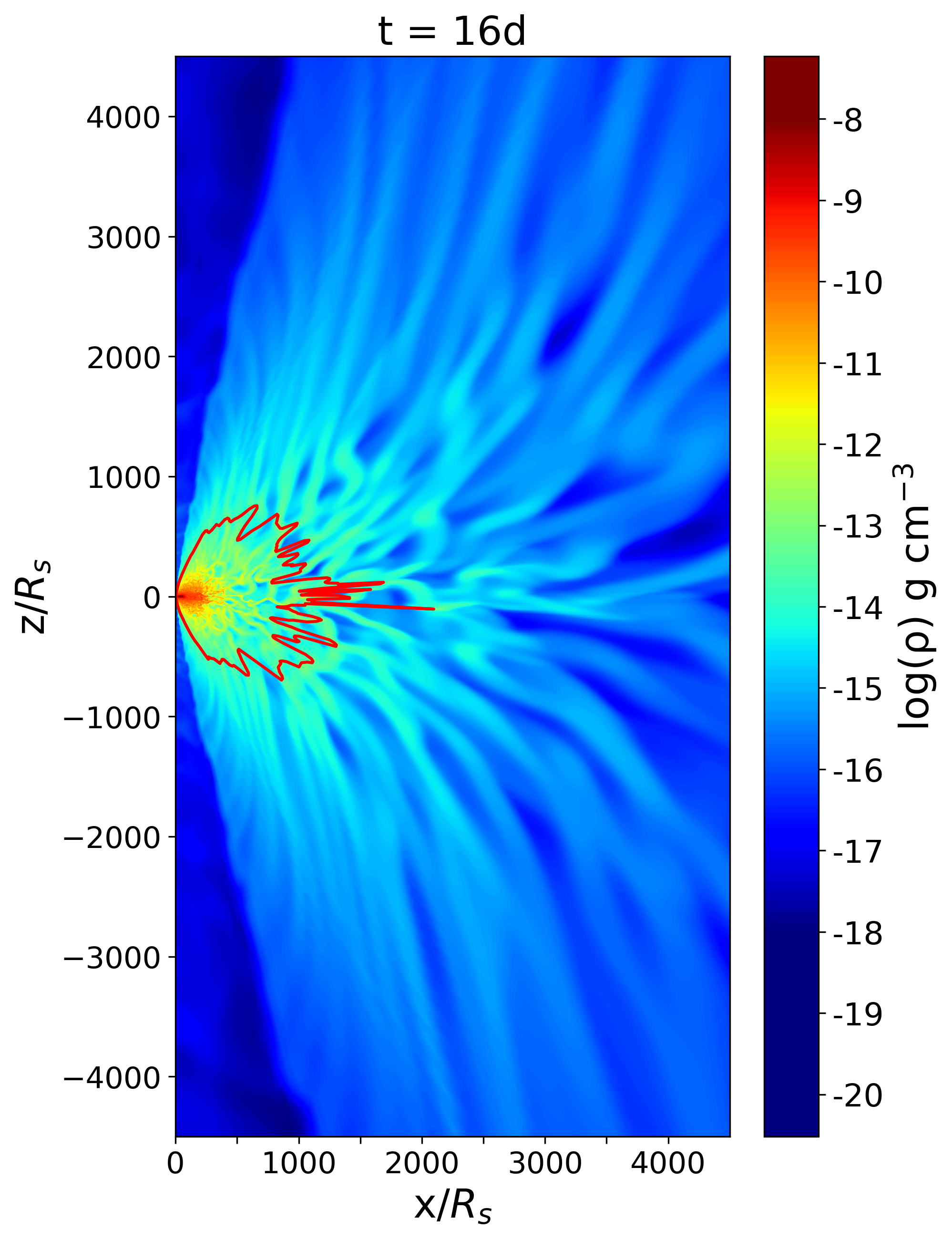}
\includegraphics[width=40mm,height=53.1mm,angle=0.0]{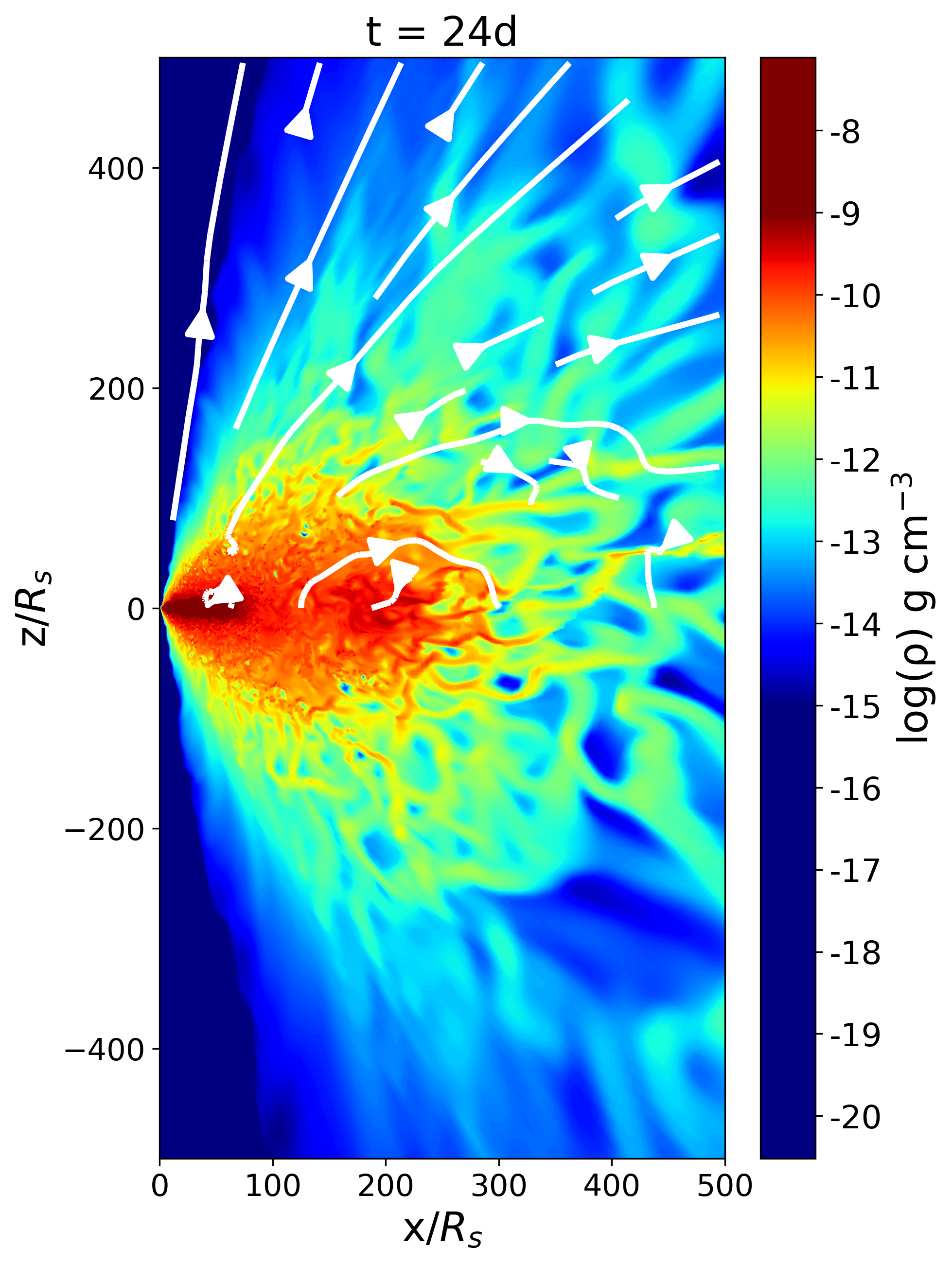}
\includegraphics[width=40mm,height=53.1mm,angle=0.0]{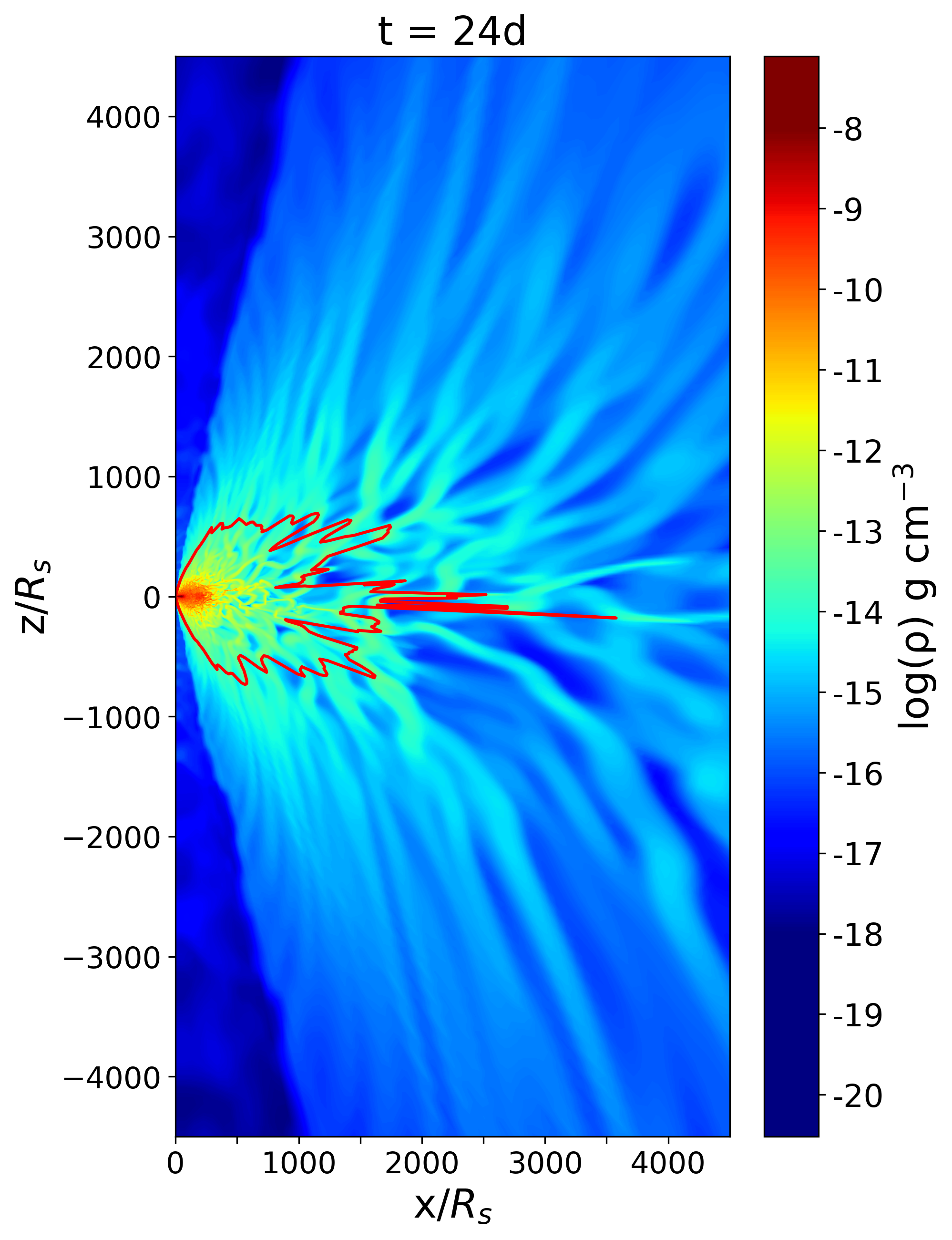}
\includegraphics[width=40mm,height=53.1mm,angle=0.0]{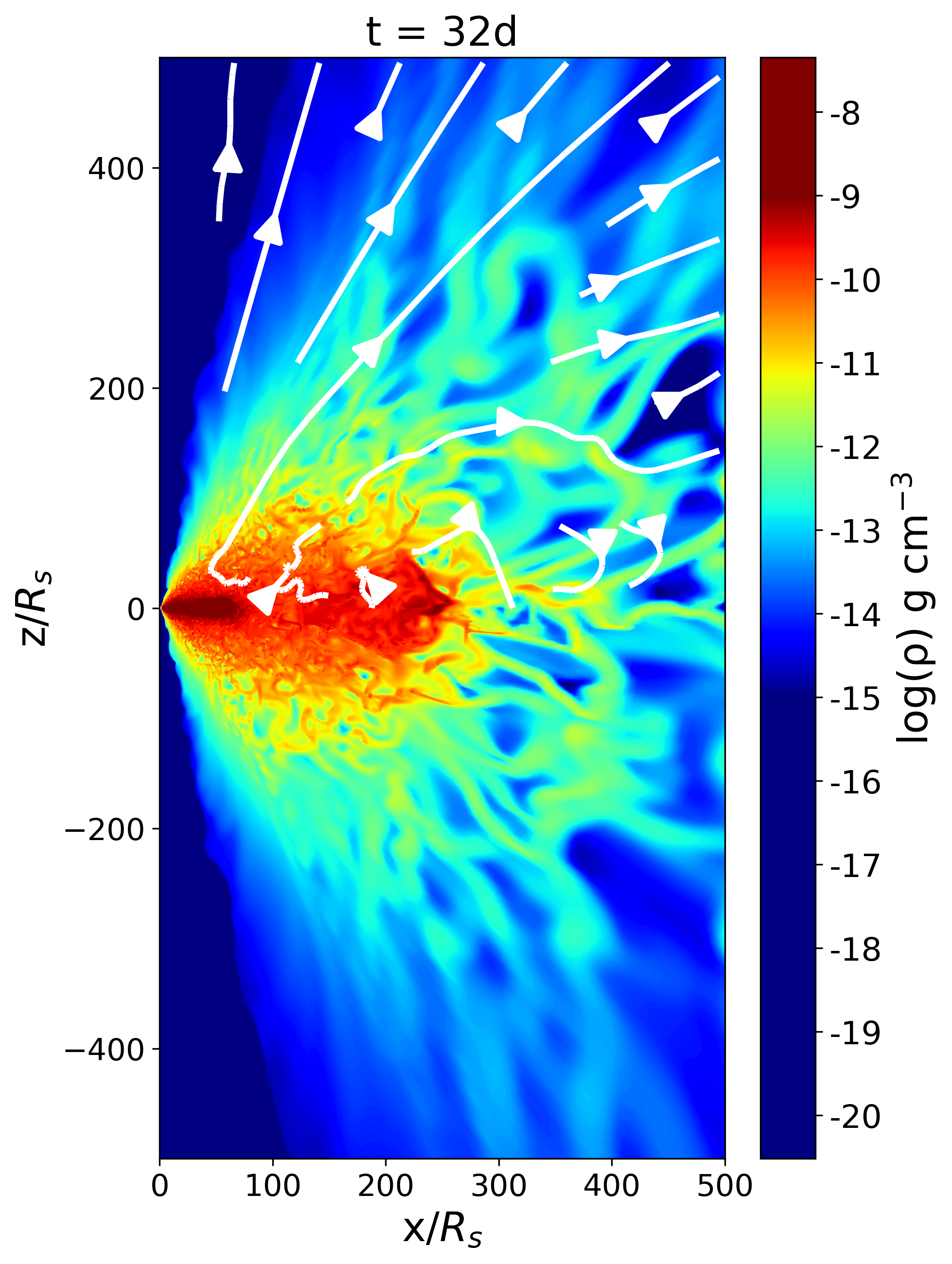}
\includegraphics[width=40mm,height=53.1mm,angle=0.0]{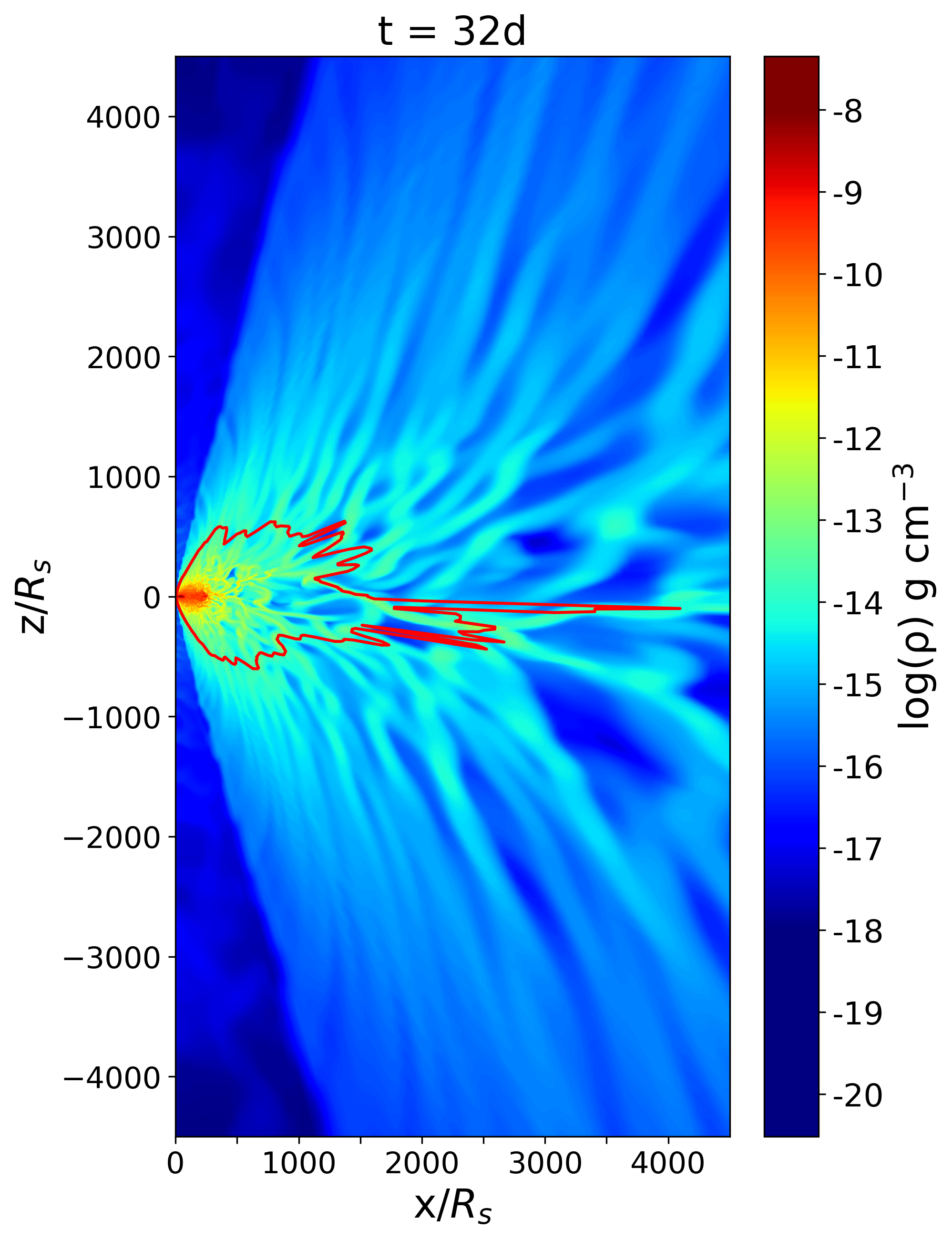}
\caption{\label{f:App_density} Snapshots of gas density at $t=8, 16, 24, 32$ day since the injection of matter at the circularization radius.
All the marks and notations in the figures are same with Fig. \ref{f:8-1-density}.}
\end{figure*}

\begin{figure*}
\includegraphics[width=40mm,height=53.1mm,angle=0.0]{v8-0.png}
\includegraphics[width=40mm,height=53.1mm,angle=0.0]{v8-1.png}
\includegraphics[width=40mm,height=53.1mm,angle=0.0]{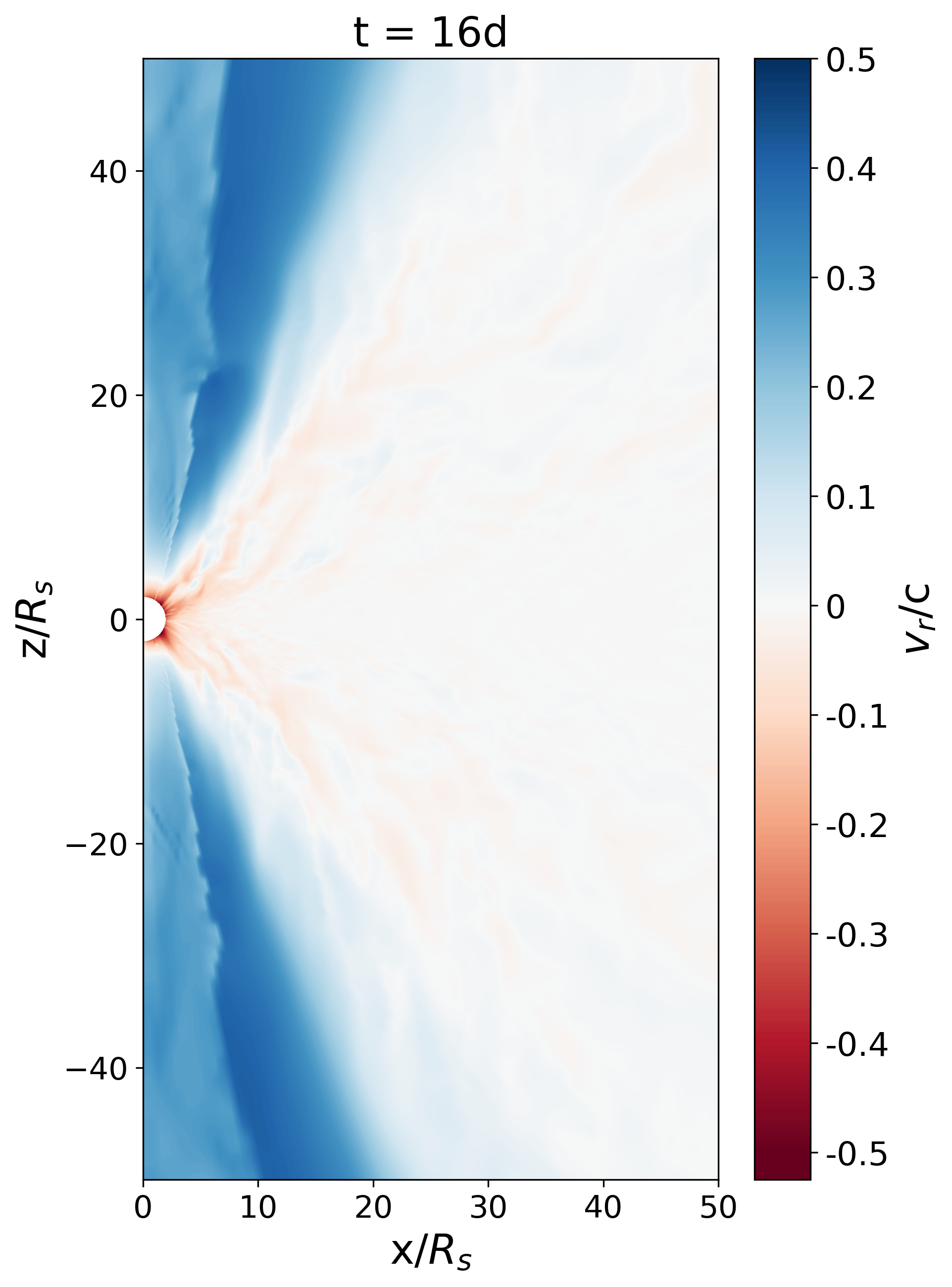}
\includegraphics[width=40mm,height=53.1mm,angle=0.0]{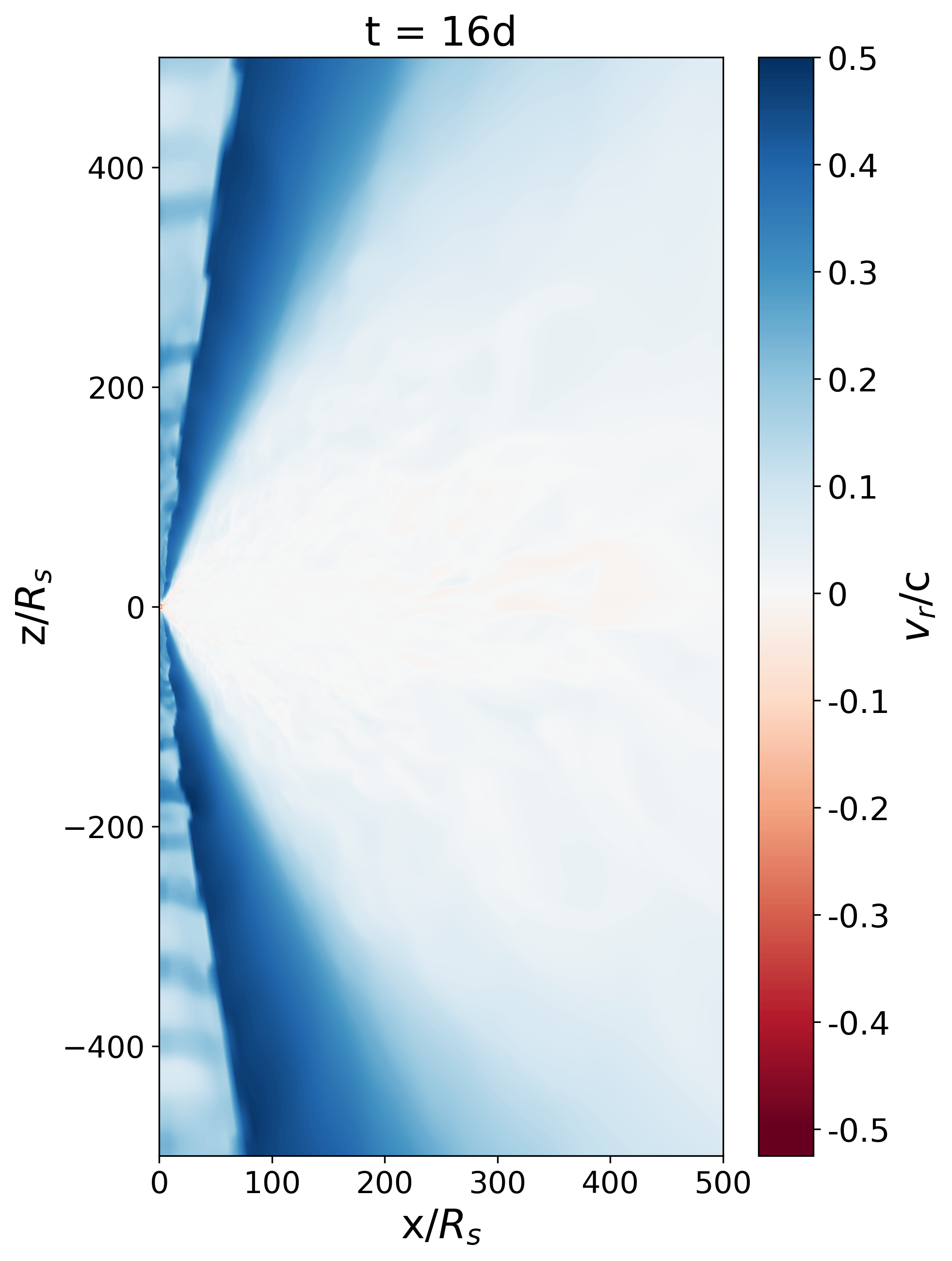}
\includegraphics[width=40mm,height=53.1mm,angle=0.0]{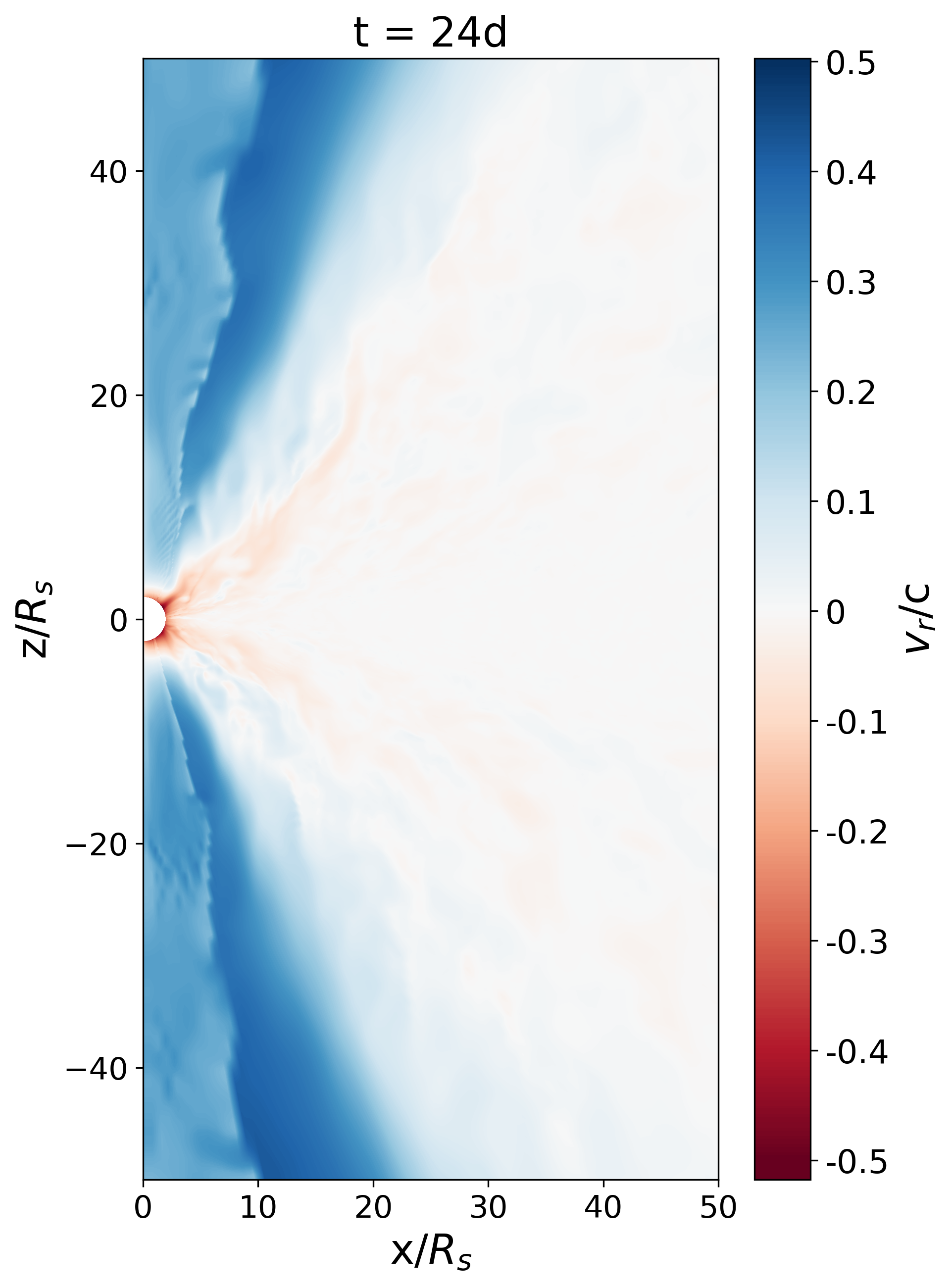}
\includegraphics[width=40mm,height=53.1mm,angle=0.0]{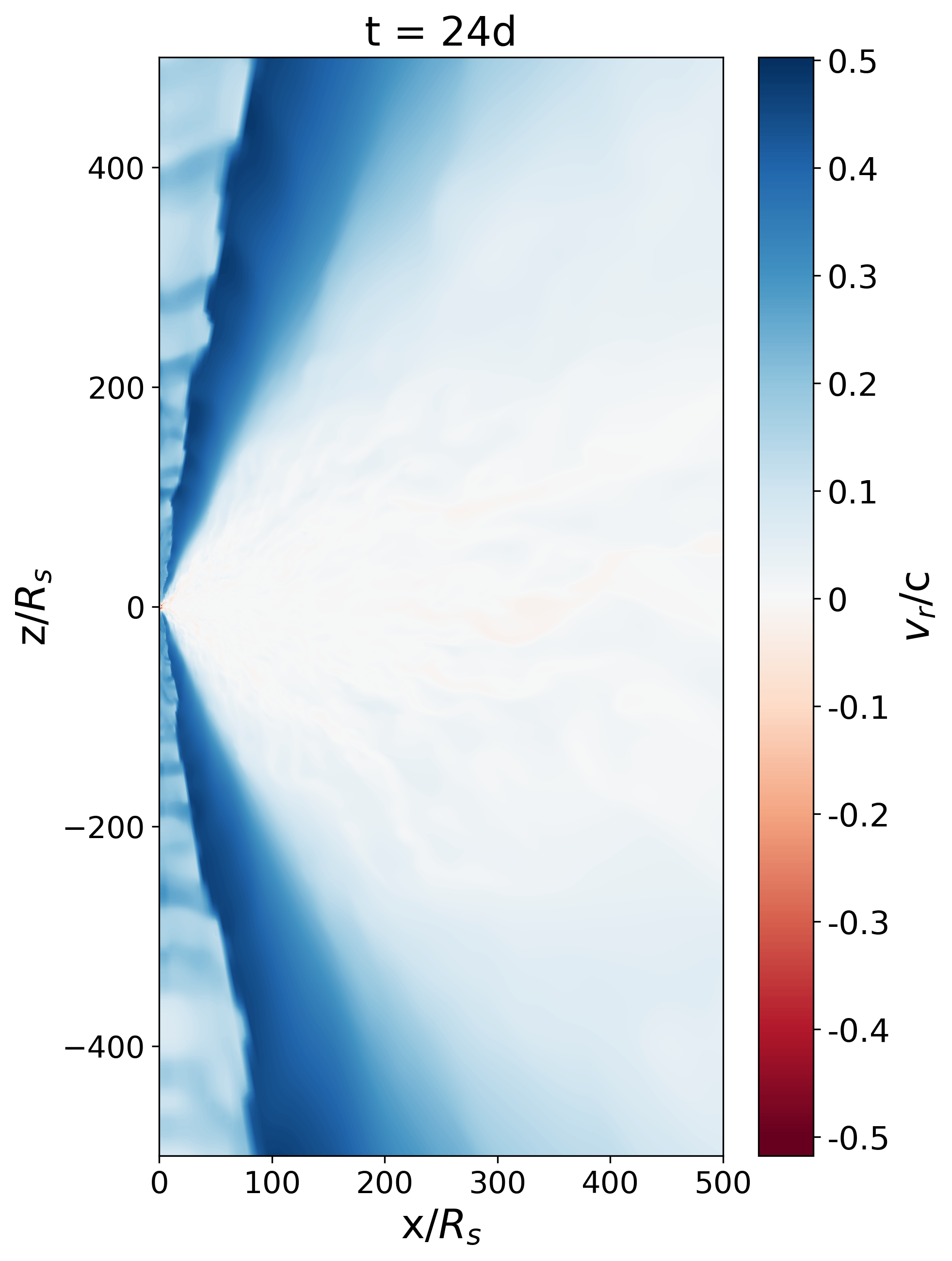}
\includegraphics[width=40mm,height=53.1mm,angle=0.0]{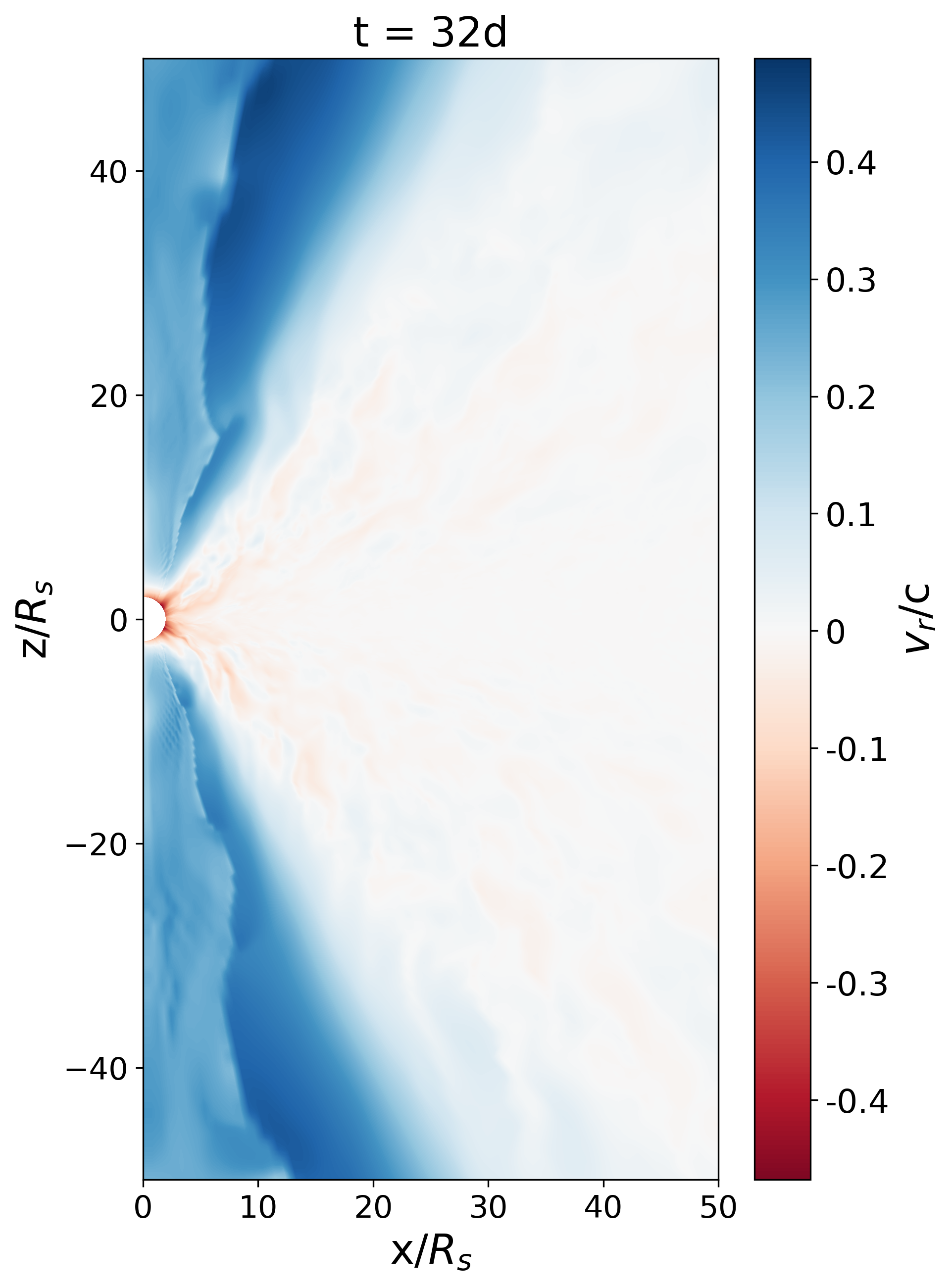}
\includegraphics[width=40mm,height=53.1mm,angle=0.0]{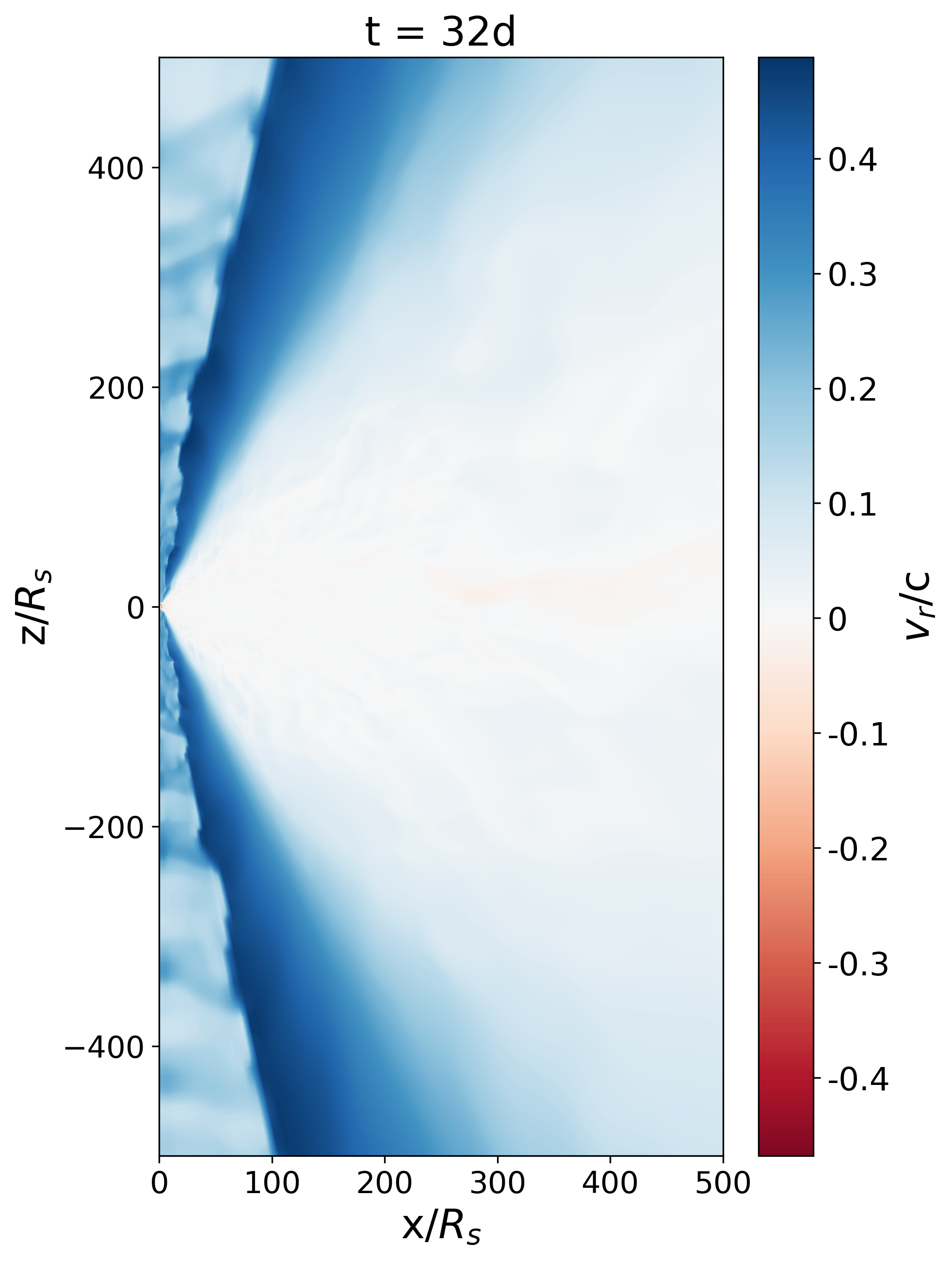}
\caption{\label{f:App_velocity} Snapshots of gas velocity in the radial direction at $t=8, 16, 24, 32$ day since the injection of matter at the circularization radius.
$v_{r}/c>0$ indicates outflow and $v_{r}/c<0$ indicates inflow. All the marks and notations in the figures are same with Fig. \ref{f:8-1-velocity}.}
\end{figure*}

\begin{figure*}
\includegraphics[width=40mm,height=53.1mm,angle=0.0]{Tr8-1.png}
\includegraphics[width=40mm,height=53.1mm,angle=0.0]{Tr8-2.png}
\includegraphics[width=40mm,height=53.1mm,angle=0.0]{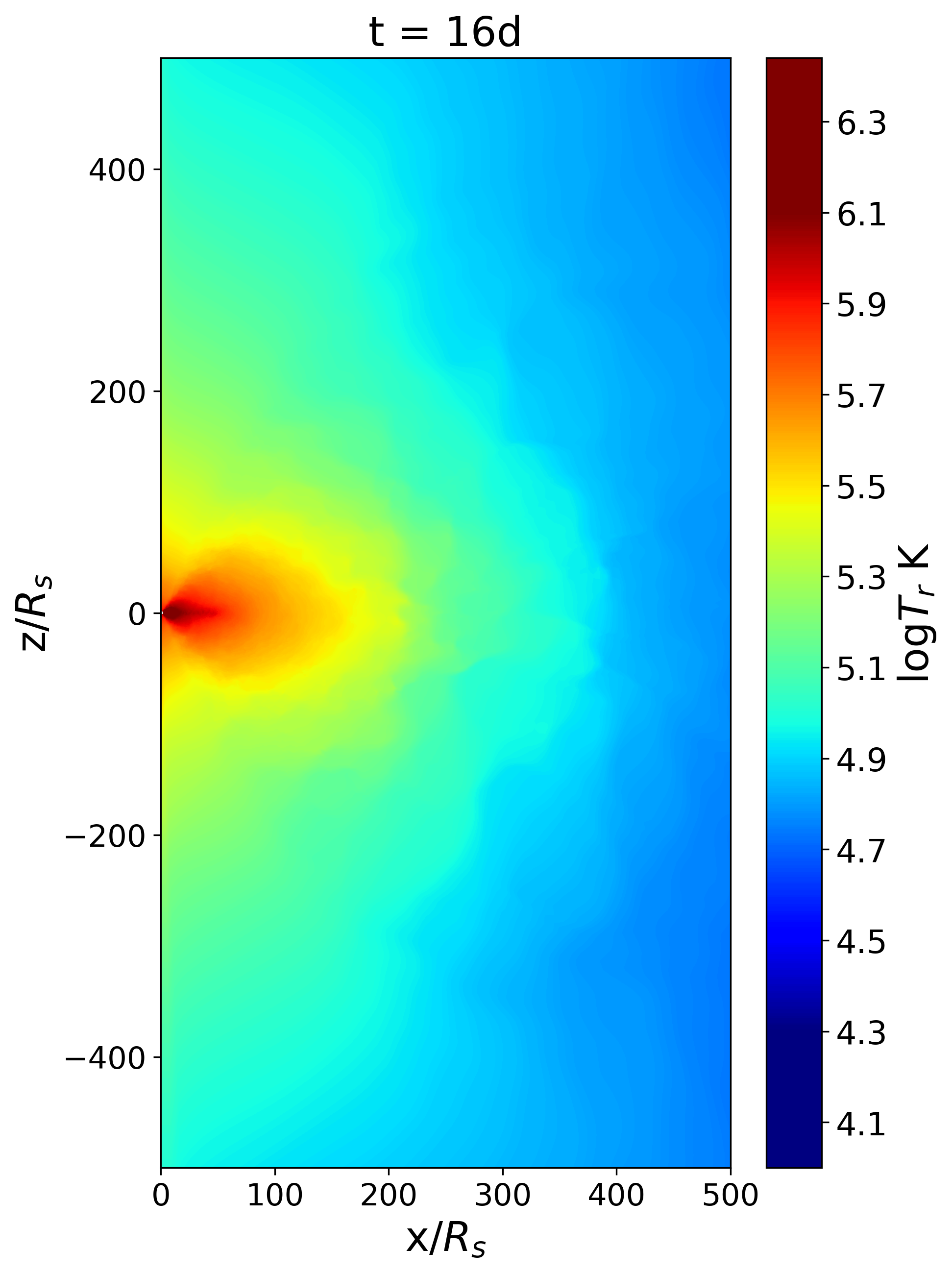}
\includegraphics[width=40mm,height=53.1mm,angle=0.0]{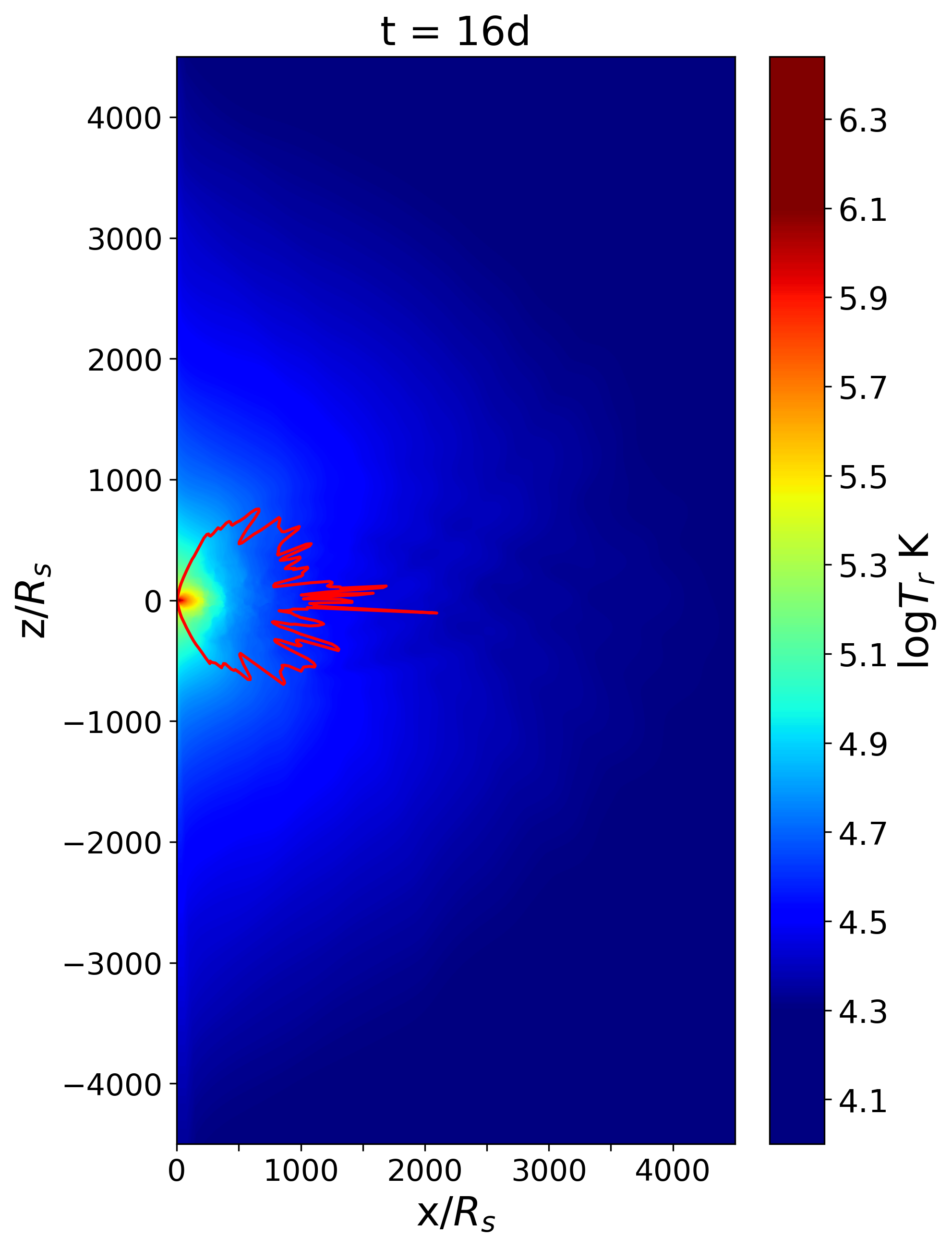}
\includegraphics[width=40mm,height=53.1mm,angle=0.0]{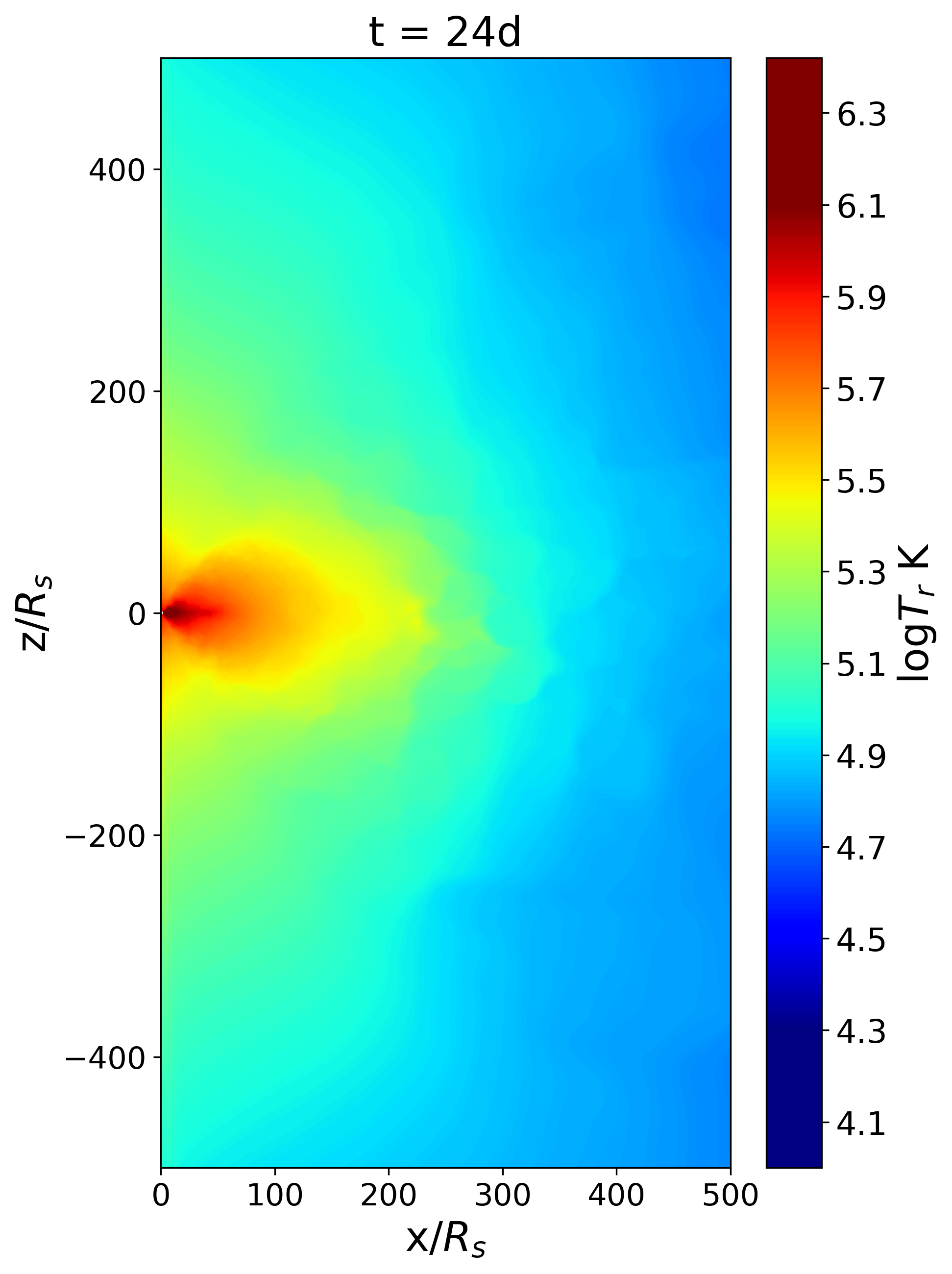}
\includegraphics[width=40mm,height=53.1mm,angle=0.0]{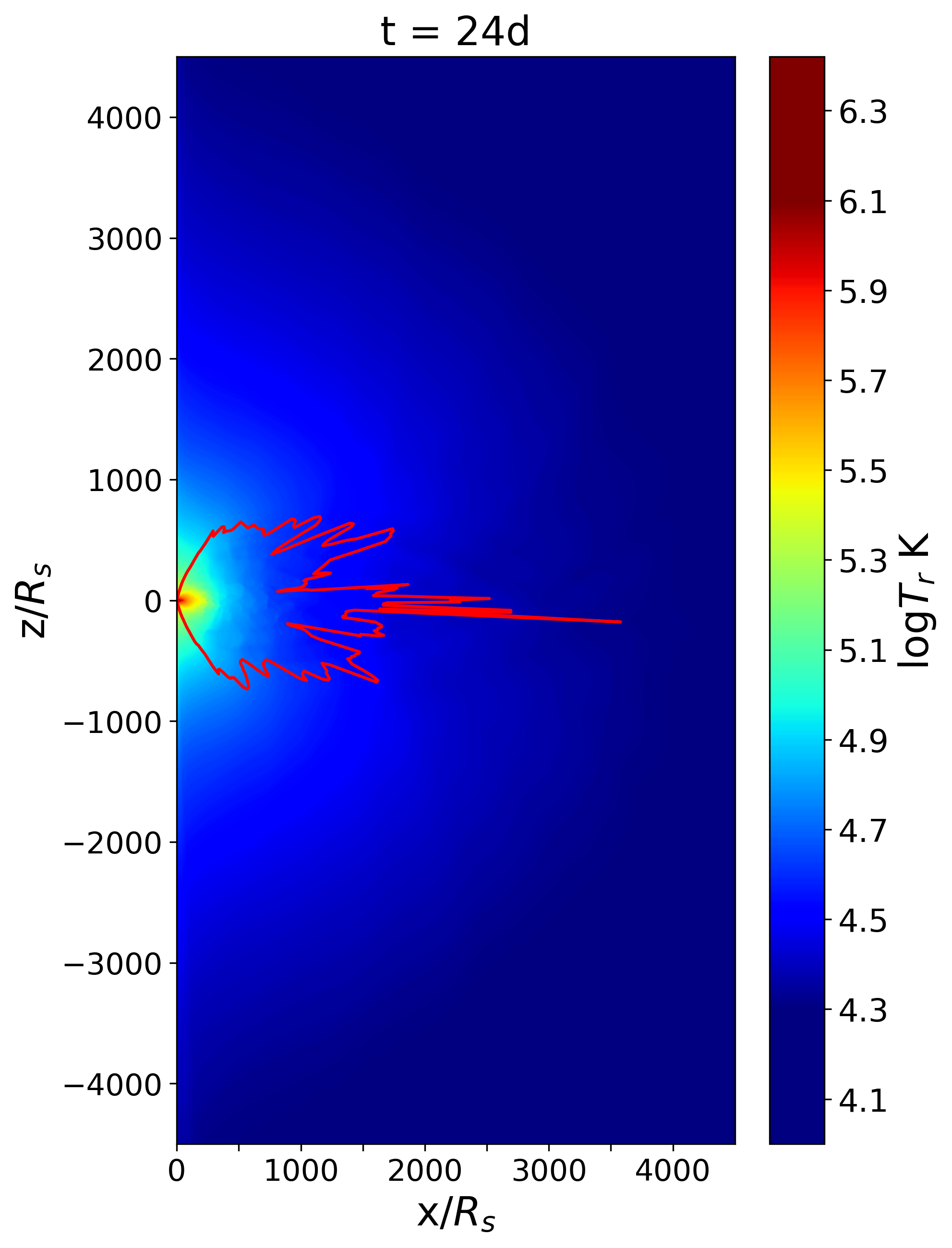}
\includegraphics[width=40mm,height=53.1mm,angle=0.0]{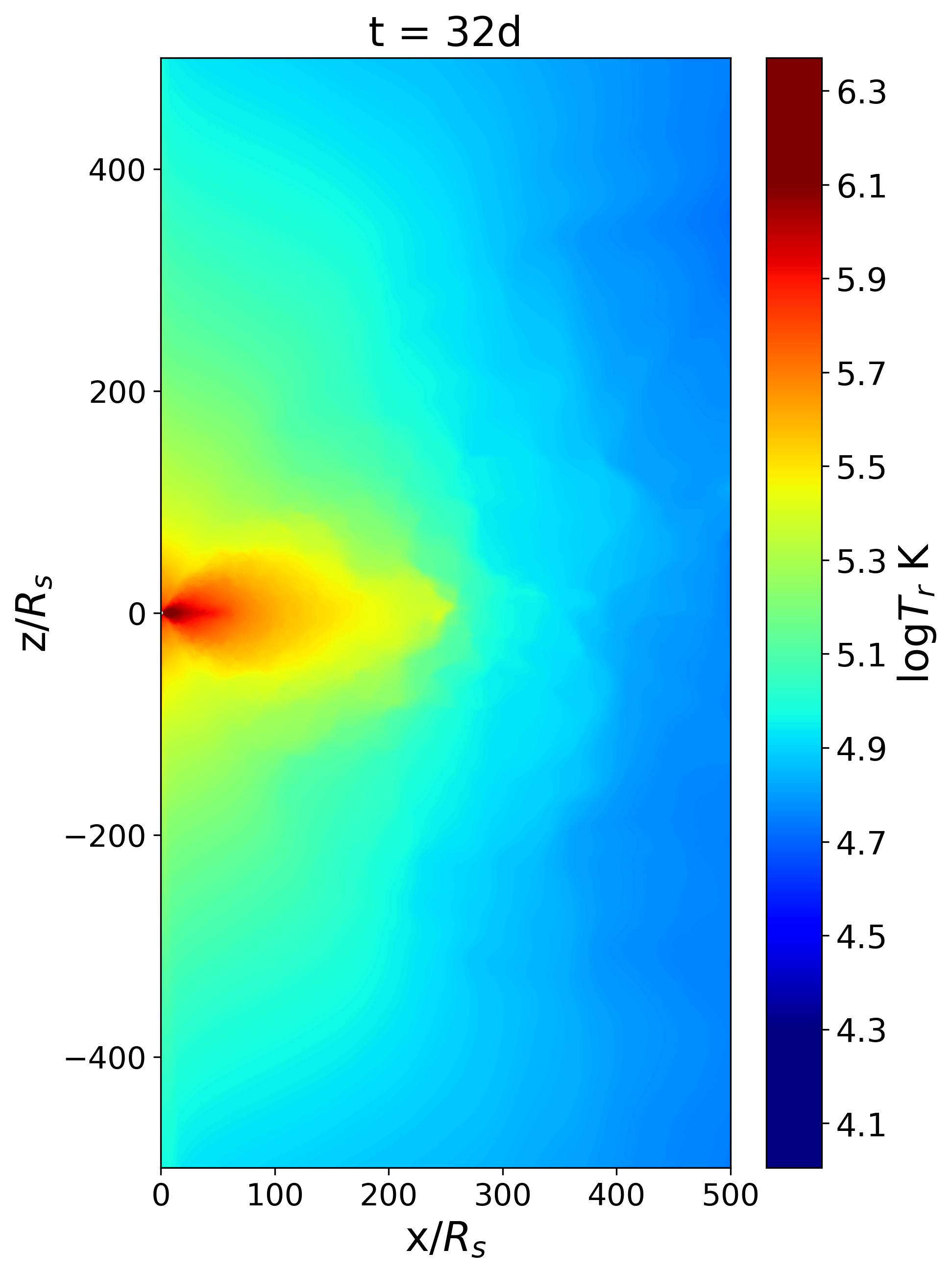}
\includegraphics[width=40mm,height=53.1mm,angle=0.0]{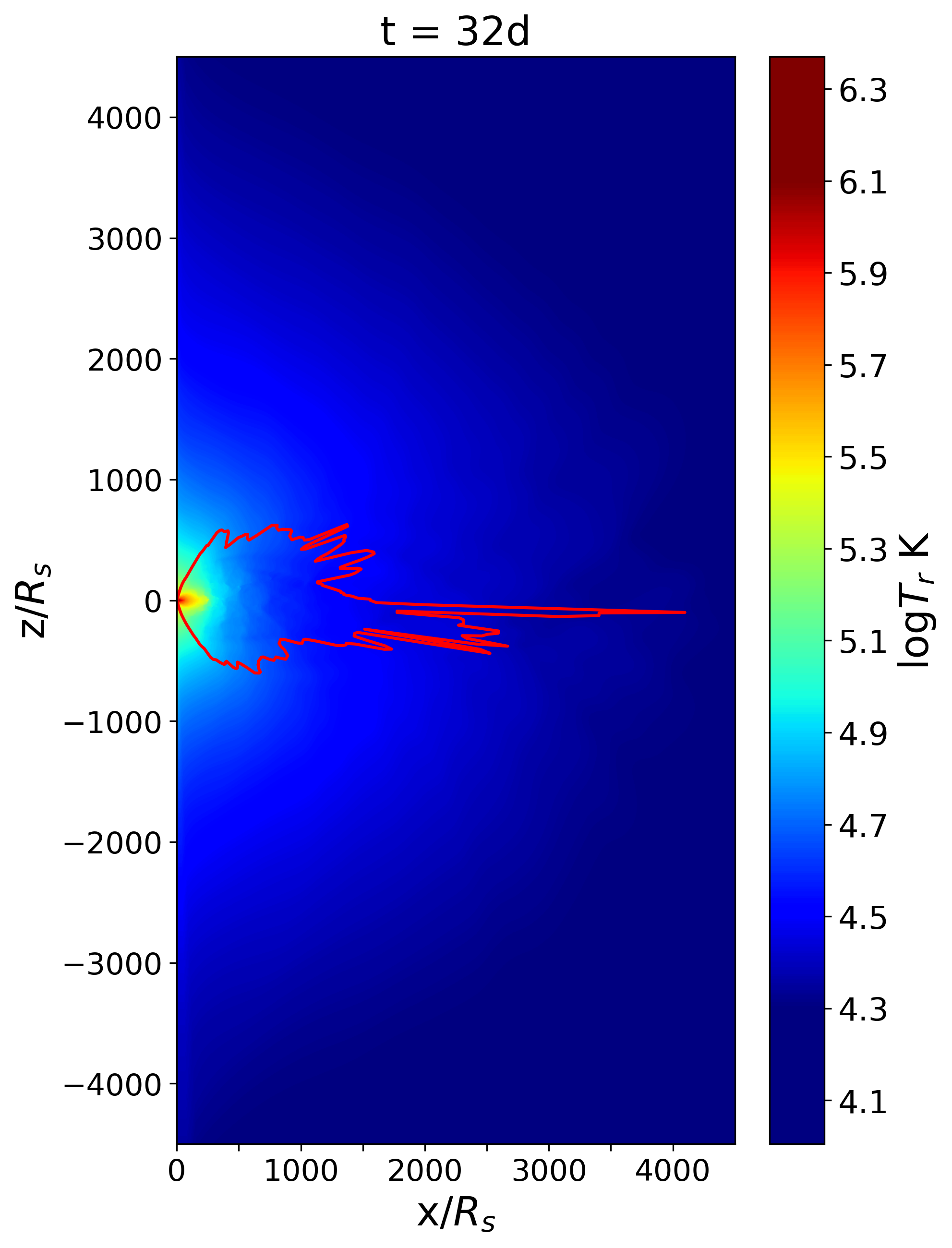}
\caption{\label{f:App_Tr} Snapshots of radiation temperature at $t=8, 16, 24, 32$ day since the injection of matter at the circularization radius.
 All the marks and notations in the figures are same with Fig. \ref{f:8-1-Tr}.}
\end{figure*}

\begin{figure*}
\includegraphics[width=80mm,height=52.6mm,angle=0.0]{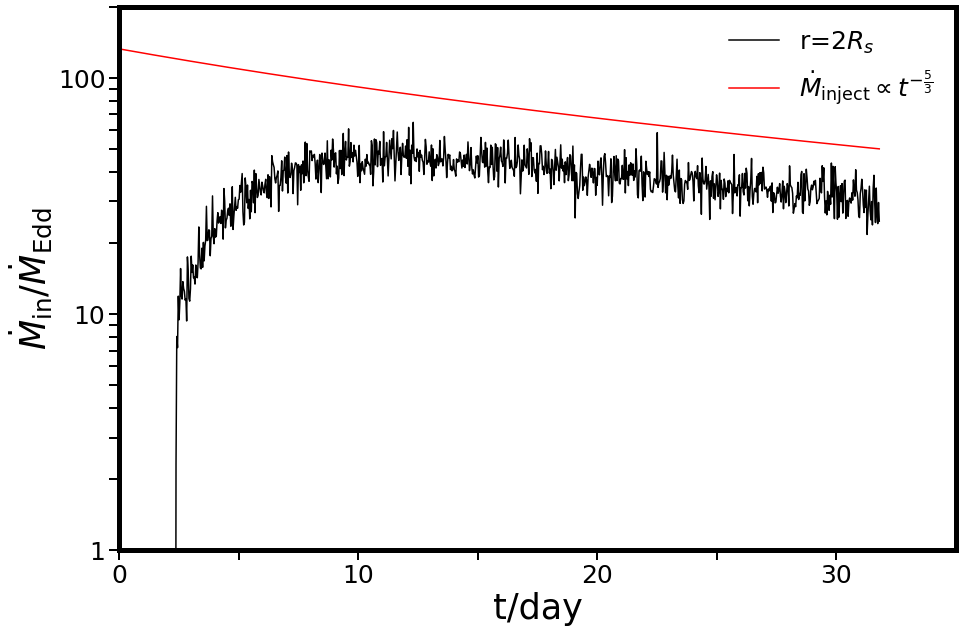}
\caption{\label{f:App_mdotin}  Mass inflow rate $\dot M_{\rm in}$ (scaled with $\dot M_{\rm Edd}$) as a function of $t$. The red solid line indicates
the mass injection rate, which has a form of $\dot M_{\rm inject}(t)=133.8\dot M_{\rm Edd}(1+t/41.3{\rm day})^{-5/3}$ as used in the simulation.
}
\end{figure*}


\begin{figure*}
\includegraphics[width=80mm,height=41.5mm,angle=0.0]{TDE_1e6_8d_spectra.jpg}
\includegraphics[width=80mm,height=41.5mm,angle=0.0]{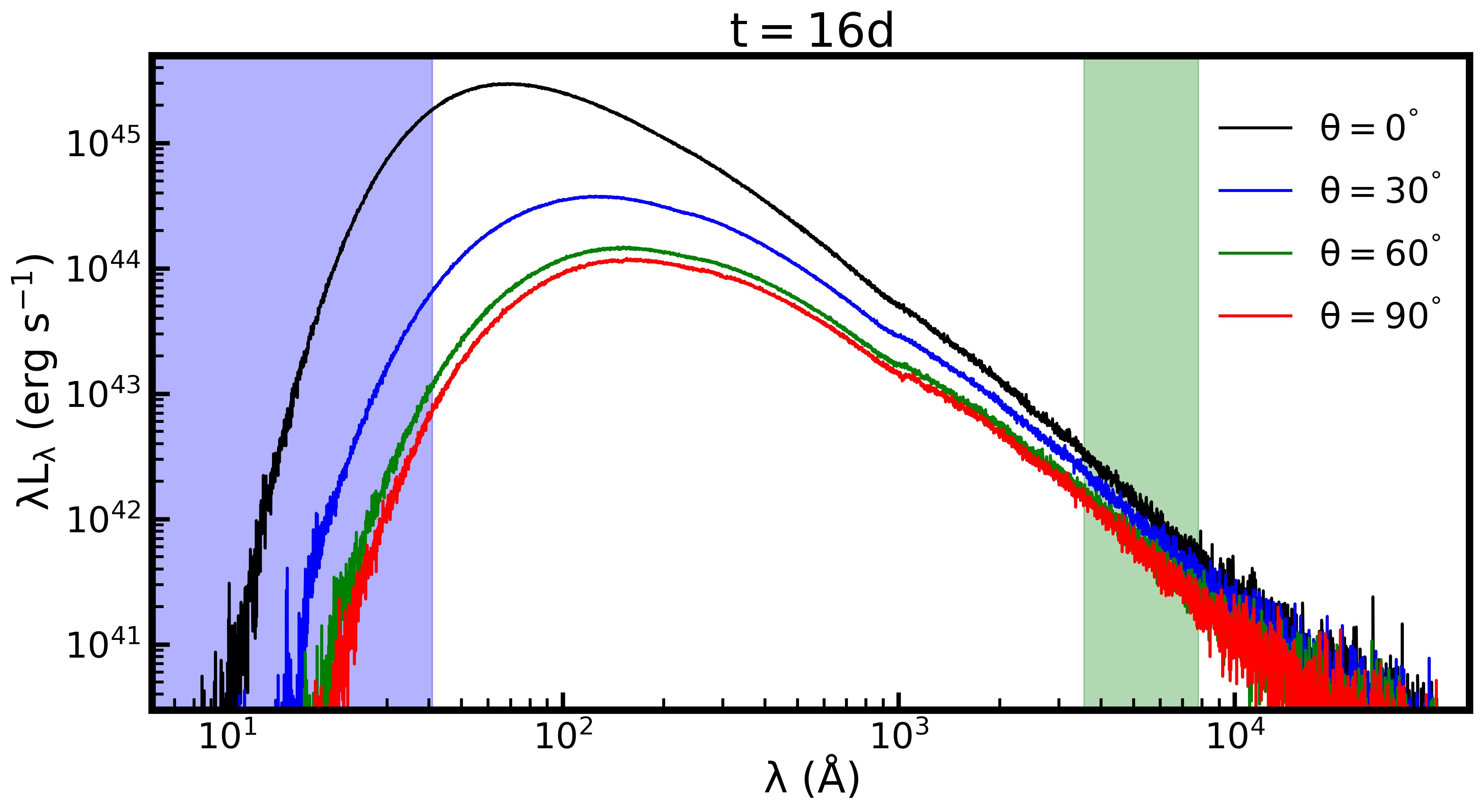}
\includegraphics[width=80mm,height=41.5mm,angle=0.0]{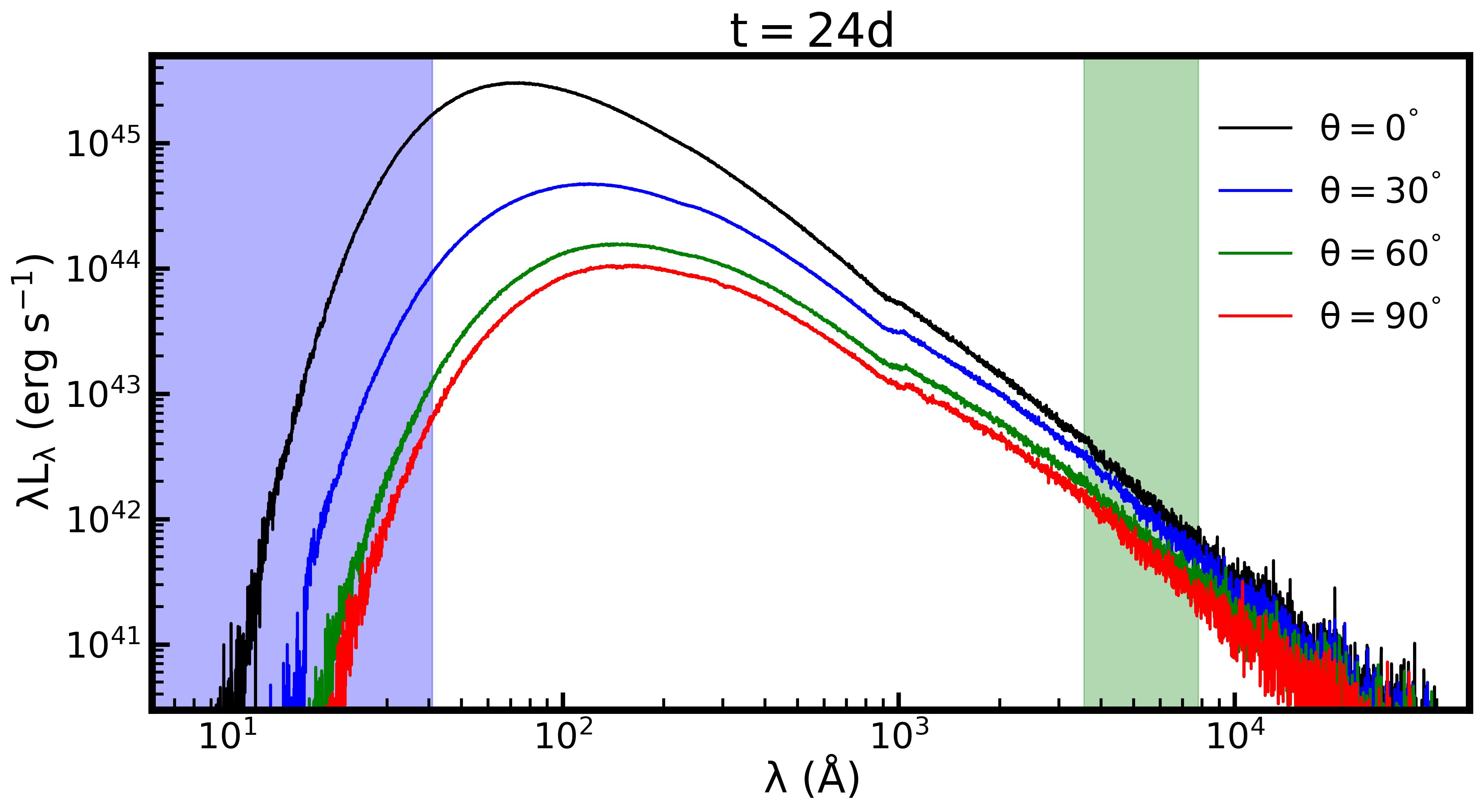}
\includegraphics[width=80mm,height=41.5mm,angle=0.0]{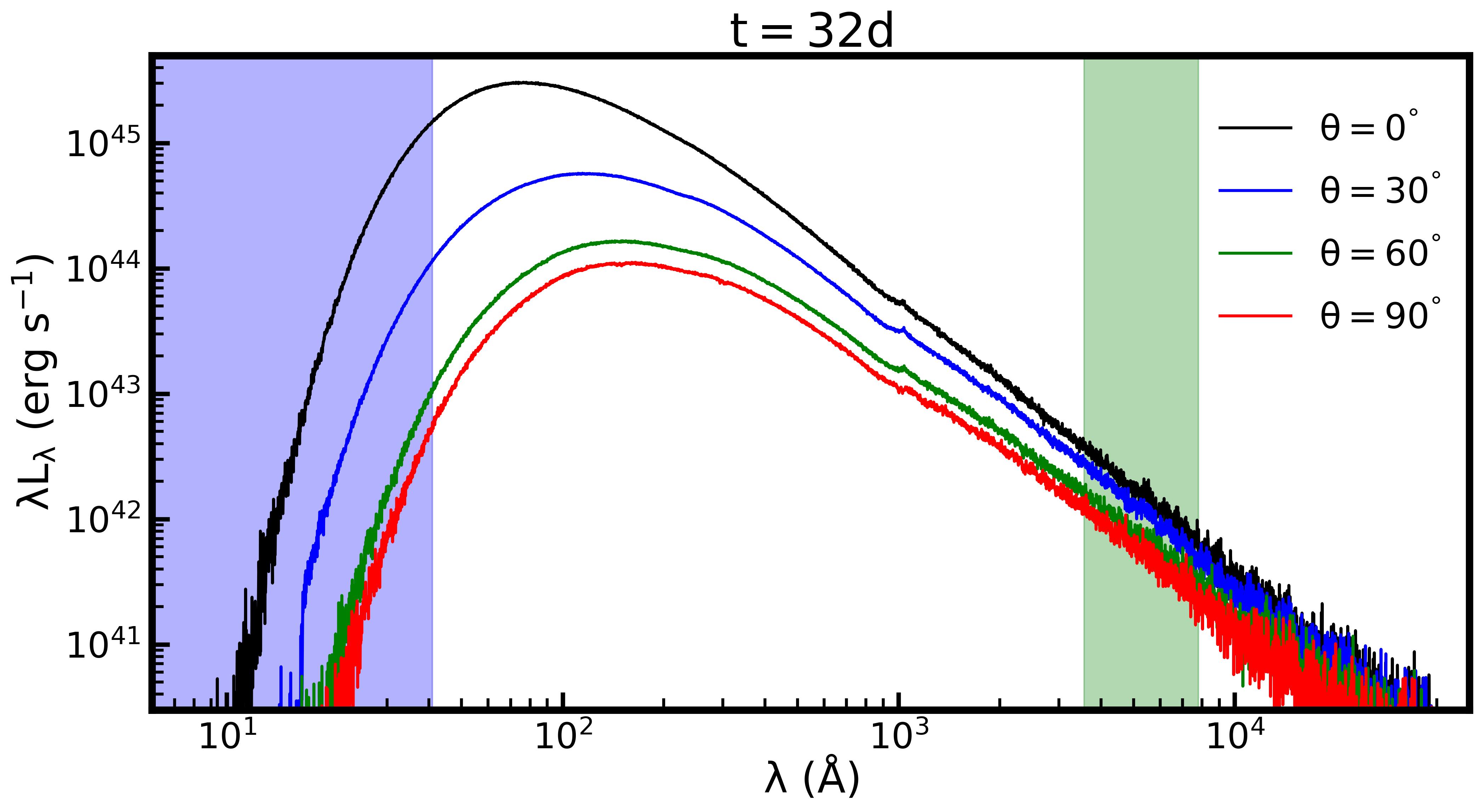}
\caption{\label{f:App_sp}Emergent spectra for different viewing angles,  i.e., $\rm \theta=0^{\rm o}, 30^{\rm o}, 60^{\rm o}$, $90^{\rm o}$ respectively
at $t=8,16, 24, 32$ day respectively since the injection of matter at the circularization radius. All the lines and the notations in the figures are same with Fig. \ref{f:sp8day}.
}
\end{figure*}


\begin{figure*}
\includegraphics[width=80mm,height=40.65mm,angle=0.0]{TDE_1e6_8d_spectra_fits.jpg}
\includegraphics[width=80mm,height=40.65mm,angle=0.0]{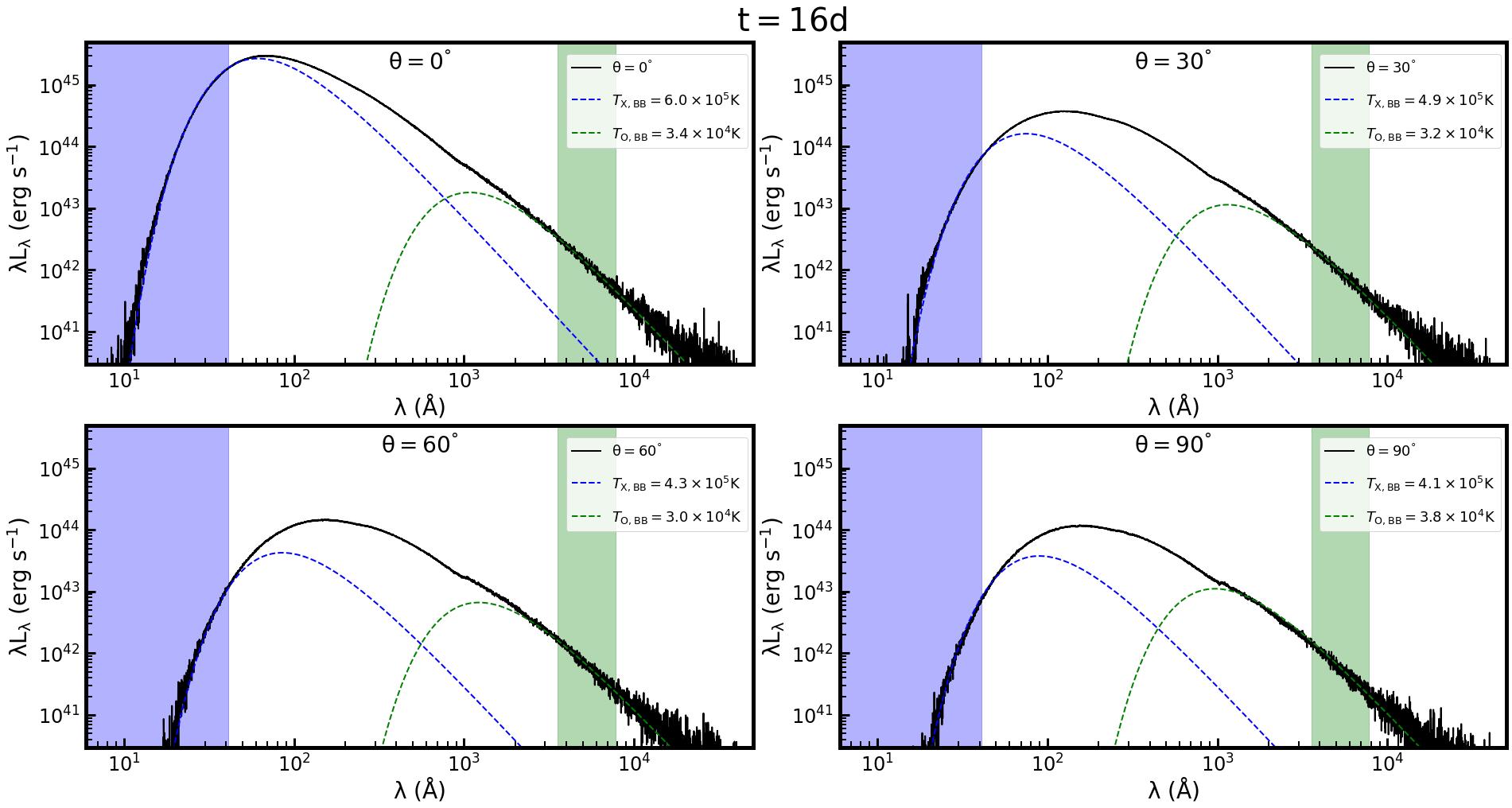}
\includegraphics[width=80mm,height=40.65mm,angle=0.0]{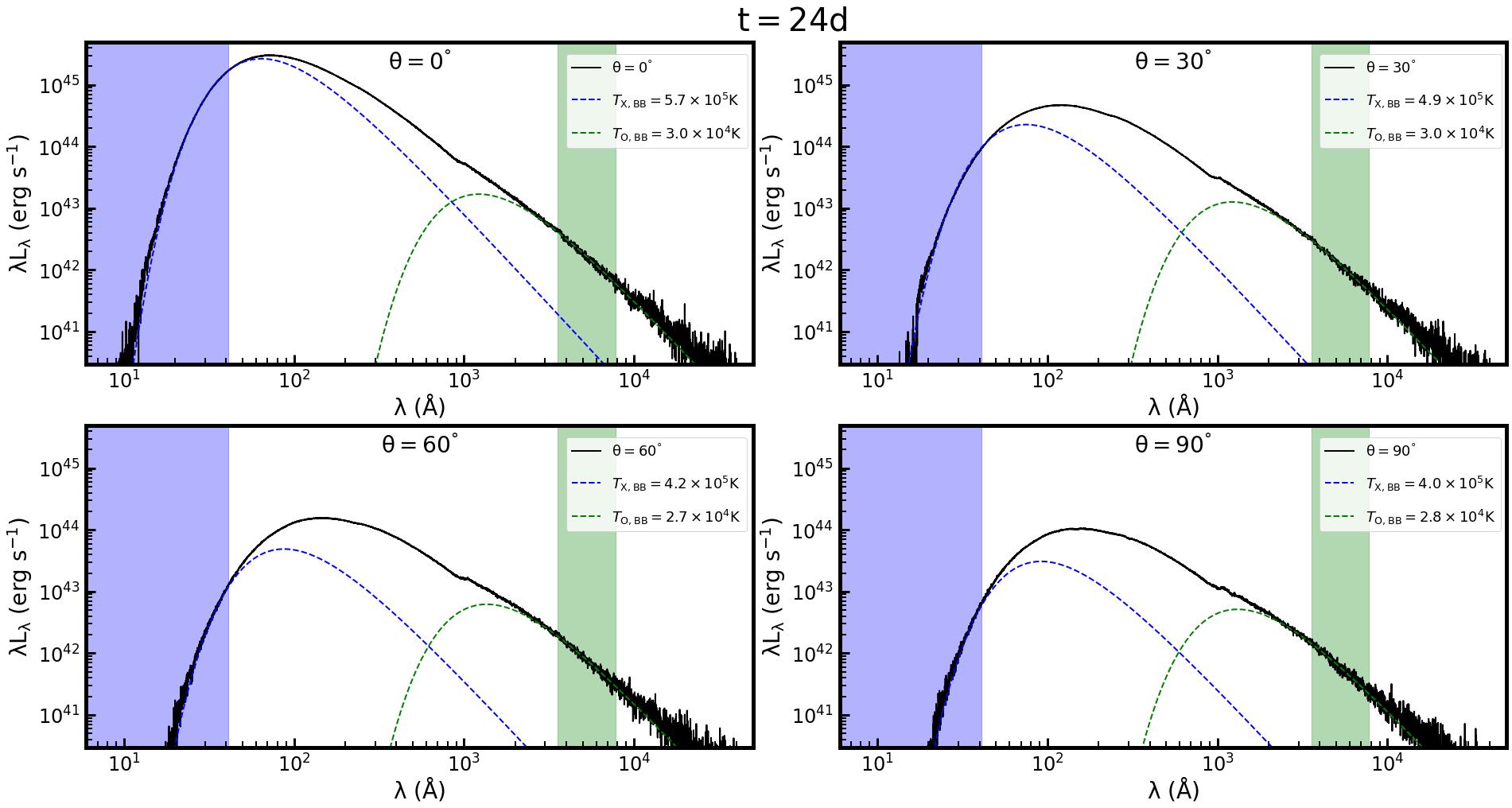}
\includegraphics[width=80mm,height=40.65mm,angle=0.0]{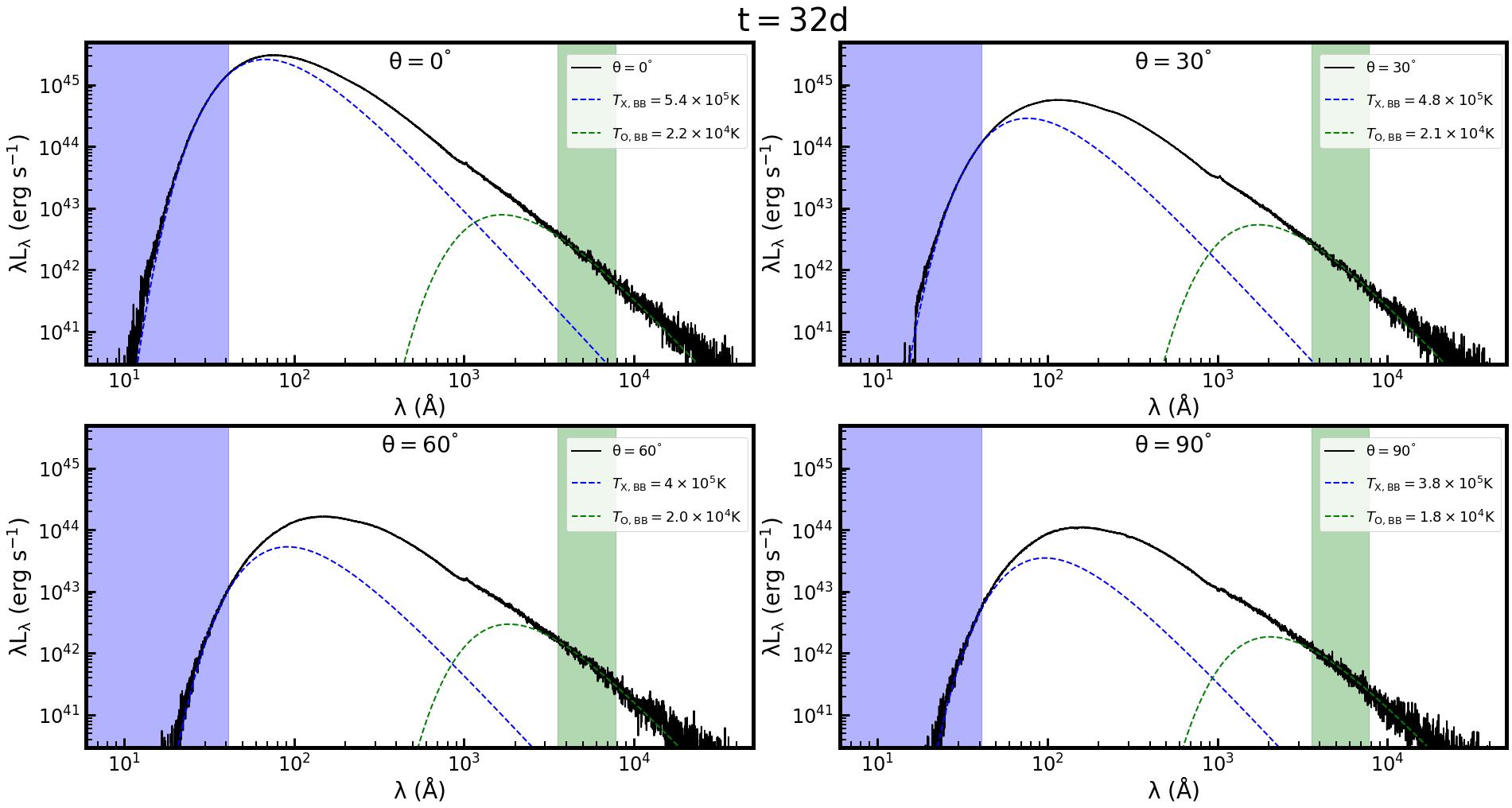}
\caption{\label{f:App_sp_fit} Emergent spectra for different viewing angles, i.e., $\rm \theta=0^{\rm o}, 30^{\rm o}, 60^{\rm o}$, $90^{\rm o}$ respectively, at $t=8,16, 24, 32$ day since the injection of matter 
at the circularization radius. All the lines and the notations in the figures are same with Fig. \ref{f:sp_fit}.
}
\end{figure*}

\begin{figure*}
\includegraphics[width=80mm,height=52.93mm,angle=0.0]{TDE_1e6_8d_properties.jpg}
\includegraphics[width=80mm,height=52.93mm,angle=0.0]{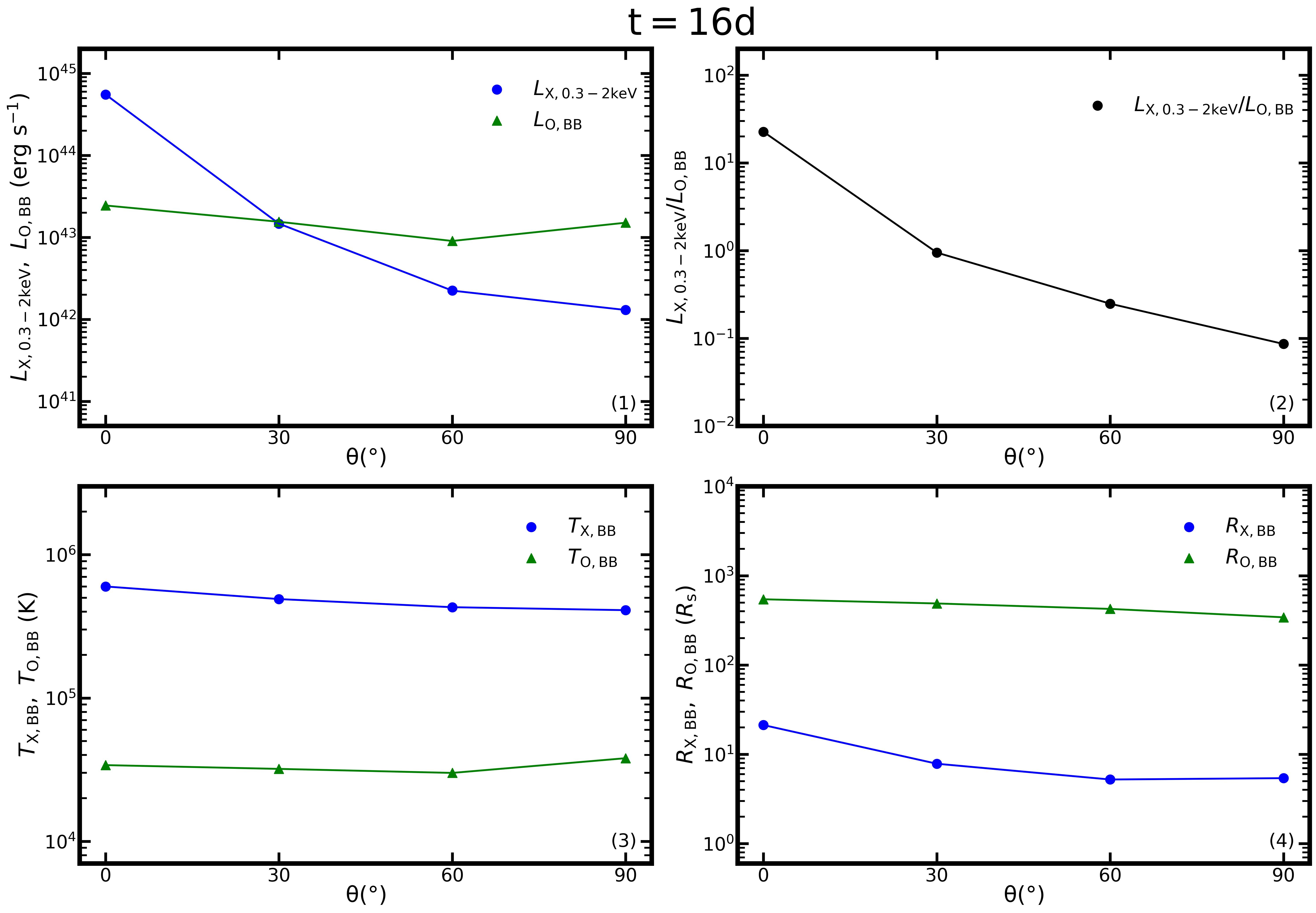}
\includegraphics[width=80mm,height=52.93mm,angle=0.0]{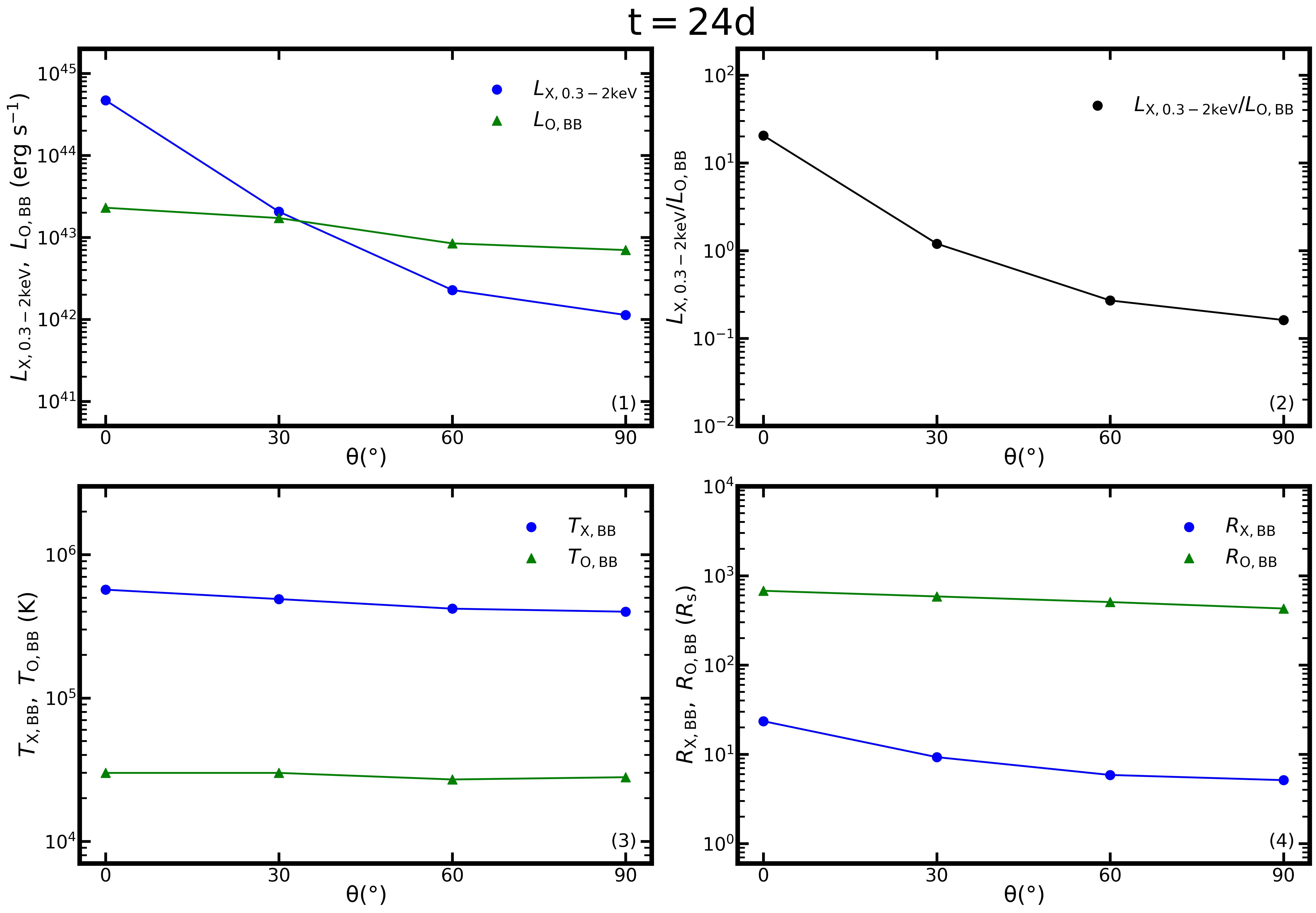}
\includegraphics[width=80mm,height=52.93mm,angle=0.0]{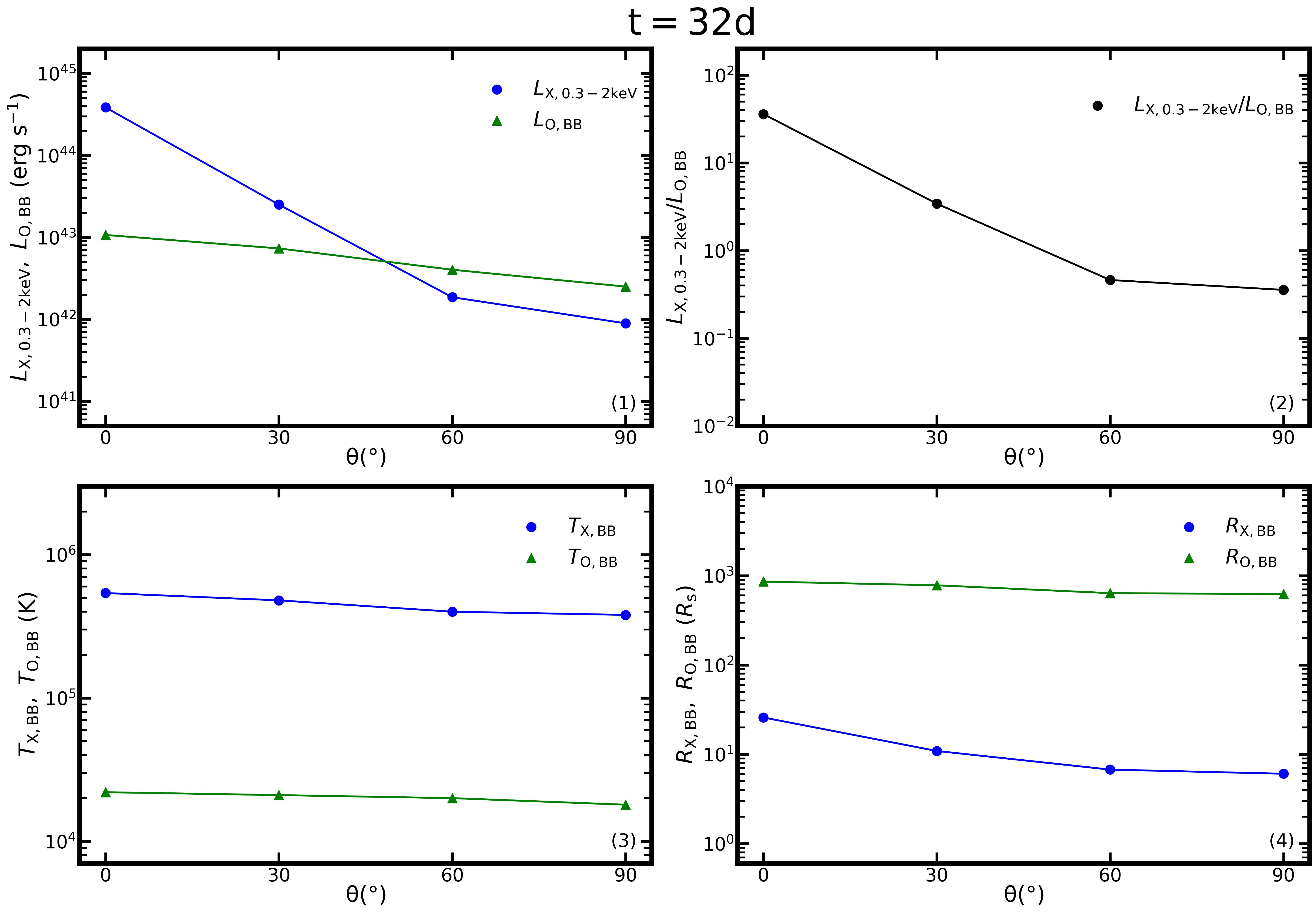}
\caption{\label{f:App_sp_parameter} Fitting parameters based on the emergent spectra as a function of viewing angle $\theta$
at $t=8, 16, 24, 32$ day since the injection of matter at the circularization radius. All the lines and the notations in the figures are same with Fig. \ref{f:sp_parameter}.
}
\end{figure*}


\bsp	
\label{lastpage}
\end{document}